\documentclass{jfm}
\usepackage{epstopdf, epsfig}
\usepackage{graphicx,footmisc}
\usepackage{subfig,color}
\usepackage{natbib,url}
\usepackage{lineno,amsmath}
\usepackage{float}
\usepackage{placeins}
\usepackage{changepage}
\usepackage{longtable}
\usepackage{xcolor}

\usepackage[counterclockwise, figuresright]{rotating}

\usepackage{adjustbox}

\newcommand*{\be}{\begin{equation}}
\newcommand*{\ee}{\end{equation}}

\def\begineq{\begin{equation}}
\def\endeq{\end{equation}}

\def\begineqn{\begin{equation*}}
\def\endeqn{\end{equation*}}

\def\beginar{\begin{eqnarray}}
\def\endar{\end{eqnarray}}
\def\beginarn{\begin{eqnarray*}}
\def\endarn{\end{eqnarray*}}

\def\lb{\left ( }
\def\rb{\right ) }

\def\ep{\epsilon}

\def\Rat{\widetilde{Ra}}
\def\Rmt{\widetilde{Rm}}

\def\ub{\mathbf{u}}

\def\Bb{\mathbf{B}}

\def\Bbp{\mathbf{B}^{\prime}}

\def\mBb{\overline{\bf B}}

\def\Jb{\mathbf{J}}

\def\Zb{\boldsymbol{\zeta}}

\def\dst{{\partial_t}}

\def\dsz{{\partial_z}}

\def\hz{{\bf\widehat z}}

\def\Ret{\widetilde{Re}}
\def\Rmt{\widetilde{Rm}}
\def\kt{\widetilde{k}}

\shorttitle{Rotating dynamos}
\shortauthor{M. Yan and M.A. Calkins}

\title{Asymptotic behaviour of rotating convection-driven dynamos in the plane layer geometry}

\author{Ming Yan$^1$ and
Michael A. Calkins$^1$
}

\affiliation{
$^1$Department of Physics, University of Colorado, Boulder, CO  80309, USA \\
}
\begin{document}

\maketitle

\begin{abstract}
\textcolor{black}{Dynamos driven by rotating convection in the plane layer geometry are investigated numerically for a range of Ekman number ($E$), magnetic Prandtl number ($Pm$) and Rayleigh number ($Ra$). The primary purpose of the investigation is to compare results of the simulations with previously developed asymptotic theory that is applicable in the limit of rapid rotation. We find that all of the simulations are in the quasi-geostrophic regime in which the Coriolis and pressure gradient forces are approximately balanced at leading order, whereas all other forces, including the Lorentz force, act as perturbations. Agreement between simulation output and asymptotic scalings for the energetics, flow speeds, magnetic field amplitude and length scales is found. The transition from large scale dynamos to small scale dynamos is well described by the magnetic Reynolds number based on the small convective length scale, $\widetilde{Rm}$, with large scale dynamos preferred when $\widetilde{Rm} \lesssim O(1)$. The magnitude of the large scale magnetic field is observed to saturate and become approximately constant with increasing Rayleigh number. Energy spectra show that all length scales present in the flow field and the small-scale magnetic field are consistent with a scaling of $E^{1/3}$, even in the turbulent regime. For a fixed value of $E$, we find that the viscous dissipation length scale is approximately constant over a broad range of $Ra$; the ohmic dissipation length scale is approximately constant within the large scale dynamo regime, but transitions to a $\widetilde{Rm}^{-1/2}$ scaling in the small scale dynamo regime.}
\end{abstract}

\section{Introduction}

%
%
%
%

The majority of solar system planets possess global scale magnetic fields. These fields are believed to be generated by the convective motion of electrically conducting fluid within the interior regions of the planets \citep{cJ11b}. For example, the geomagnetic field of the Earth is thought to originate in the liquid metal outer core where buoyancy-driven flows continuously generate electric currents and associated magnetic field \citep{pR13}. One particular physical ingredient that is thought to be critical for sustaining large, or global, scale magnetic field is the Coriolis force \citep{eP55,mS66a}. Although many previous investigations have shown the tendency for rotating convection to generate magnetic fields (beginning with \cite{sC72} and \cite{aS74}), it is still not completely understood how the various input parameters influence: (1) the characteristic length scales of the velocity and magnetic fields; (2) the strength of the resulting magnetic fields; and (3) the prevailing force balance that characterises the dynamics. \textcolor{black}{However, the asymptotic theory developed by \cite{mC15b} provides predictions for these various properties. In this regard, we utilize direct numerical simulation (DNS) in the plane layer geometry to better understand how rotation, buoyancy and the relative influence of diffusion influence system behavior. The simulation output is used to test various predictions from the asymptotic theory.}

Natural dynamo systems such as the Earth's outer core are characterised by several small (or large) physical parameters \citep{cJ11b,gS11}. In particular, the Ekman number, defined as $E = \nu/(\Omega H^2)$ (where $\nu$ is the kinematic viscosity, $\Omega$ is the rotation rate and $H$ is the characteristic length scale), is very small; estimates suggest $E =O(10^{-15})$ in the core. The Rossby number, representing the ratio of inertia to the Coriolis force, is also small in the core; $Ro = U/(\Omega H)=O( 10^{-6} )$, where $U$ is a characteristic speed. It is therefore of interest to understand how dynamos and rotating convection behave in the dual limit $(E , Ro) \rightarrow 0$. 

\textcolor{black}{Rotating convection has been investigated extensively in both the spherical and plane layer geometries, though computational restrictions prevent the use of realistic values of $E$. \textcolor{black}{Significant progress has been made toward understanding the asymptotic theory of plane layer rotating convection}; output from simulations of the fully nonlinear asymptotic model developed by \cite{kJ98a} agrees well with the corresponding output from DNS \citep{sS14,mP16}. In particular, this theory predicts that the flow will be geostrophically balanced to leading order -- i.e.~the Coriolis and pressure gradient forces should be approximately balanced and all other forces act as perturbations. \cite{aG21} explicitly computed the forces in plane layer rotating convection and have confirmed the dynamics are quasi-geostrophic (QG), provided the Rossby number remains small.} 

In comparison to the purely hydrodynamic rotating convection problem, the asymptotic behavior of the dynamo problem is less well understood. In particular, accurately diagnosing the force balance remains an ongoing effort, and to complicate matters it appears that the force balance depends on the geometry, as well as the particular length scale on which the dynamics are analyzed. Recent work in the spherical geometry points to a leading order geostrophic balance on the small-scale motions \citep{rY16,jA17,nS17,tS21} and a semi-magnetostrophic force balance on the large scales \citep{jA05,mC21,rO21}. Here the term `semi-magnetostrophic' is used because the Lorentz force only enters the leading order force balance in \textcolor{black}{the zonal component} of the large scale momentum equation, and it is of smaller magnitude than the mean buoyancy force. For the small-scale dynamics in a spherical geometry, i.e.~on the scale of the convection, the Lorentz force acts as a perturbation to the leading order geostrophic balance \citep{rY16}.

Another important parameter in natural dynamos is the magnetic Prandtl number, $Pm = \nu /\eta$, where $\eta$ is the magnetic diffusivity. Estimates suggest values that range from $Pm=O(10^{-3})$ in the interior regions of stars to as small as $Pm=O(10^{-6})$ in planetary interiors and the outer regions of stars \citep{mO03}. Self sustaining dynamos require sufficiently large flow speeds to counteract the effects of ohmic dissipation, as characterised by the magnetic Reynolds number, $Rm = U H/\eta$. The magnetic and hydrodynamic Reynolds numbers are related by $Rm = Pm Re$, where $Re = UH/\nu = Ro/E$. Rotating dynamos typically require at least $Rm = O(10)$ to sustain dynamo action. Thus, for $Pm=O(10^{-6})$, Reynolds numbers in excess of $Re = O(10^7)$ would be required to generate self-sustaining magnetic fields. Such values of $Re$ are well beyond the reach of current DNS and studies therefore must use unrealistically large values of $Pm$ -- typically $Pm=O(1)$. Recent state-of-the-art simulations in spherical geometry have used values as small as $Pm=0.05$ \citep{aS18}, though such simulations require enormous computational resources and therefore parameters cannot be varied systematically. The disparity in parameter values between DNS and natural dynamos leads to reasonable suspicion as to how the dynamics of these two systems relate to one another.



Dynamos are often distinguished by the typical length scale of the magnetic field \citep[e.g.][]{sT21}. \textcolor{black}{Large scale dynamos have a component of the magnetic field that varies on a global length scale}. In contrast, small scale dynamos have a negligible large scale component and are instead dominated by length scales comparable to that of the velocity field. \textcolor{black}{Using weakly nonlinear asymptotic theory,} \cite{sC72} and \cite{aS74} showed that rapidly rotating convection in the plane layer geometry \textcolor{black}{can readily drive} large-scale dynamo action near the onset of convection. In this geometry, the large scale magnetic field is often defined as the horizontally averaged component of the field, \textcolor{black}{which is required to be purely horizontal due to the solenoidal constraint on the magnetic field. Using the methods of \cite{mS06}, \cite{mC15b} developed a fully nonlinear extension of the Childress-Soward dynamo model -- this extended model suggests that large scale dynamo action is achievable for arbitrarily large forcing, so long as the convection remains geostrophically balanced. This model also predicts that the large scale magnetic field remains energetically dominant relative to the small scale magnetic field provided the magnetic Reynolds number based on the small, horizontal convective length scale is less than unity \citep{mC17b,mC18} -- this prediction is supported by recent numerical simulations \citep{mY22}. Other aspects of this asymptotic theory, including the scaling behaviour of the various physical quantities and the associated force balance remain untested.}


\textcolor{black}{Previous DNS investigations of dynamos in the plane layer geometry have confirmed that large-scale dynamo action is achievable, provided that the convection is rotationally constrained and $Rm$ is not too large \citep[e.g.][]{cJ00b,sS04,pK09,cG17}. As the buoyancy forcing is increased at a fixed value of $E$, the dynamo transitions from a large scale dynamo to a small scale dynamo \citep[e.g.][]{aT12}, though the influence of rotation must be sufficiently strong to observe this transition \citep[e.g.][]{fC06,bF13b}. A similar relationship between the effects of rotation, buoyancy force and magnetic field morphology is well known in spherical geometries \citep{cK02,uC06,kS12}. For certain values of $Pm$ there exists an intermediate regime in the plane layer geometry in which no dynamo is observed \citep{aT12,cG17}. Moreover, \cite{cG17} found that significant large-scale magnetic energy could be generated for strongly supercritical flows if large-scale vortices (LSVs) are present. Such LSVs are the consequence of an inverse kinetic energy cascade in which kinetic energy is transferred from small-scale convective motions to domain-scale flows \citep{kJ12,aR14,cG14,bF14}. In a Cartesian domain of square cross-section the vortices are depth-invariant, and for sufficiently small Rossby numbers are dipolar in structure \citep{sS14}, but become predominantly cyclonic at larger Rossby numbers \citep{cG14,bF14}. Magnetic fields of sufficient magnitude can damp the vortices and prevent their formation \citep{cG17,pB18,sM19}. }

%
%

In the present study we investigate convection-driven dynamos in the rapidly rotating regime. One of our primary goals is to understand the asymptotic behaviour of the resulting dynamos, including heat transfer, flow speed, magnetic field strength, length scales and force balances. \textcolor{black}{This analysis has not been performed previously,} and we find that the simulations are consistent with much of the theory of \cite{mC15b}, including the asymptotic scaling behavior and the prediction that large scale dynamo action is achieved for arbitrarily large forcing as long as $E \rightarrow 0$ and $Pm \rightarrow 0$. In section \ref{S:methods} we present the non-dimensional equations, physical parameters and the numerical methods employed. The asymptotic theory of rapidly rotating dynamos in the plane layer geometry is \textcolor{black}{briefly summarised} in section \ref{S:theory}, results are discussed in section \ref{S:results}, and a discussion is given in section \ref{S:conclusions}.

\section{Governing equations and methods}
\label{S:methods}

We consider an electrically conducting Boussinesq fluid layer of depth $H$ contained between two plane parallel boundaries. The system is  heated from the bottom and cooled from the top, with the temperature difference $\Delta T = T_{bot} - T_{top} > 0$, where $T_{bot} $ and $T_{top}$ are the temperatures at the bottom and top boundaries, respectively. A constant gravitational acceleration is used and given by $\bold{g}=-g\bold{\hat{z}}$, where $\bold{\hat{z}}$ points from the bottom boundary to the top boundary. The rotation of the system is characterised by a constant rotation rate, $\boldsymbol \Omega=\Omega\bold{\hat{z}} $. The fluid has density $\rho$, kinematic viscosity $\nu$, thermal diffusivity $\kappa$, thermal expansion coefficient $\gamma$, magnetic diffusivity $\eta$ and vacuum permeability $\mu_0$. The governing equations are non-dimensionalised with depth $H$, time scale $2\Omega^{-1}$ and magnetic field scale $\mathcal{B} = 2\Omega H \sqrt{\rho \mu_0}$, and are given by
\be
D_t \ub + \hz \times \ub = - \nabla p + \left(\bold{\nabla}\times\bold{B}\right)\times\bold{B} + \frac{E^2Ra}{Pr}\theta \bold{\hat{z}} + E\nabla^2 \ub,\\\\
\label{eq:mom}
\ee  
\be
D_t \theta =  \frac{E}{Pr}\nabla^2 \theta,\\\\
\label{eq:heat}
\ee
\be
\dst \Bb = \bold{\nabla}\times\left(\bold{u}\times\bold{B}\right) + \frac{E}{Pm}\nabla^2 \Bb,\\\\
\label{eq:induction}
\ee
\be
\nabla\cdot\bold{u} = 0,\\\\
\ee
\be
\nabla\cdot\bold{B} = 0,\\\\
\ee
where the material derivative is denoted by $D_t ( \cdot )  = \dst ( \cdot ) + \ub \cdot \nabla ( \cdot )$. We denote the velocity field  as $\bold{u}=(u,v,w)$, the magnetic field as  $\bold{B}=(B_x,B_y,B_z)$, the temperature as $\theta$, the pressure as $p$ and the Cartesian coordinate system as $(x,y,z)$. 

The non-dimensional control parameters are the Rayleigh number ($Ra$), the thermal Prandtl number ($Pr$), the magnetic Prandtl number ($Pm$) and the Ekman number ($E$), defined as
\be
Ra = \frac{ g \gamma \Delta T H^3}{\nu \kappa}, \quad Pr = \frac{\nu}{\kappa}, \quad Pm = \frac{\nu}{\eta}, \quad E = \frac{\nu}{2\Omega H^2}.
\ee
For simplicity we fix the thermal Prandtl number at $Pr=1$ for all simulations.

\textcolor{black}{In the limit of asymptotically small Ekman number the dynamics depend not on $Ra$ and $E$ independently, but on the combination \citep[e.g.~see][]{kJ12}
\be
\Rat \equiv E^{4/3} Ra,
\ee
which is consistent with the asymptotic scaling for the critical Rayleigh number, i.e.~$Ra_c = O(E^{-4/3})$ \citep{sC61}. This asymptotically rescaled Rayleigh number will be used throughout. }

We apply impenetrable and stress-free velocity boundary conditions at the top and bottom boundaries,
\be
w = \frac{\partial u}{\partial z } = \frac{\partial v}{\partial z } = 0 \quad \text{at} \quad z = 0, 1.
\ee
We use vertical magnetic field boundary conditions such that
\be
B_x = B_y  =  \frac{\partial B_z}{\partial z } =  0 \quad \text{at} \quad z = 0, 1.
\ee
The thermal boundary conditions are isothermal,
\be
\theta = 1 \quad \text{at} \quad z = 0, \quad \text{and} \quad \theta = 0 \quad \text{at} \quad z = 1.
\ee
Periodic boundary conditions are used in the horizontal directions.
\textcolor{black}{For a detailed discussion on the influence of thermal, mechanical and electromagnetic boundary conditions on dynamos we refer the reader to \cite{pK22}. In addition, \cite{pR00} have shown that the choice of electromagnetic boundary conditions can have a strong influence on the linear properties of magnetoconvection.}

\subsection{Diagnostic quantities}

The volumetric and time-averaged kinetic energy density $E_{kin}$ and magnetic energy density $E_{mag}$ are calculated as
\be
E_{kin}  \equiv \frac{1 }{2E^2} \langle    \ub^2\rangle ,
\ee
and
\be
E_{mag} \equiv  \frac{1 }{2E^2} \langle  \Bb^2 \rangle,
\ee
where angled brackets, $\langle \cdot \rangle$, denote a volumetric and time-average. 

We decompose the magnetic field into horizontally averaged (mean) and fluctuating components according to $\Bb(x,y,z,t) = \mBb(z,t) + \Bbp(x,y,z,t)$, where the overline $\overline{\lb \cdot  \rb}$ denotes a horizontal average. The corresponding mean and fluctuating magnetic energy are then defined by, respectively,
\be
\overline{E}_{mag}=  \frac{1 }{2E^2} \int_{0}^{1} \overline{\bold{B}}^2 dz,
\ee
and 
\be
{E}_{mag}^{\prime}=  E_{mag} -  \overline{E}_{mag} .
\ee
\textcolor{black}{Note that while this particular definition of the mean magnetic field is simple, it has the disadvantage that any motion characterized by a horizontal wavenumber greater than zero is defined as small-scale.}

The Reynolds number $Re$ is used to quantify the non-dimensional flow speed and is computed by 
\be
Re = \sqrt{2 E_{kin}} .
\ee

Decomposing the temperature into mean and fluctuating components, $\theta(x,y,z,t) = \overline{\theta}(z,t) + \theta'(x,y,z,t) $, we define the Nusselt number $Nu$ according to 
\be
Nu = - \frac{\partial \overline{\overline{\theta}}}{\partial z} \bigg |_{z=0} ,
\ee
where the additional overline indicates the quantity is also averaged in time.
The exact relationship between the Nusselt number and the two sources of dissipation can be derived from the governing equations to give \citep[e.g.][]{mY21}
 \be
\frac{Ra}{Pr^2} \lb Nu - 1 \rb =  \varepsilon_u + \varepsilon_B.
\label{E:Nu_diss}
\ee
The viscous dissipation and ohmic dissipation are defined by, respectively, 
\be
\varepsilon_u = \frac{1}{E^2} \langle \Zb^2 \rangle, \quad \varepsilon_B = \frac{1}{PmE^2} \langle \Jb^2 \rangle ,
\ee
where the vorticity and current density vectors are denoted by $\Zb = \nabla \times \ub$ and $\Jb = \nabla \times \Bb$, respectively.

 \subsection{Numerical methods}
 
The velocity and magnetic field vectors are represented in terms of poloidal and toroidal scalars such that the solenoidal conditions are satisfied exactly \citep[e.g.][]{cJ00b,pM16}. The resulting variables are expanded in Fourier series in the horizontal dimensions and Chebyshev polynomials in the vertical dimension. The non-linear terms are de-aliased with the standard 2/3-rule. The equations are discretized in time with a third-order implicit-explicit Runge-Kutta scheme \citep{pS91}. The code has been used in previous investigations \citep[e.g.][]{mY19,mY21}, and the dynamo model used for the present study has been benchmarked with the work of \cite{sS04}.

An important non-dimensional parameter characterizing the geometry is the aspect ratio $\Gamma$, defined as
\be
\Gamma=\frac{L}{H} ,
\ee
where $L$ is the periodicity length in the horizontal direction. In the present work only domains of square cross section are used.
We scale the horizontal periodicity length with the (non-dimensional) critical horizontal wavelength, $\lambda_c$, so that
\be
L=n\lambda_c H,
\ee
where $n$ is the integer number of critical horizontal wavelengths. This normalization for the horizontal dimensions is done to ensure that the most unstable wavelength is present near the onset of convection. Thus, for our simulations, the aspect ratio is given by
\be
\Gamma= n \lambda_c .
\ee
All of the results presented use $n=10$ since this value was found to be sufficient for observing convergence of the primary global diagnostic quantities. \textcolor{black}{The critical wavelength is determined from the relationship \citep{sC61}
\be
2k_c^6+3k_c^4\pi^2 - \pi^6 -\pi^2E^{-2} = 0 ,
\ee
where $k_c = 2 \pi/\lambda_c$ is the critical horizontal wavenumber. The above relationship can be derived from the governing equations upon linearizing about a state of rest.}

\section{Asymptotic theory}
\label{S:theory}

One of the main purposes of the present investigation is to shed light on the behaviour of rotating convection-driven dynamos in the asymptotic limit $E \rightarrow 0$, since this limit is relevant for natural systems. Since the asymptotic scalings will be used in the analysis of the results, we briefly summarise the theory here. \textcolor{black}{This approach is general and can be used to analyze the asymptotic behaviour of DNS output, though the observed scalings will be dependent on the particular non-dimensionalisation that is employed.}

If both the Rossby number and the Ekman number are small, the leading order force balance will be geostrophic and the resulting QG dynamics are the result of small perturbations away from this balance. As previously mentioned, QG dynamo theory can be considered a fully nonlinear generalization of the weakly nonlinear theory developed by \cite{sC72}, and studied in detail by \cite{aS74}. We note that the hydrodynamic QG theory shows excellent agreement with DNS results \citep{sS14,mP16}; in the present work we demonstrate that the corresponding QG dynamo theory also shows excellent agreement with DNS.

The aim of QG theory is to understand the dynamics of rapidly rotating convection-driven dynamos based on a perturbation expansion. The relevant small parameter is the Rossby number based on the small scale horizontal convective length scale, $\ell$,  
\be
\ep \equiv \frac{U}{2 \Omega \ell} .
\ee
\textcolor{black}{We recall that $U$ is the characteristic dimensional scale for the velocity field. Here we scale} the velocity in small scale viscous diffusion units, $U = \nu/\ell$, such that
\be
\ep = \frac{\nu}{2 \Omega \ell^2} = \lb \frac{\nu}{2 \Omega H^2} \rb \lb \frac{H}{\ell} \rb^2 .
\ee
\textcolor{black}{Using $\ell  = H E^{1/3}$ the above relationship becomes}
\be
\epsilon = E^{1/3}.
\ee
This scaling is the relevant distinguished limit relating the small scale Rossby number and the large scale Ekman number, and allows for a self consistent set of asymptotically reduced equations \citep{kJ98a, mC15b}.

The large scale Rossby number and the small scale Rossby number are related via
\be
Ro = \frac{U}{2 \Omega H}  = \frac{U}{2 \Omega \ell} \frac{\ell}{H} = E^{2/3} = \ep^2 .
\ee
This relationship is useful for scaling the various terms in the momentum equation, since our particular choice of non-dimensionalisation implies that flow speeds are in units of the large scale Rossby number. Thus, denoting the magnitude of the velocity as $u$, we have
\be
u = O\lb \ep^2 \rb ,
\ee
\textcolor{black}{where we note that all three velocity components have the same asymptotic scaling behavior.} With this asymptotic scaling for the flow speed, and assuming that all horizontal derivatives are $O(\ep^{-1})$ and vertical derivatives are order one, i.e.~$\dsz = O(1)$, we can estimate the asymptotic size of various terms in the momentum equation and other quantities that will be useful for analyzing the results of the numerical simulations. 

\textcolor{black}{The leading order geostrophic force balance implies that the} Coriolis force and the pressure gradient force are the largest terms in the momentum equation. The asymptotic size of the Coriolis force is then
\be
|\hz \times \ub| = O(\ep^2) = O(E^{2/3}) .
\ee
With the exception of the leading order pressure gradient force, all other terms are asymptotically smaller than the above scaling, \textcolor{black}{though the relative size of these asymptotically subdominant terms is crucial for the resulting QG dynamics}. The viscous force is of size
\be
| E \nabla^2 \ub | = O( \ep^3 ) = O(E) .
\ee
The buoyancy force must be of the same asymptotic size as the viscous force; this requires that the temperature perturbation $\theta' = \theta - \overline{\theta} = O(\ep)$ such that 
\be
\left | \frac{RaE^2}{Pr} \theta^{\prime} \right | = \frac{\Rat E^{2/3}}{Pr} \left | \theta^{\prime} \right | = O(\ep^3) = O(E).
\ee
The asymptotic size of the nonlinear momentum advection term is then
\be
| \ub \cdot \nabla \ub | = O( \ep^3 ) = O( E ).
\ee
If we assume the Lorentz force is also of the same size as these asymptotically subdominant terms then we have
\be
| \lb \nabla \times \Bb \rb \times \Bb | = O(\ep^3) = O( E ) \quad \Rightarrow \quad B = O(\ep^2) = O(E^{2/3}),
\ee
where $B$ is the magnitude of the magnetic field vector. \textcolor{black}{Like the velocity field, the asymptotic scaling is the same for all three components of the magnetic field.} To summarize, the Coriolis and pressure gradient forces are both of size $O(E^{2/3})$, and all other forces are smaller by a factor of $E^{1/3}$. \textcolor{black}{We note that the magnitude of the mean magnetic field can become asymptotically larger than the magnitude of the fluctuating magnetic field, though this requires $Rm E^{1/3} \ll 1$ \citep{mC15b}. Our estimate for the asymptotic size of the magnetic field given above assumes that $Rm E^{1/3} =O(1)$.}

With the above scalings we can provide asymptotic estimates for other quantities of interest. The viscous dissipation scales as
\be
\varepsilon_u = \frac{1}{E^2} \langle (\nabla \times \ub)^2 \rangle = O(\ep^{-4}) = O(E^{-4/3}).
\ee
Similarly, the ohmic dissipation scales as 
\be
\varepsilon_B = \frac{1}{Pm E^2} \langle (\nabla \times \Bb)^2 \rangle = O(\ep^{-4}) = O(E^{-4/3}) .
\ee
\textcolor{black}{An alternative derivation of these scalings is to use equation \eqref{E:Nu_diss} and to substitute $Ra = \ep^{-4} \Rat$ such that
\be
\frac{\Rat}{Pr^2} \lb Nu - 1 \rb = \ep^{4} \lb  \varepsilon_u + \varepsilon_B \rb .
\ee
In order for $Nu$ to be independent of the Ekman number, this requires $\ep^4 \varepsilon_u = O(1)$ and $\ep^4 \varepsilon_B = O(1)$, and therefore both forms of dissipation must scale as $\ep^{-4}$.}

The kinetic and magnetic energy scale as 
\be
E_{kin} = O\lb \frac{u^2}{E^2} \rb = O\lb \ep^{-2} \rb = O\lb E^{-2/3} \rb,
\label{E:kin_scale}
\ee
and 
\be
E_{mag} = O\lb \frac{B^2}{E^2} \rb = O\lb \ep^{-2} \rb = O\lb E^{-2/3} \rb.
\ee

We emphasize that the above asymptotic relationships only specify the dependence on the Ekman number. In general, the various quantities still depend on the reduced Rayleigh number.

\section{Results}
\label{S:results}

In the simulations we vary the Ekman number, the Rayleigh number and the magnetic Prandtl number. The thermal Prandtl number is fixed at $Pr=1$ for all simulations  -- we refer the reader to \cite{jmA18} and \cite{tV21} for detailed investigations on the influence of small values of $Pr$, as relevant to liquid metals and plasmas. The Ekman number is varied from $E = 10^{-8}$ to $E=10^{-4}$. The Rayleigh number is varied from $1.3 Ra_c$ up to $9 Ra_c$. For simplicity we will often translate these supercritical values to reduced Rayleigh numbers $\Rat$; we reach up to $\Rat \approx 78$. The magnetic Prandtl number is varied less extensively, but we consider values of $Pm = \lb 1, 0.3, 0.2, 0.1, 0.05 \rb$. For $Pm=1$ we consider Ekman numbers from $E=10^{-4}$ down to $E=10^{-6}$. \textcolor{black}{Since smaller values of $E$ allow us to reach smaller values of $Pm$, we consider $Pm<1$ for $E \le 10^{-6}$}. Various input and output quantities, along with numerical parameters, are listed in the Appendix.

\textcolor{black}{
\subsection{Dynamo and flow regimes}
}

 \begin{figure*}
 \begin{center}
                  \subfloat[]{
      \includegraphics[width=0.45\textwidth]{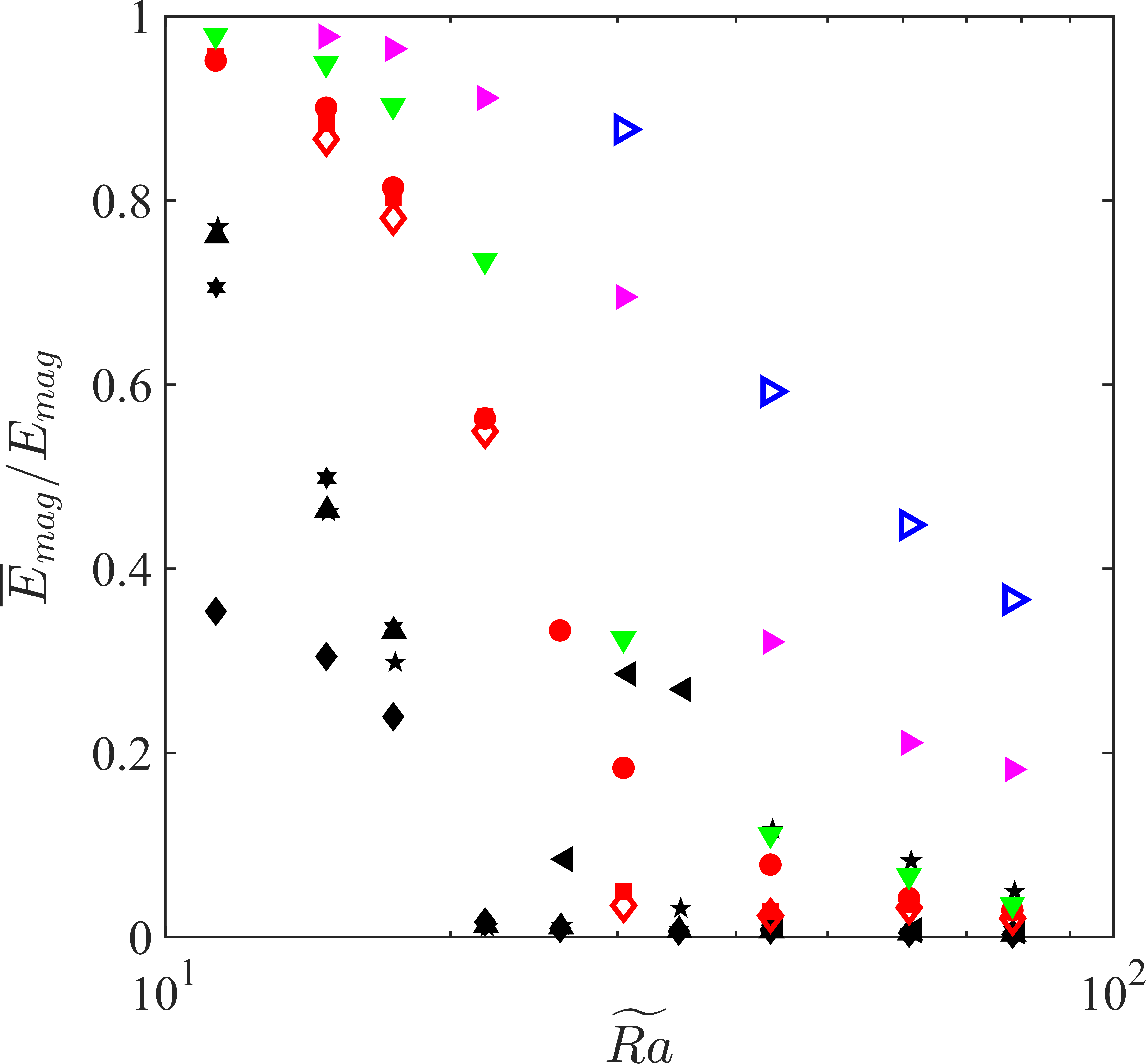} } \quad
                 \subfloat[]{
      \includegraphics[width=0.45\textwidth]{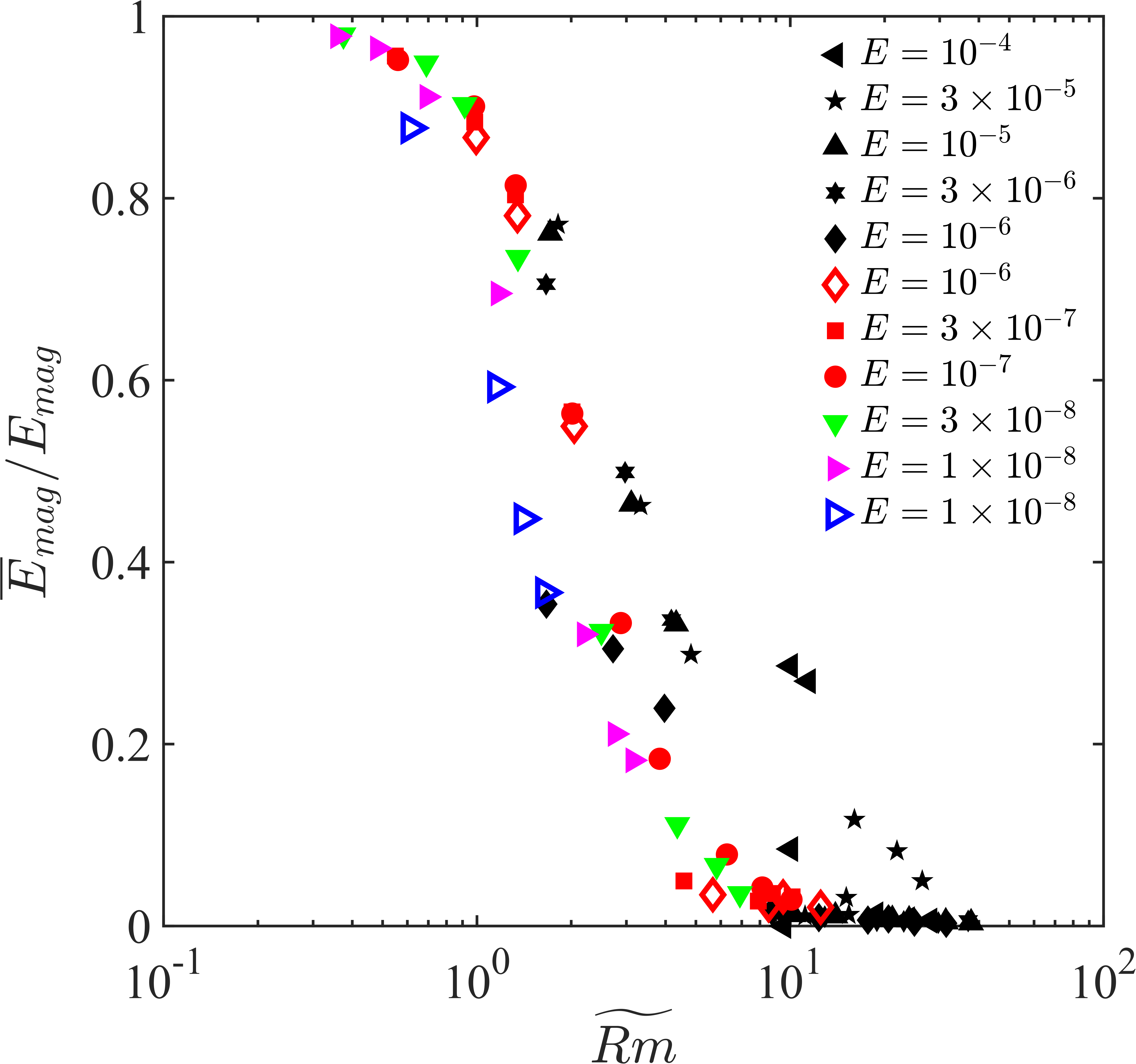} }         
 \end{center}
\caption{ \textcolor{black}{Overview of the dynamo regimes, as characterised by the fraction of the mean magnetic energy to the total magnetic energy, $\overline{E}_{mag}/E_{mag}$, in all simulations: 
(a) $\overline{E}_{mag}/E_{mag}$ versus reduced Rayleigh number, $\Rat$;
(b) $\overline{E}_{mag}/E_{mag}$ versus reduced magnetic Reynolds number, $\Rmt$.
Symbol shape represents different values of the Ekman number ($E$) and color represents different values of the magnetic ($Pm$): black indicates $Pm=1$; red indicates $Pm=0.3$; green indicates $Pm=0.2$; magenta indicates $Pm=0.1$; blue indicates $Pm=0.05$. }
\label{F:Emfraction}}
\end{figure*}

\textcolor{black}{The different dynamo regimes can be characterised by the fraction of the mean magnetic energy to the total magnetic energy, $\overline{E}_{mag}/E_{mag}$, as shown in Figures~\ref{F:Emfraction}(a) and (b) where this quantity is plotted as a function of $\Rat$ and the reduced magnetic Reynolds number, $\Rmt = E^{1/3} Rm$, respectively. For the purpose of discussion we classify dynamos as large scale if $\overline{E}_{mag}/E_{mag} \ge 0.5$; dynamos with smaller values are then referred to as small scale dynamos. For fixed values of $E$ and $Pm$ the mean energy fraction generally decreases with increasing $\Rat$. In general we find that smaller values of $Pm$ allow for larger values of $\overline{E}_{mag}/E_{mag}$, with some cases approaching unity. As shown in panel (a), larger values of $\overline{E}_{mag}/E_{mag}$ exist over a larger range in $\Rat$ for decreasing values of $Pm$, indicating that magnetic diffusion plays an important role in the relative strength of the large scale magnetic field \cite[e.g.][]{mY22}. When the energy fraction is plotted as a function of $\Rmt$ in panel (b), we find a collapse of the data. In agreement with the theory of \cite{mC15b}, the mean magnetic energy is predicted to dominate when $\Rmt \ll 1$. Although the simulations do not reach very small values of $\Rmt$, there is nevertheless an observed trend of increasing $\overline{E}_{mag}/E_{mag}$ as $\Rmt$ is decreased.}

\textcolor{black}{\cite{cG17} have found that it is possible to generate large scale dynamos with the aid of LSVs for $E = 5 \times 10^{-6}$, provided that the magnetic Reynolds number is within the range $100 \lesssim Rm \lesssim 550$ and $Pm < 1$; in terms of the small scale magnetic Reynolds number this range becomes $1.7 \lesssim \Rmt \lesssim 9.4$. \cite{cG17} do not list values of $\overline{E}_{mag}/E_{mag}$, though visual inspection of their magnetic energy spectra shows that they do not observe large scale dynamos in which the large scale magnetic field is energetically dominant relative to the small scale magnetic field. This finding is consistent with Figure~\ref{F:Emfraction}(b) for comparable parameter values in which $\overline{E}_{mag}/E_{mag} \lesssim 0.5$. For hydrodynamic convection with $E \ll 1$ and $Pr=1$, LSVs become energetically dominant (relative to the small-scale velocity field) for $\Rat \gtrsim 20$ \citep{sM21}, so that they are certainly present in many of the simulations presented here. However, our data and the asymptotic theory \cite[e.g.~see][]{mY22} indicate that large scale dynamo action is more directly controlled by the small scale magnetic Reynolds number. Moreover, kinematic investigations of QG convection suggest that LSVs do not appear to alter the onset of dynamo action \citep{mC16b}. Additional insight into the role that LSVs play in dynamo action could be made by performing a set of simulations in which the depth-averaged flow is set to zero at each timestep \citep[e.g.][]{sM21}. }

The flow regimes observed in the present study are broadly consistent with the regimes identified in previous studies of non-magnetic rotating convection \citep[e.g.][]{kJ12}, though in the present study we do not attempt to characterise precisely the location (in parameter space) of possible transitions. The presence of magnetic field can certainly influence the regimes, as discussed, for instance, by \cite{cG17} and \cite{sM19}. Given the broad range of parameter values used here, we simply highlight some of the main effects of varying parameters on the observed flow regimes. 

\begin{figure*}
 \begin{center}
                  \subfloat[]{
      \includegraphics[width=0.33\textwidth]{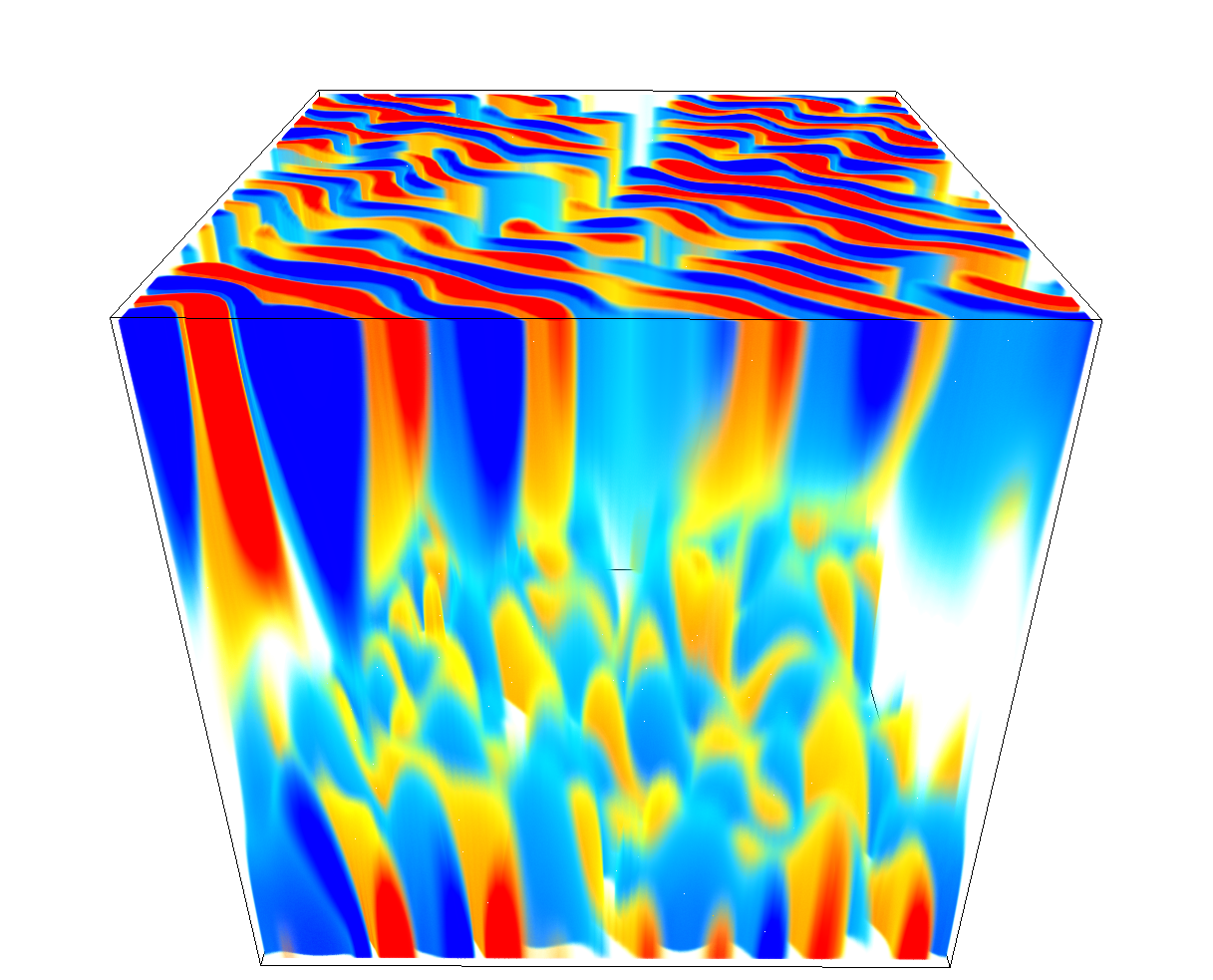} }
                  \subfloat[]{
      \includegraphics[width=0.33\textwidth]{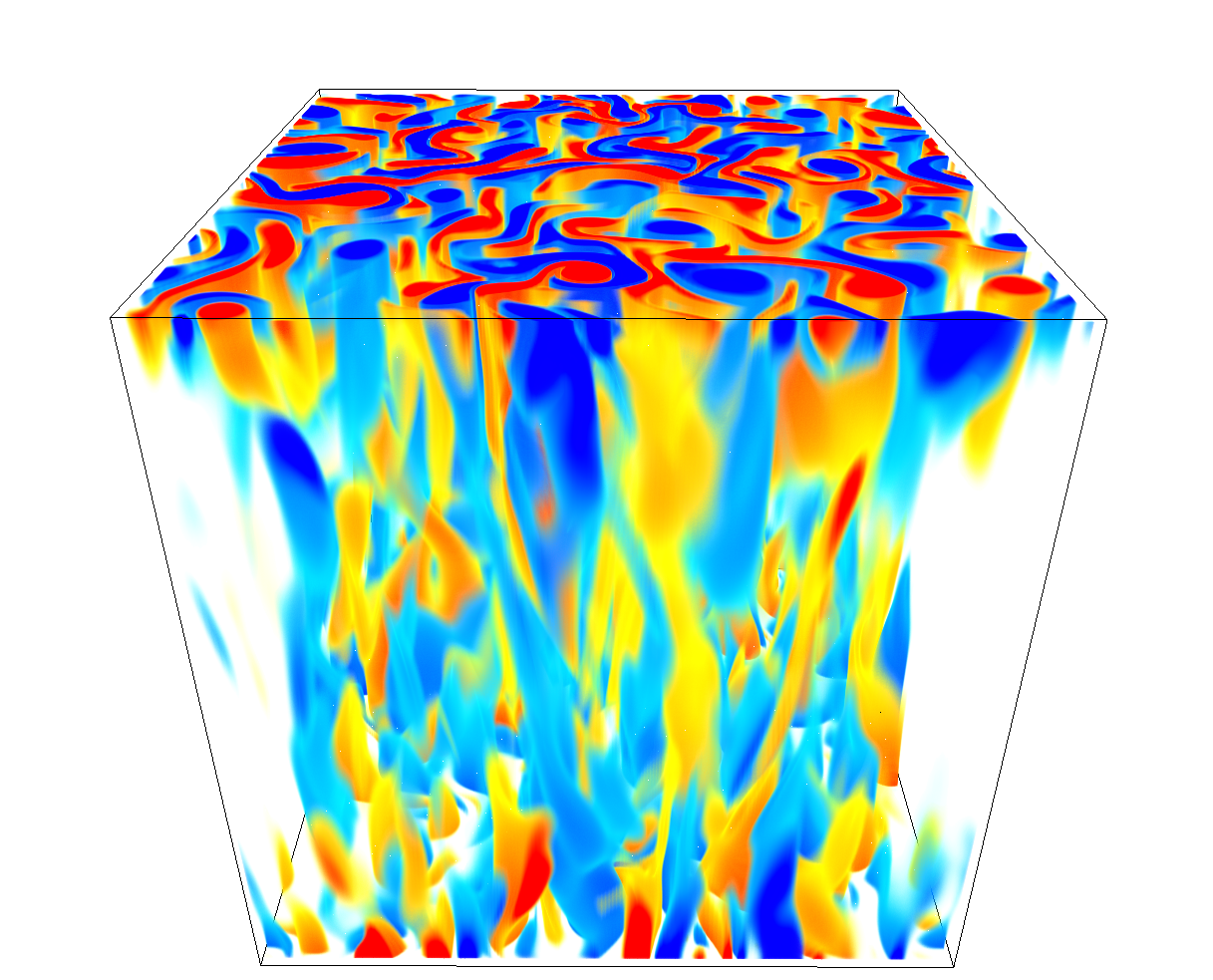} }
                        \subfloat[]{
      \includegraphics[width=0.33\textwidth]{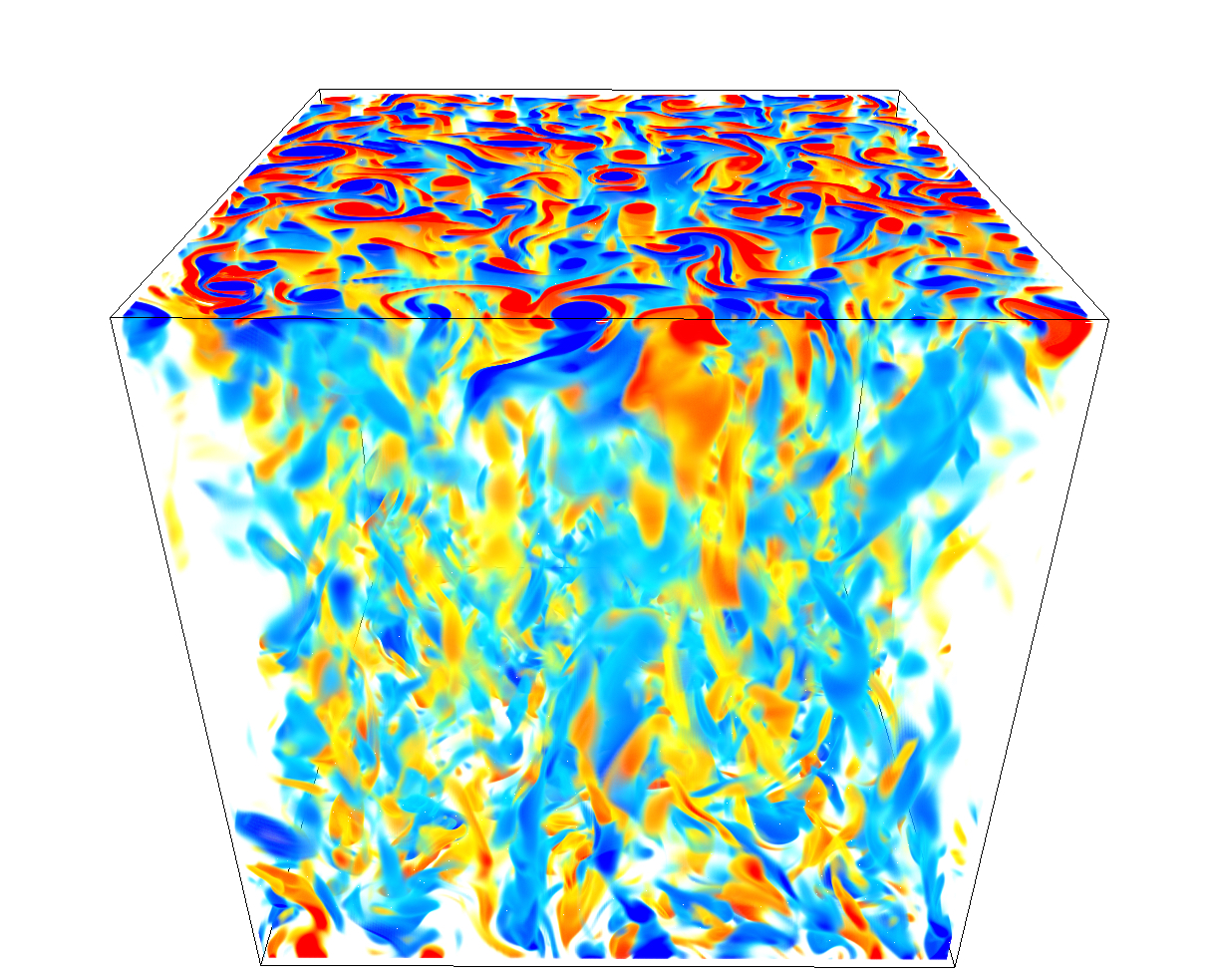} }
 \end{center}
\caption{Volumetric renderings of the instantaneous vertical vorticity for $E=10^{-8}$, $Pm=0.1$ and increasing Rayleigh number from left to right: 
(a) $Ra=1.7Ra_c$ ($\Rat\approx 15$);
(b) $Ra=3.5Ra_c$ ($\Rat\approx 30$);
(c) $Ra=9Ra_c$ ($\Rat\approx 78$). Red denotes positive (cyclonic) vorticity and blue denotes negative (anti-cyclonic) vorticity.
\label{F:vorticityrendering_E1e8}}
\end{figure*}

Figure \ref{F:vorticityrendering_E1e8} shows perspective views of the instantaneous vertical vorticity for three values of the Rayleigh number at the lowest Ekman number considered here, $E=10^{-8}$, and $Pm=0.1$. \textcolor{black}{The corresponding mean magnetic energy fraction for panels (a), (b) and (c) is, respectively, $\overline{E}_{mag}/E_{mag} \approx (0.98,  0.7, 0.18)$. Thus, cases shown in (a) and (b) are well within the large scale dynamo regime.} For relatively small reduced Rayleigh numbers ($\Rat\approx 15$) we find cellular structures, in the sense that the horizontal structure of the flow is relatively simple and dominated by the critical wavenumber. These cells become less coherent as $\Rat$ is increased, and geostrophic turbulence, \textcolor{black}{as characterized by a leading order geostrophic balance with a broad range of length scales that lack significant vertical coherence} \citep[e.g.][]{kJ12}, is observed for sufficiently large ($\Rat \gtrsim 40$) Rayleigh number, as shown in Figure \ref{F:vorticityrendering_E1e8}(c). For the particular set of parameters shown in Figure \ref{F:vorticityrendering_E1e8}, we do not observe an obvious LSV at any value of $\Rat$, even for $\Rat \approx 78$ in which $\Ret \approx 30$. \textcolor{black}{As discussed later, cases with energetically dominant LSVs are characterized by horizontal kinetic energy spectra that show a peak at the smallest wavenumber, though these LSVs can become damped by the magnetic field \citep{cG17, sM19}}. Previous studies of hydrodynamic convection show that LSVs become energetically dominant when $\Ret \gtrsim 6$ \citep{sM21}, though this criterion is no longer valid when magnetic field is present. 

\begin{figure*}
 \begin{center}
                  \subfloat[$E=10^{-4}$, $\Rat\approx 31$]{
      \includegraphics[width=0.45\textwidth]{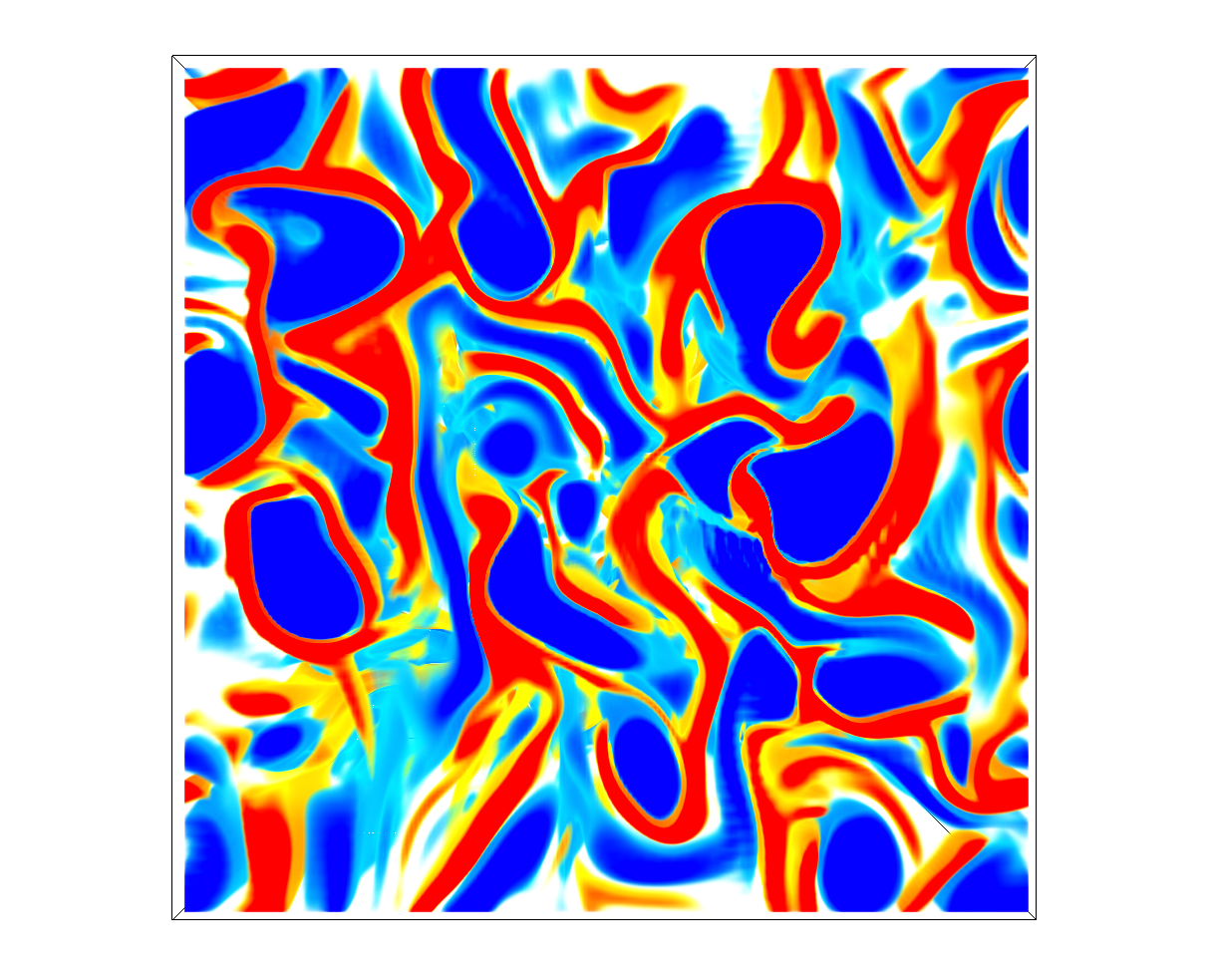} }
                        \subfloat[$E=10^{-4}$, $\Rat\approx 79$]{
      \includegraphics[width=0.45\textwidth]{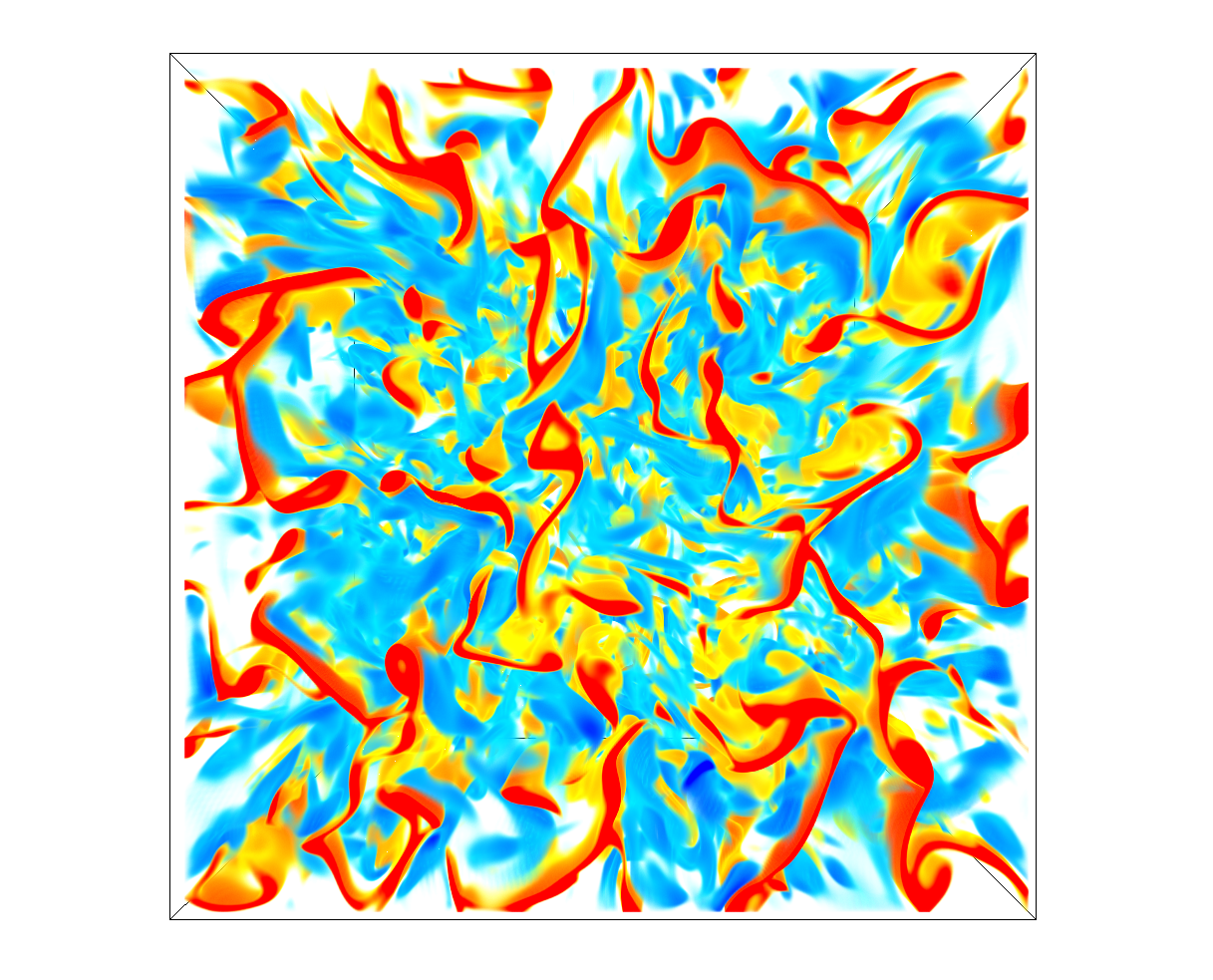} }
      \quad
                  \subfloat[$E=10^{-6}$, $\Rat\approx 26$]{
      \includegraphics[width=0.45\textwidth]{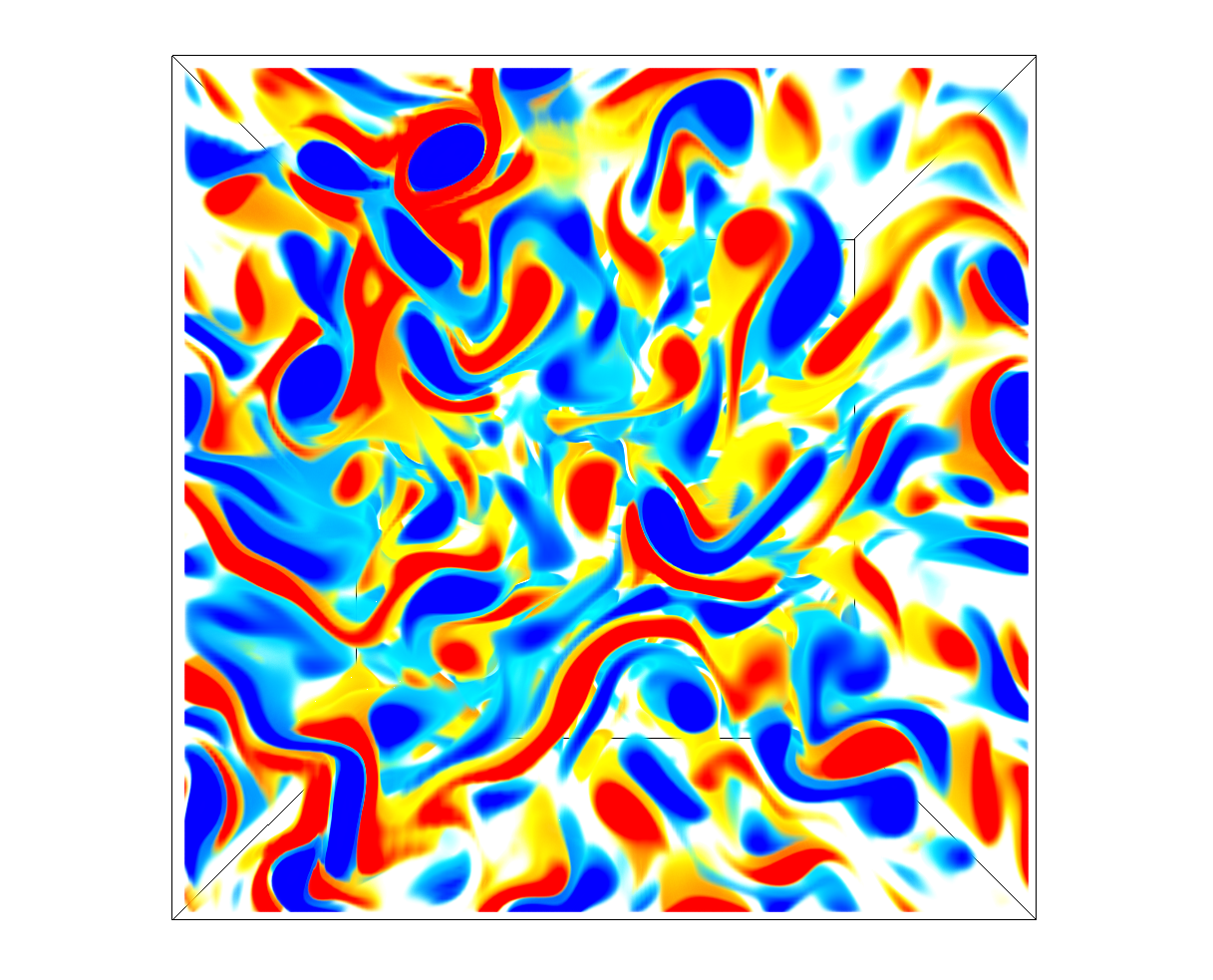} }
                        \subfloat[$E=10^{-6}$, $\Rat\approx 78$]{
      \includegraphics[width=0.45\textwidth]{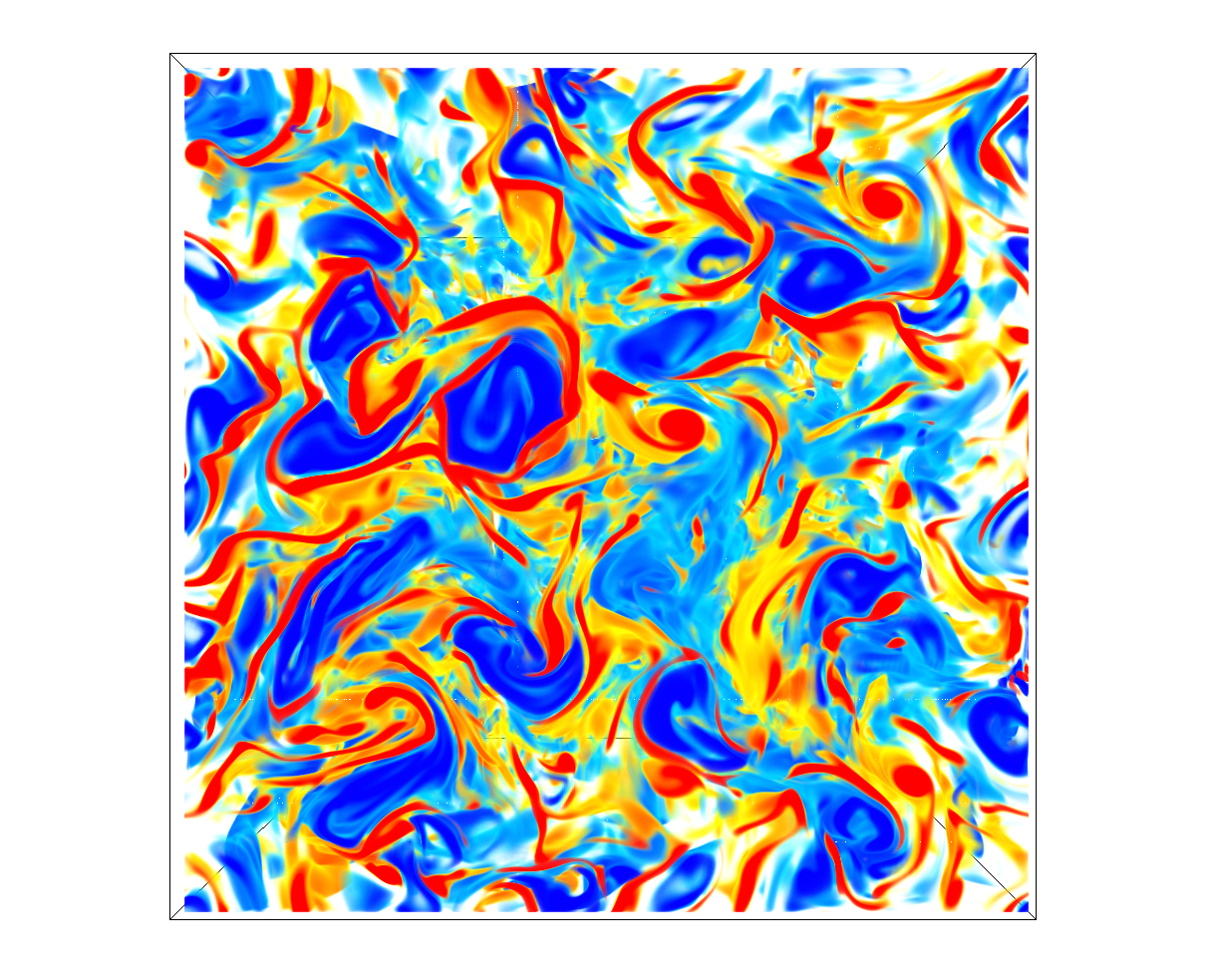} }
            \quad
                  \subfloat[$E=10^{-8}$, $\Rat\approx 30$]{
      \includegraphics[width=0.45\textwidth]{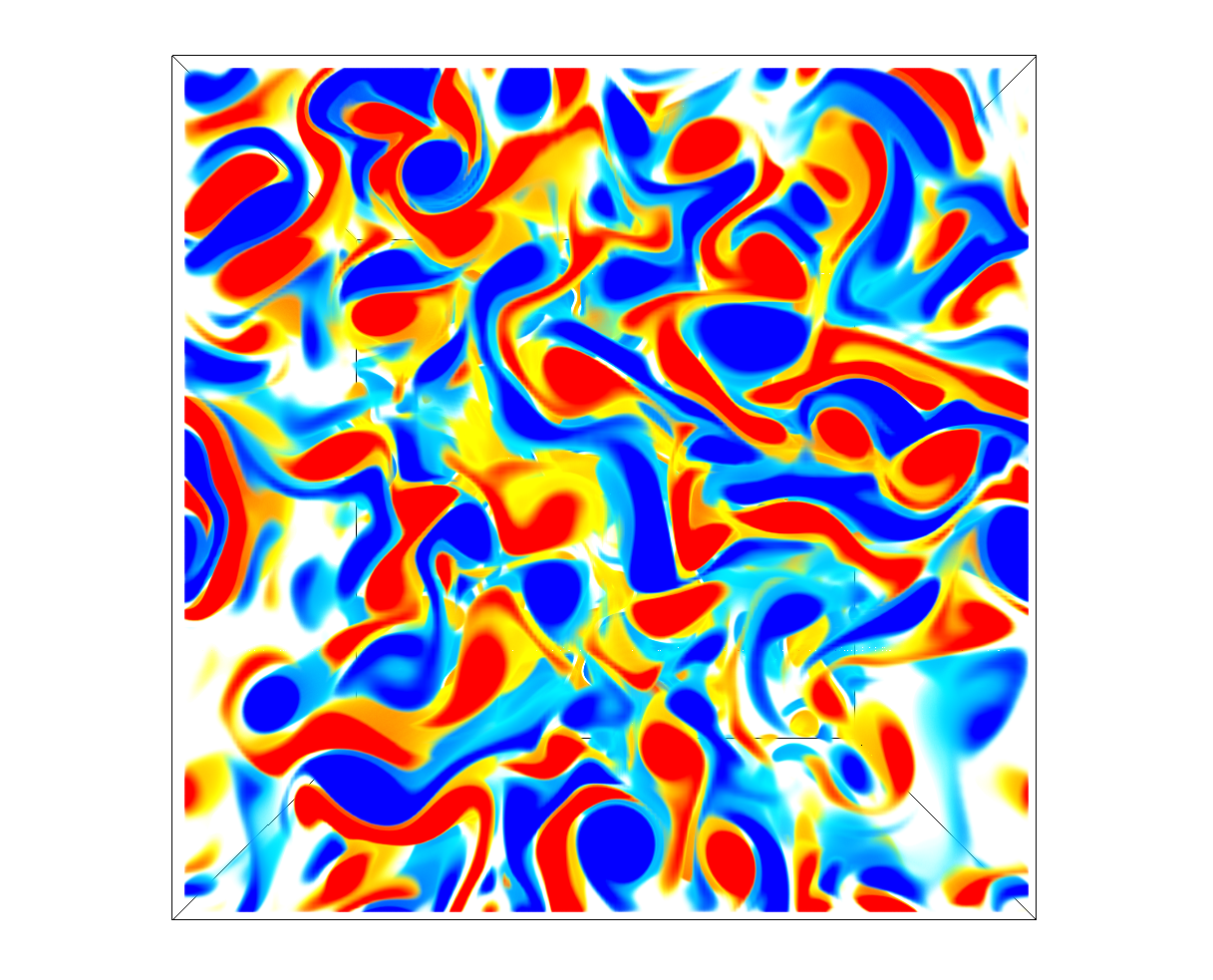} }
                        \subfloat[$E=10^{-8}$, $\Rat\approx 78$]{
      \includegraphics[width=0.45\textwidth]{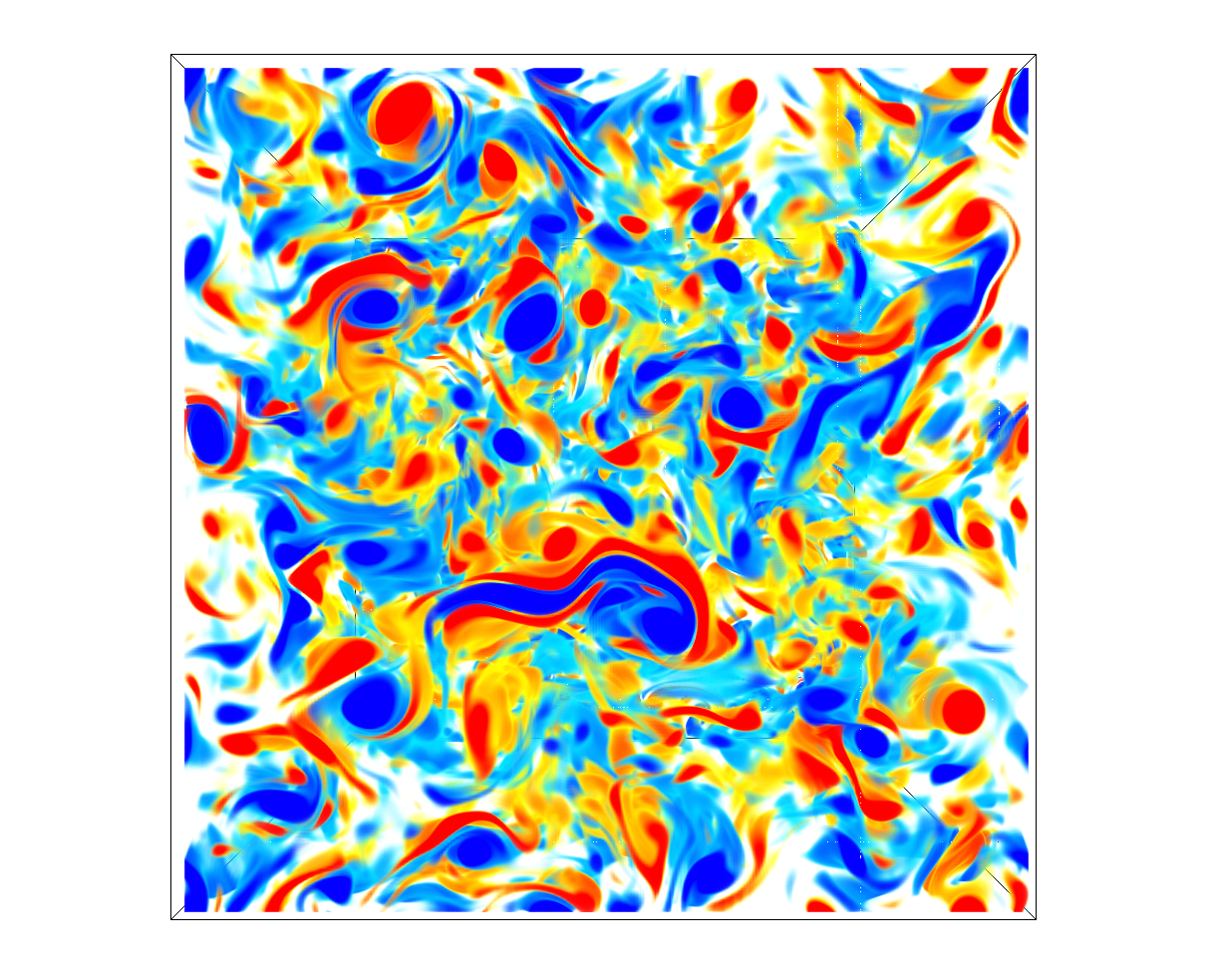} }
 \end{center}
\caption{Top-down view of volumetric renderings of the instantaneous vertical vorticity for a range of Ekman and Rayleigh numbers. The Rayleigh number $\Rat$ increases from left to right and the Ekman number $E$ decreases from top to bottom: 
(a) $E=10^{-4}$, $Ra=3.5Ra_c$ ($\Rat\approx 31$) and $Pm=1$;
(b) $E=10^{-4}$, $Ra=9Ra_c$ ($\Rat\approx 79$) and $Pm=1$;
(c) $E=10^{-6}$, $Ra=3Ra_c$ ($\Rat\approx 26$) and $Pm=1$;
(d) $E=10^{-6}$, $Ra=9Ra_c$ ($\Rat\approx 78$) and $Pm=1$;
(e) $E=10^{-8}$, $Ra=3.5Ra_c$ ($\Rat\approx 30$) and $Pm=0.1$;
(f) $E=10^{-8}$, $Ra=9Ra_c$ ($\Rat\approx 78$) and $Pm=0.1$.
\label{F:vorticityrendering}}
\end{figure*}

A comparison of the flow morphology observed for different Ekman numbers is presented in  Figure \ref{F:vorticityrendering}, where top-down views of the vertical component of the vorticity are shown for $E=(10^{-4}, 10^{-6}, 10^{-8})$ (top to bottom in the Figure) and $\Rat \approx 30$ (left column) and $\Rat \approx 78$ (right column). \textcolor{black}{In terms of the dynamo regimes represented by these visualisations, only case (e) is within the large scale dynamo regime, and all other cases are classified as small scale dynamos.} Whereas similar flow structures are observed for $E=10^{-6}$ and $E=10^{-8}$ when $\Rat \approx 30$ in which both positive (cyclonic, red) and negative (anti-cyclonic, blue) vorticity show similar structure, the case with $E=10^{-4}$ and $\Rat \approx 30$ (Figure \ref{F:vorticityrendering}(a)) shows significant asymmetry between cyclonic and anti-cyclonic vorticity. For this latter case we find that cyclonic structures occur in thin sheets that surround anti-cyclonic vortices. As the Rayleigh number is increased to $\Rat \approx 78$, Figure \ref{F:vorticityrendering}(b) shows that the anti-cyclonic structures near the top boundary become weak and the cyclonic sheets are more distinct for $E=10^{-4}$.  We note that this particular case for $E=10^{-4}$ and $\Rat \approx 78$ is in the small scale dynamo regime in which the large scale magnetic field represents less than $0.6\%$ of the total magnetic energy.

For $\Rat \approx 78$, we observe differences in the flow morphology between $E=10^{-6}$ and $E=10^{-8}$, as shown in Figures \ref{F:vorticityrendering}(d) and (f). The case shown in panel (d) appears similar in structure to panel (a); this similarity hints at a finite Rossby number effect, suggesting that such structure might also be observed for $E=10^{-8}$ if the Rayleigh number could be extended to larger values beyond those accessible in the present study. In contrast, panel (f) shows a more symmetric state. Whereas the case shown in (d) is within the small scale dynamo regime, the large scale magnetic field for the case shown in (f) remains significant ($\approx 20\%$ of the total magnetic energy). \textcolor{black}{We note that the asymptotic scalings presented in the previous section are strictly valid only when the Rossby number remains small and symmetry is preserved.}

\subsection{Heat transfer and dissipation}

\begin{figure*}
 \begin{center}
   \subfloat[]{
      \includegraphics[width=0.45\textwidth]{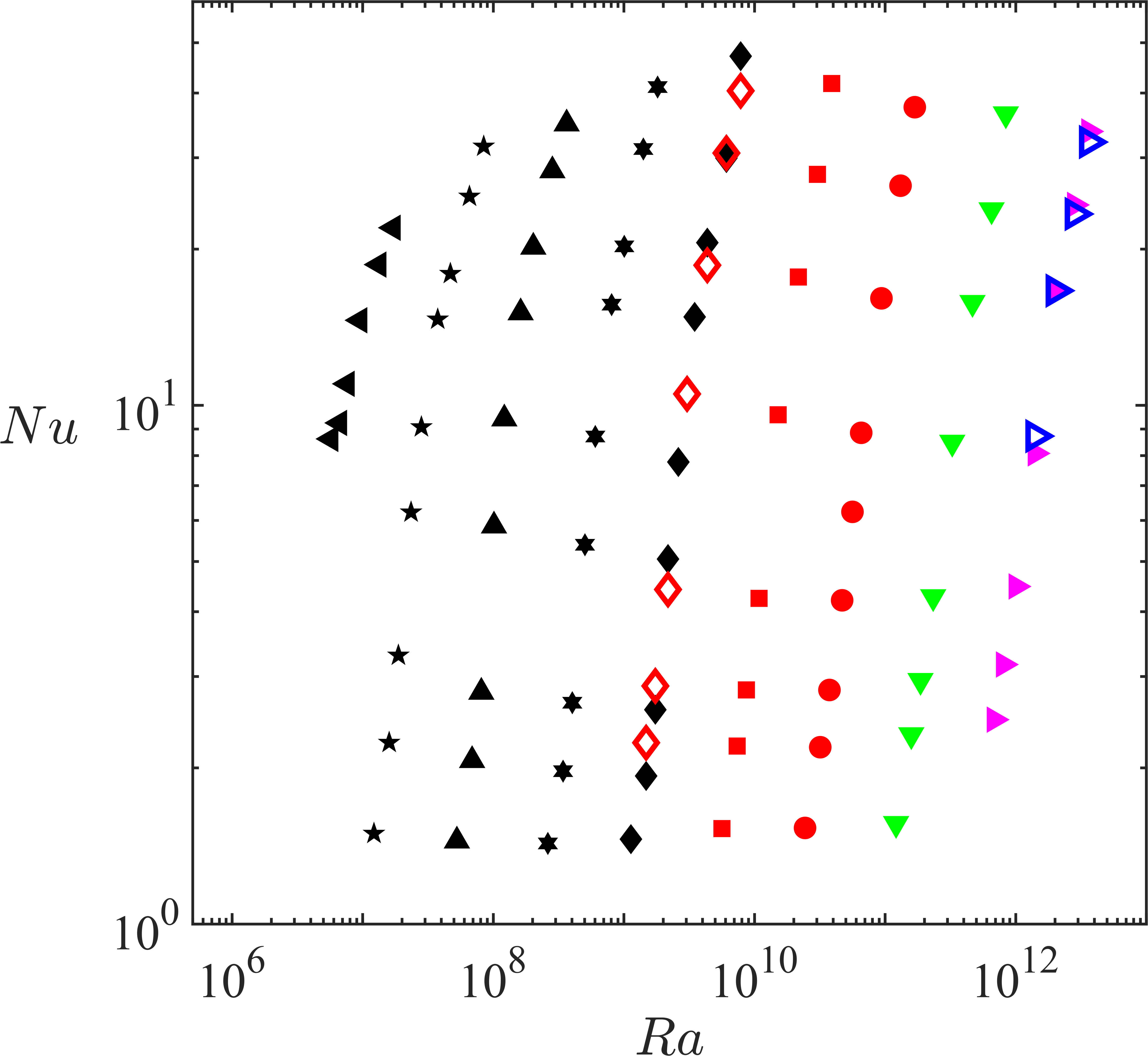}}    \qquad
   \subfloat[]{
      \includegraphics[width=0.45\textwidth]{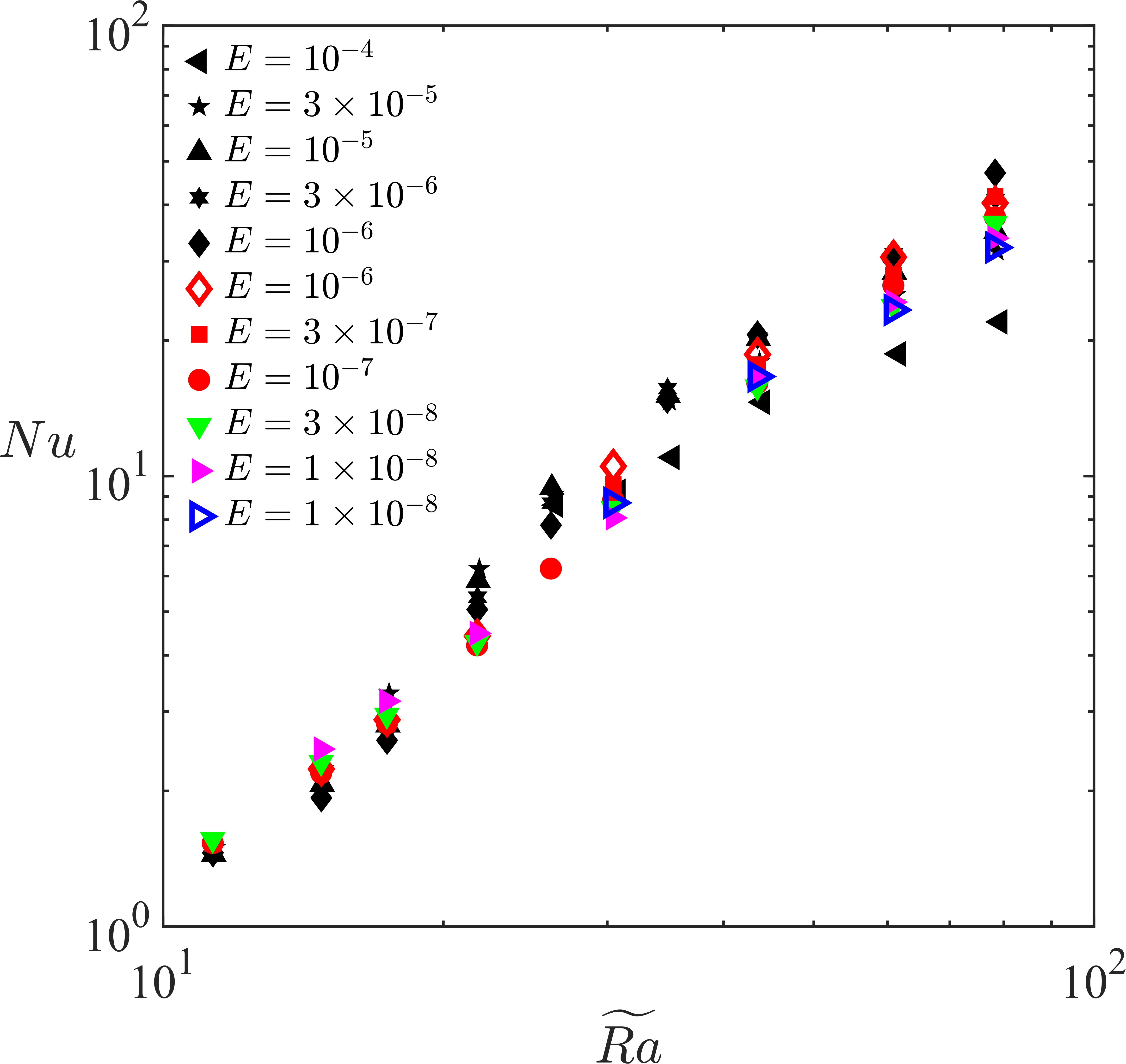}}            
 \end{center}
\caption{ Heat transfer data: (a) Nusselt number, $Nu$, versus Rayleigh number, $Ra$; (b) $Nu$ versus reduced Rayleigh number, $\Rat = E^{4/3} Ra$. Symbol shape represents different values of the Ekman number ($E$) and color represents different values of the magnetic ($Pm$): black indicates $Pm=1$; red indicates $Pm=0.3$; green indicates $Pm=0.2$; magenta indicates $Pm=0.1$; blue indicates $Pm=0.05$. \label{F:Nu_scaling}}
\end{figure*} 

The Nusselt number is shown as a function of the Rayleigh number in Figure \ref{F:Nu_scaling}(a) for all cases; it is shown as a function of $\Rat$ in panel (b). \textcolor{black}{As with the flow regimes discussed in the previous subsection, we observe heat transport behavior that is similar to hydrodynamic rotating convection \citep[e.g.][]{jmA15}. } While there is some variation in the data for different values of $E$, all of the cases show broadly similar behavior when plotted as a function of the reduced Rayleigh number, indicating the simulations are in an asymptotic dynamical regime. The Nusselt number shows a characteristic `s'-shaped dependence on $\Rat$ that is well-known in the literature for rotating convection \citep[][e.g.]{jC15}; for the smallest values of $\Rat$, $Nu$ rises steeply at first, then slows at larger values of $\Rat$ as the Rossby number increases. \textcolor{black}{For a fixed value of $\Rat$, the variation in $Nu$ may be due to the differences in observed dynamo behavior. For instance, whereas some of the cases shown have significant mean magnetic fields, other cases are within the small scale dynamo regime and therefore generate mean magnetic fields with negligible amplitude.}



  \begin{figure*}
 \begin{center}
                        \subfloat[]{
      \includegraphics[width=0.47\textwidth]{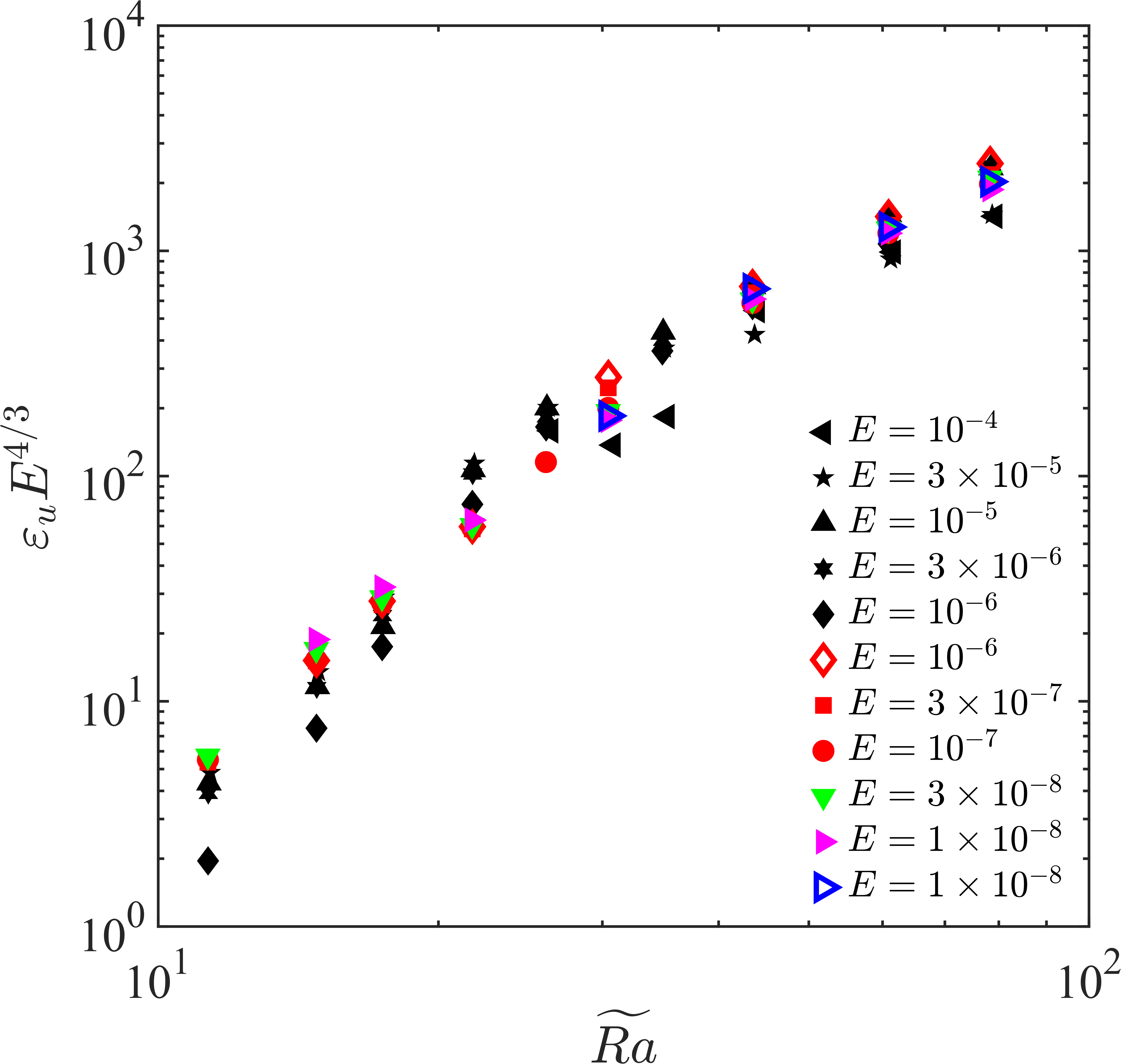} }
      \quad
                  \subfloat[]{
      \includegraphics[width=0.47\textwidth]{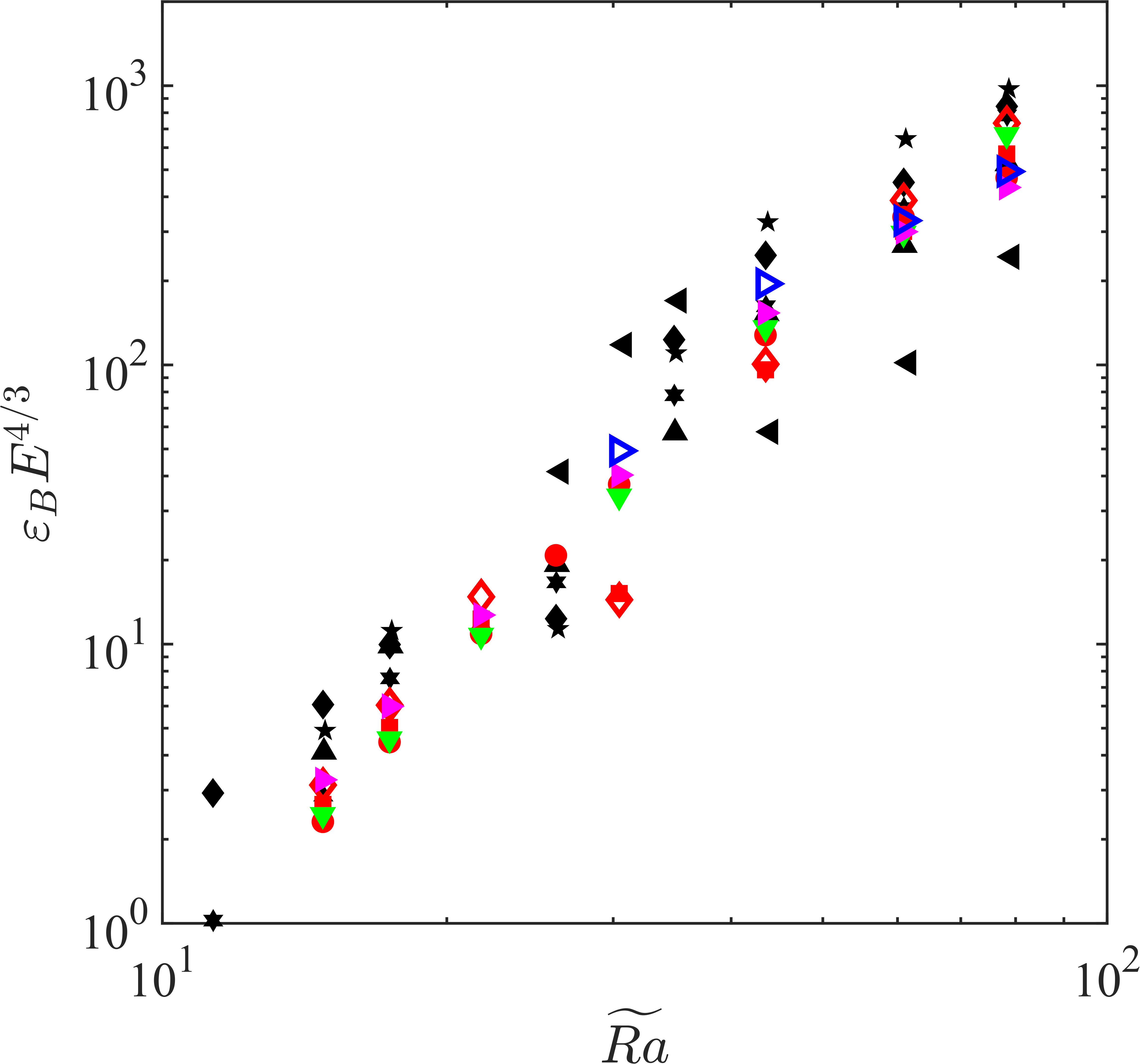} }
         \quad
  \subfloat[]{
      \includegraphics[width=0.5\textwidth]{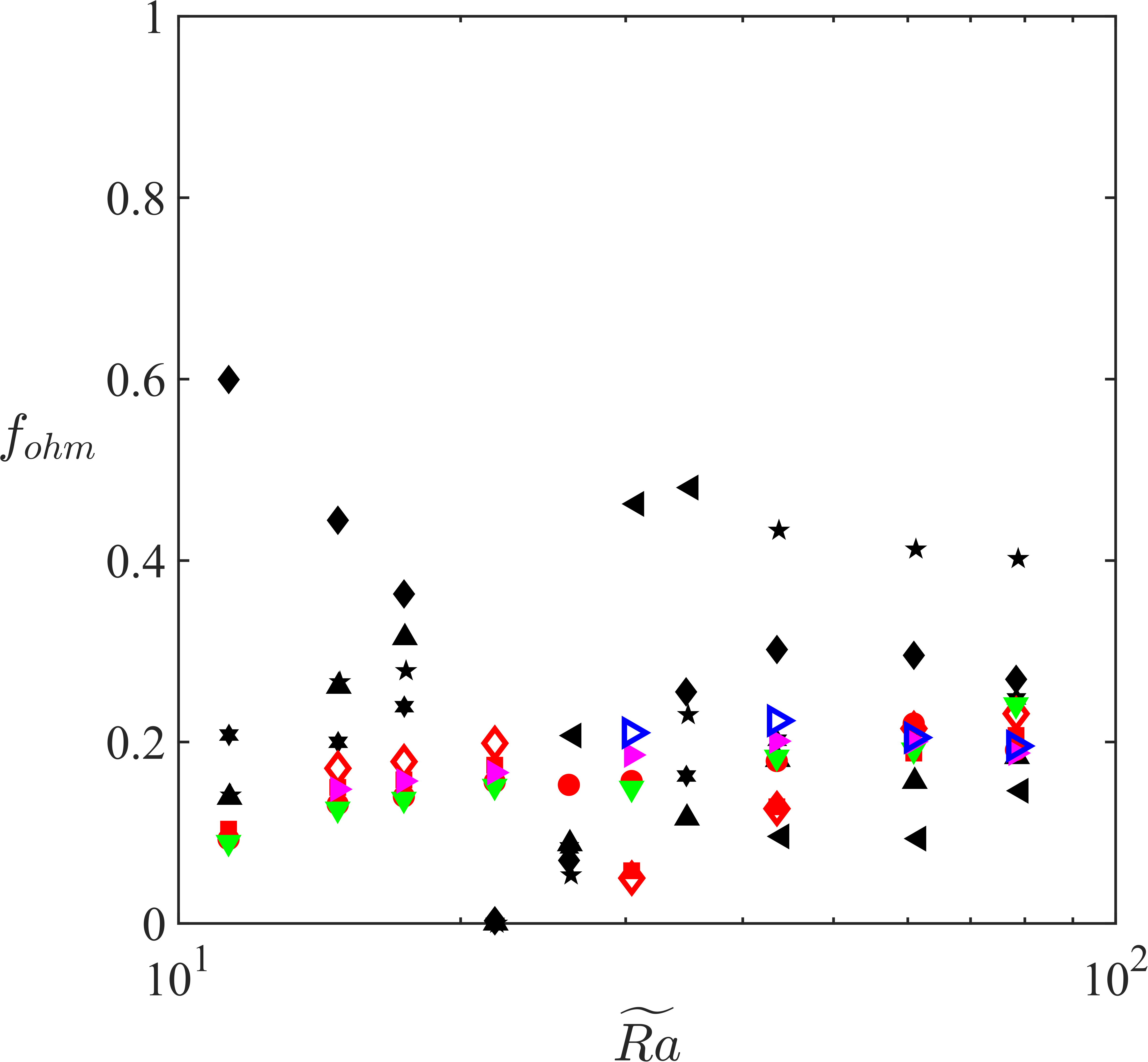} }
 \end{center}
\caption{ Dissipation as a function of reduced Rayleigh number $\Rat$ for all simulations: 
(a) rescaled viscous dissipation, $\varepsilon_u E^{4/3}$;
(b) rescaled ohmic dissipation, $\varepsilon_B E^{4/3}$;
(c) fraction of ohmic dissipation, $f_{ohm}=\varepsilon_B / (\varepsilon_B+\varepsilon_u)$. Symbol shape represents different values of the Ekman number ($E$) and color represents different values of the magnetic ($Pm$): black indicates $Pm=1$; red indicates $Pm=0.3$; green indicates $Pm=0.2$; magenta indicates $Pm=0.1$; blue indicates $Pm=0.05$.
\label{F:diss}}
\end{figure*}

Figures~\ref{F:diss}(a) and (b) show the asymptotically rescaled viscous dissipation $\varepsilon_u E^{4/3}$ and the asymptotically rescaled ohmic dissipation $\varepsilon_B E^{4/3}$ plotted as a function of the reduced Rayleigh number $\Rat$. Consistent with the energy relationship $\eqref{E:Nu_diss}$, both forms of the dissipation show behaviour that is consistent with the Nusselt number. Sudden changes in $\varepsilon_u$, and discontinuities $\varepsilon_B$, appear when (large-scale) dynamo action ceases; this behaviour is perhaps most noticeable for $E= 10^{-4}$ where the viscous dissipation shows a plateau in the vicinity of $\Rat \approx 25$, in which different values of $\Rat$ yield similar values of $\varepsilon_u$. This plateau is associated with a rapid increase in $\varepsilon_B$ as the small-scale dynamo becomes activated. At a fixed value of $\Rat$ we observe scatter for different values of $Pm$ which is related to the magnetic Reynolds number and the corresponding magnetic field behaviour (i.e.~see Figures 5 and 7).


We compute the relative size of the ohmic dissipation to the total dissipation using the fraction of ohmic dissipation, $f_{ohm}$, defined as
\be
f_{ohm}=\frac{\varepsilon_B}{\varepsilon_B+\varepsilon_u}.
\ee
 Figure~\ref{F:diss}(c)  shows $f_{ohm}$ versus $\Rat$. 
For the majority  of our cases the flow is dominated by viscous dissipation ($\varepsilon_u > \varepsilon_B$), which is consistent with previous plane layer dynamo studies \citep{aT14}.  In the small scale dynamo regime, the majority  of our cases  have a fraction of  ohmic dissipation $f_{ohm}\sim 0.2$. \textcolor{black}{In comparison to previous studies of dynamos in spherical geometries, we find $f_{ohm}$ values that are relatively small. For instance, \cite{nS17} find $f_{ohm}=0.86$ with $Pm=0.1$ and $E=5\times10^{-8}$. The differences in these values may be due to the differences in both magnetic field structure and saturation mechanisms in the two geometries.}

\subsection{Velocity and magnetic field scaling}

\begin{figure*}
 \begin{center}
                 \subfloat[]{
      \includegraphics[width=0.45\textwidth]{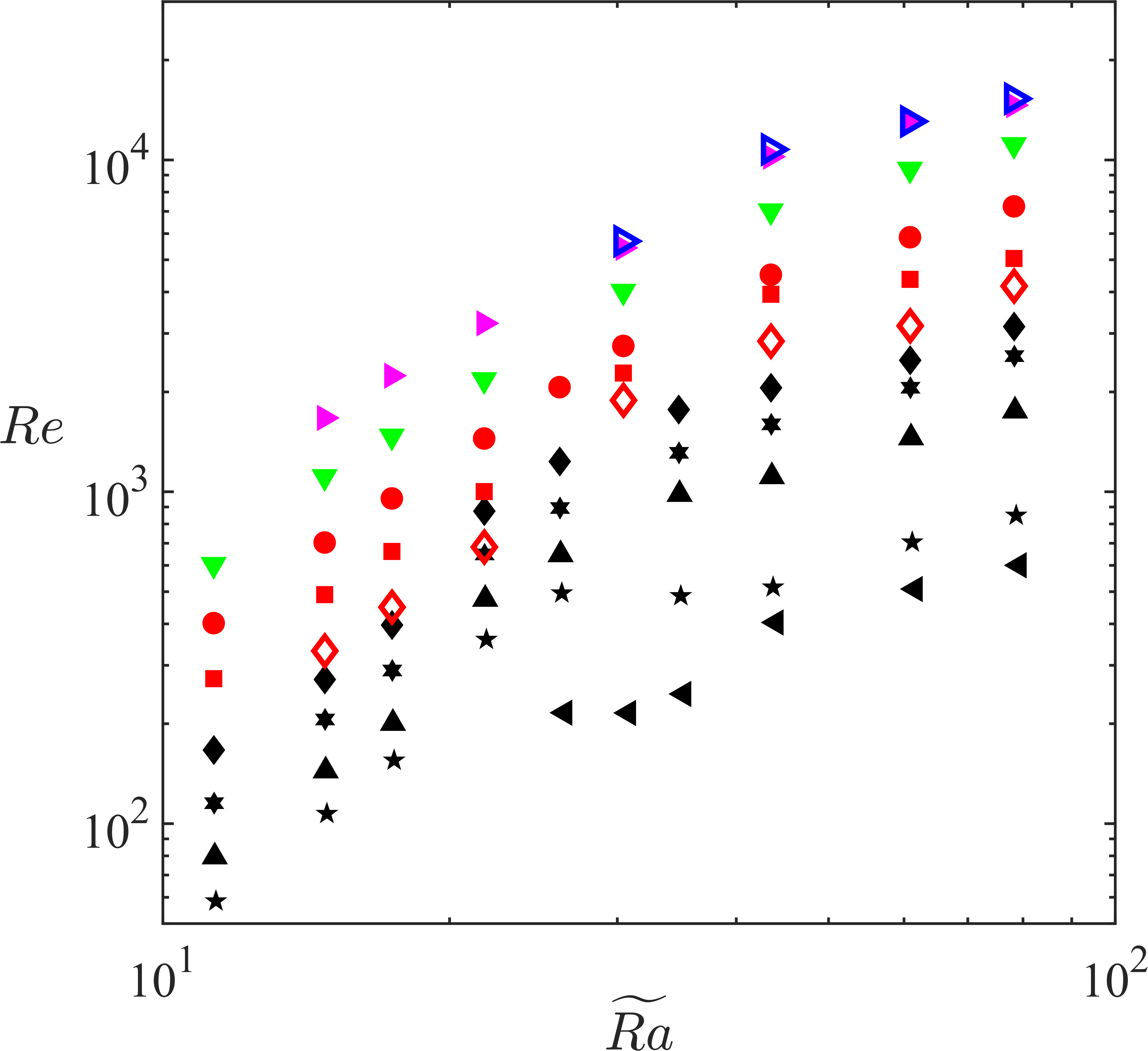} } \quad
           \subfloat[]{
      \includegraphics[width=0.45\textwidth]{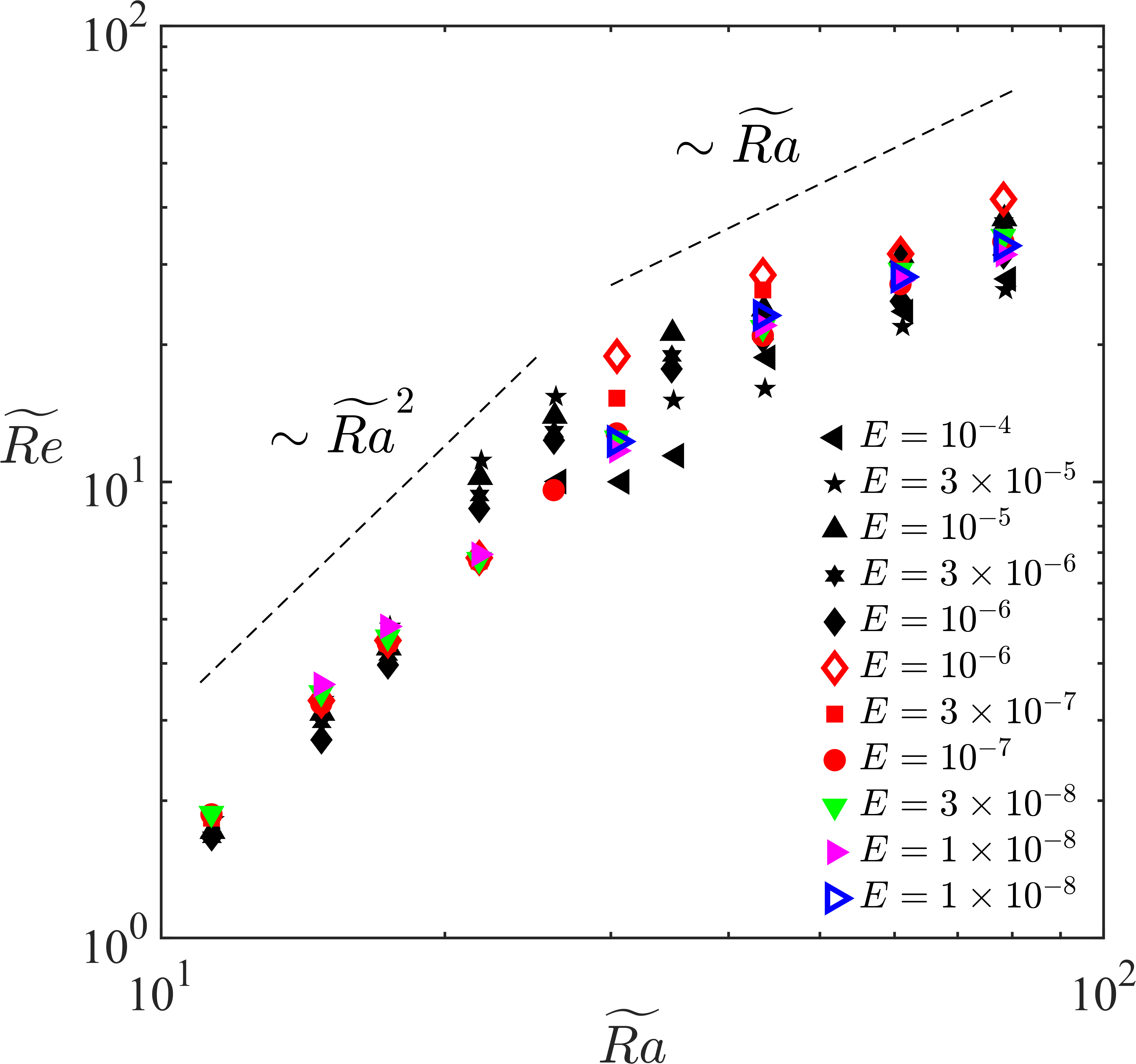} }
       \quad
          \subfloat[]{
             \includegraphics[width=0.45\textwidth]{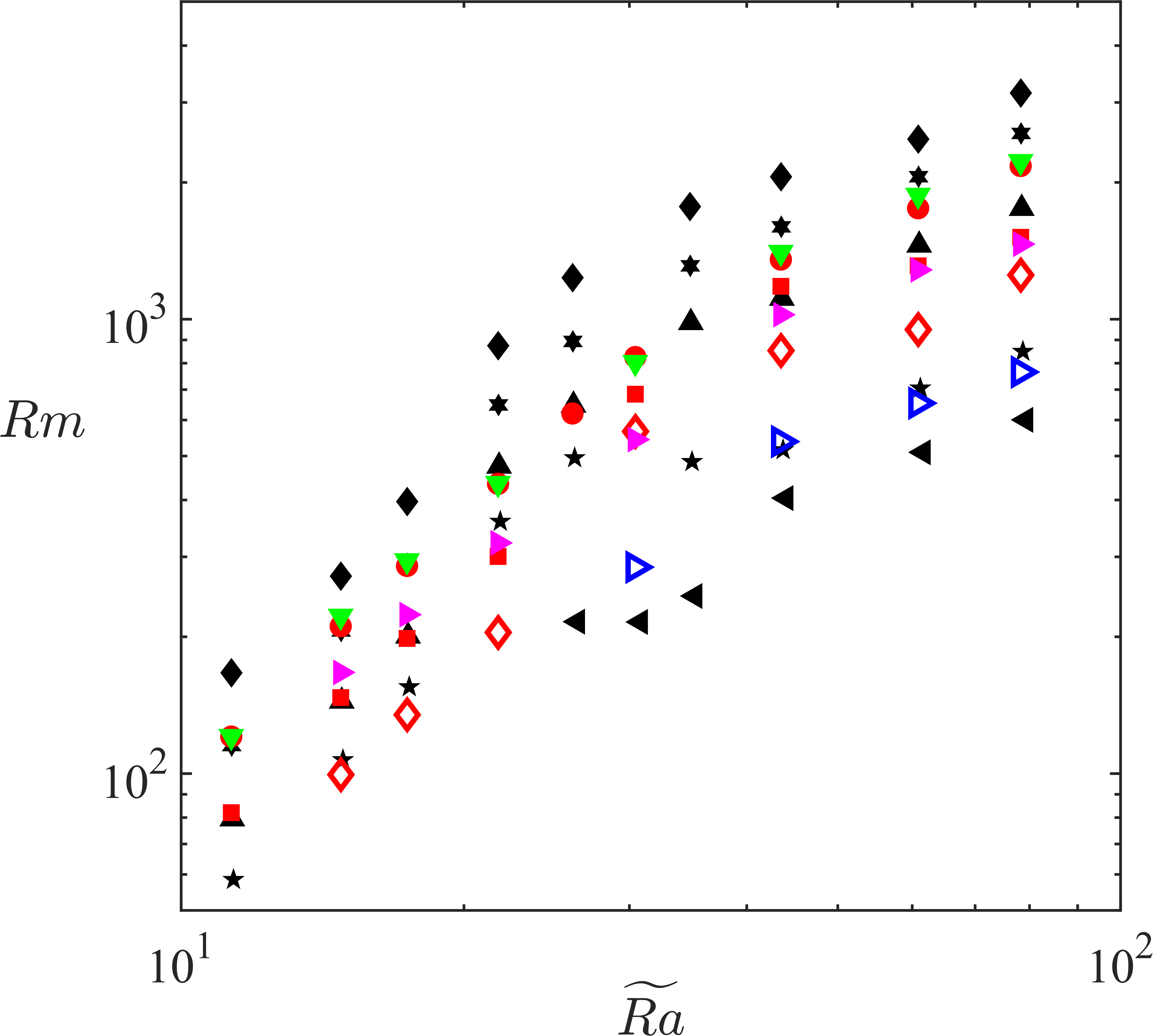} } \quad
           \subfloat[]{
      \includegraphics[width=0.45\textwidth]{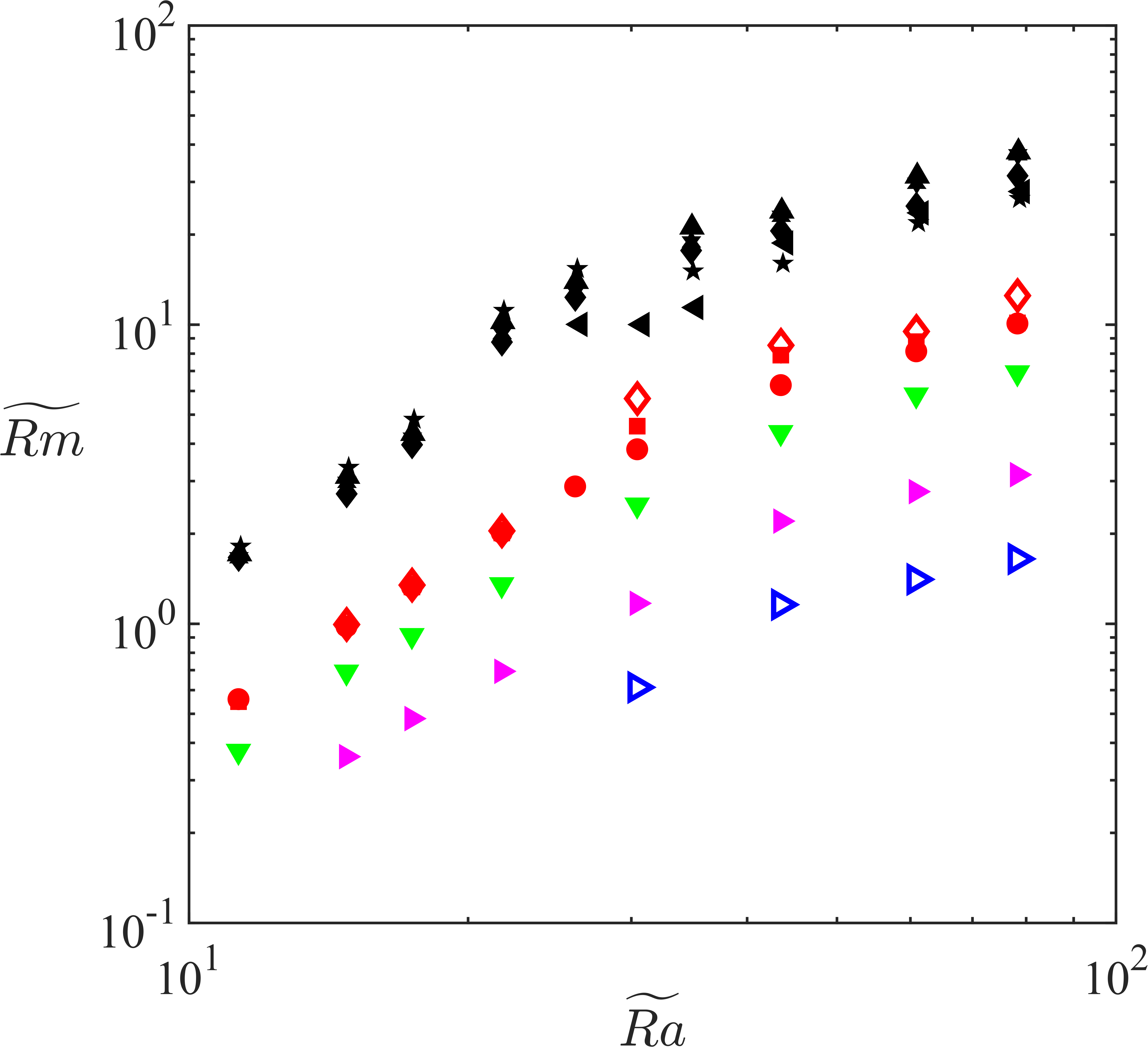} }
 \end{center}
\caption{ Reynolds number and magnetic Reynolds number versus reduced Rayleigh number $\Rat$:  
(a) large scale Reynolds number, $Re$;
(b) small scale Reynolds number, $\Ret =ReE^{1/3}$; 
(c) large scale magnetic Reynolds number, $Rm$; 
(d) small scale magnetic Reynolds number,  $\Rmt =RmE^{1/3}$.  
Symbol shape represents different values of the Ekman number ($E$) and color represents different values of the magnetic ($Pm$): black indicates $Pm=1$; red indicates $Pm=0.3$; green indicates $Pm=0.2$; magenta indicates $Pm=0.1$; blue indicates $Pm=0.05$.
\label{F:scaling}}
\end{figure*}

Figure~\ref{F:scaling}(a)  shows the scaling behaviour of the Reynolds number, $Re$. 
For a fixed value of $\Rat$, the flow speed shows a systematic increase with decreasing Ekman number. We note that for $E=10^{-4}$  and $3\times 10^{-5}$ cases, we observe a regime near $Ra\sim 3Ra_c$  ($\Rat\approx 26$) where the Reynolds number increases slowly (or remains constant) as $Ra$ increases, while the Nusselt number still increases with increasing the $Ra$. \textcolor{black}{Over this same parameter range we also observe a relatively large increase in the ohmic dissipation (as shown in Figure~\ref{F:diss}(b)) and the magnetic energy (discussed in the next subsection) suggesting that the energy from the thermal forcing is transformed into magnetic energy very efficiently for these cases}. The reduced Reynolds number, $\Ret = Re E^{1/3}$, is plotted in figure~\ref{F:scaling}(b) where a collapse is observed. Moreover, that these dynamos are characterised by $\Ret=O(1)$ for a wide range of Ekman numbers suggests that \textcolor{black}{the $Re = O\lb E^{-1/3} \rb$ scaling is the appropriate asymptotic relationship for describing the characteristic flow speeds.} A change in the scaling behavior of $\Ret$ is observed around $Ra\sim 3Ra_c$  ($\Rat \sim 30$). For cases  just above the onset of convection ($Ra < 3Ra_c$), an approximate scaling relation of $\Ret \sim \widetilde{Ra}^ {2}$ is found; while for cases with relatively high  supercriticality ($Ra > 3Ra_c$), \textcolor{black}{the scaling becomes weaker and a trend of $\Ret \sim \widetilde{Ra}$ is shown for reference}.

The magnetic Reynolds number $Rm$ and rescaled magnetic Reynolds number $\Rmt = RmE^{1/3}$ are shown in Figures~\ref{F:scaling}(c) and (d). Values of the magnetic Reynolds number up to $Rm \approx 3000$ are reached for $E=10^{-6}$ and $Pm=1$. The use of $\Rmt$ in Figure~\ref{F:scaling}(d) shows how different values of $Pm$ collapse onto different curves, though we find similar scaling behaviour with $\Rat$. For $Pm < 1$, the $\Rmt < 1$ regime that is relevant to planetary interiors is accessible provided that the Ekman number is also reduced. 
We find that the regime where large scale dynamo action is no longer important occurs at $\Rmt \approx 5$; a value of $\Rmt \approx 13.5$ for the transition was identified by \cite{aT12}, though larger values of $Pm$ were employed in that investigation.

\begin{figure*}
 \begin{center}

                  \subfloat[]{
      \includegraphics[width=0.45\textwidth]{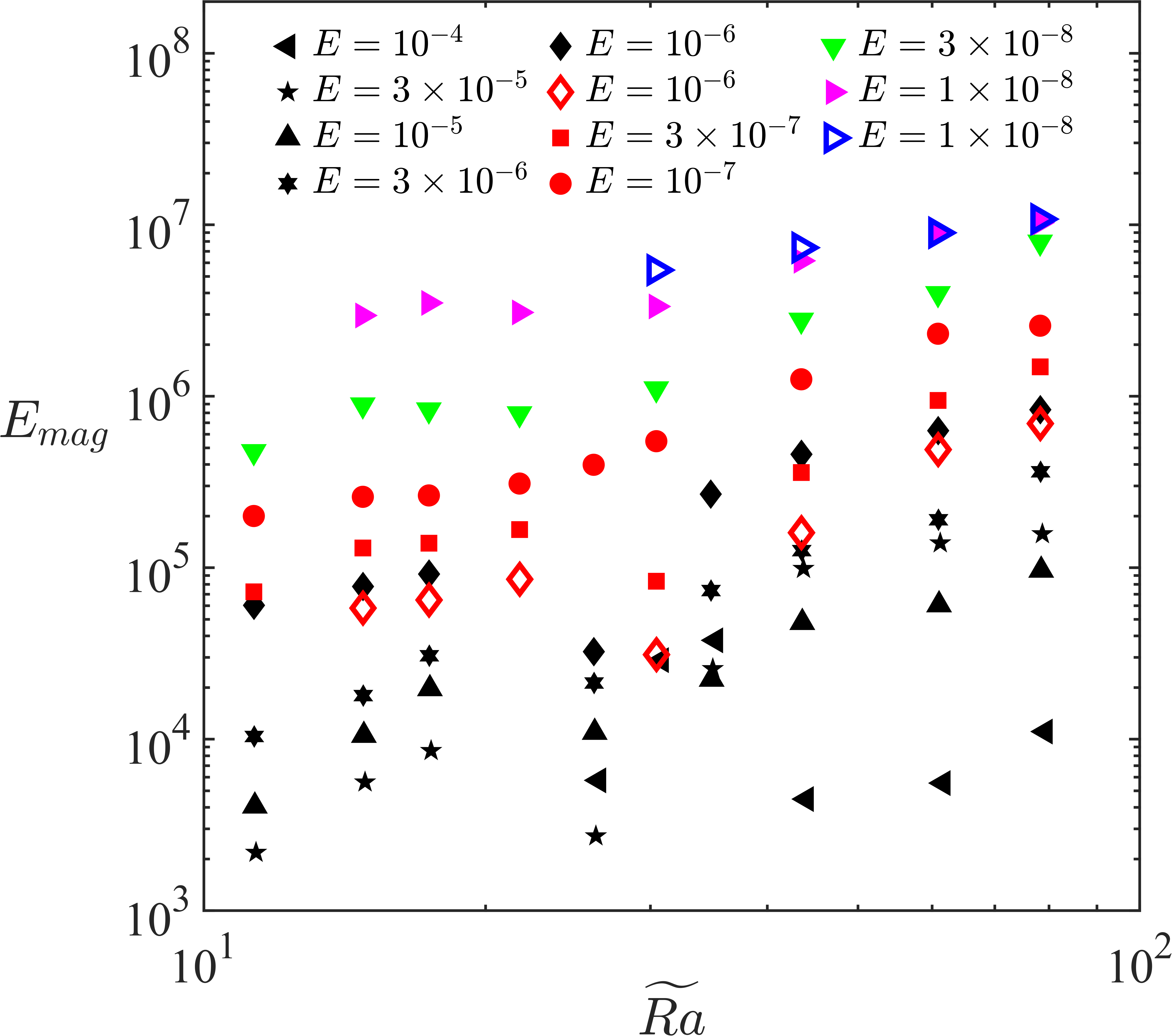} } \quad
                 \subfloat[]{
      \includegraphics[width=0.43\textwidth]{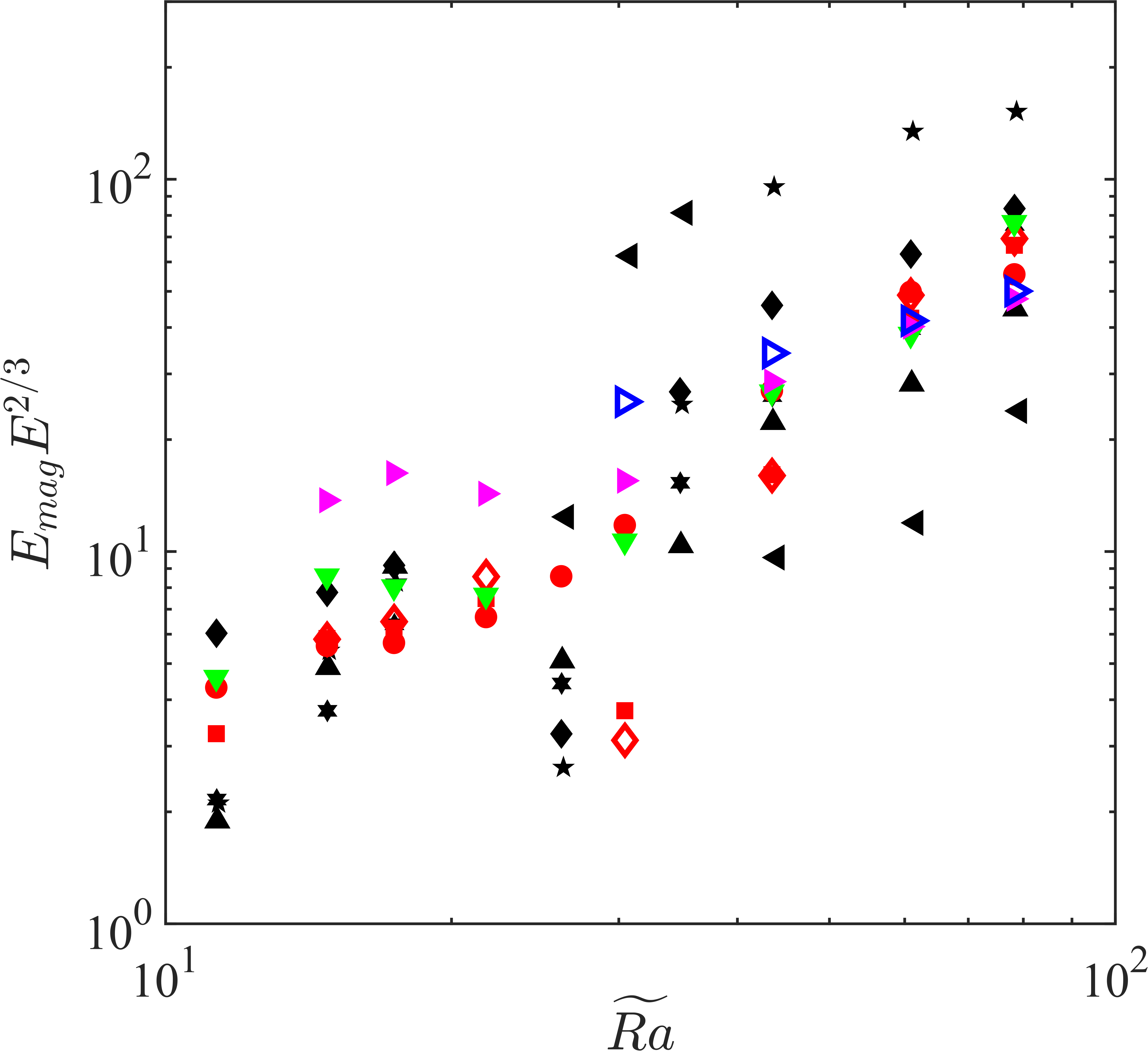} }
      \quad
                    \subfloat[]{
      \includegraphics[width=0.45\textwidth]{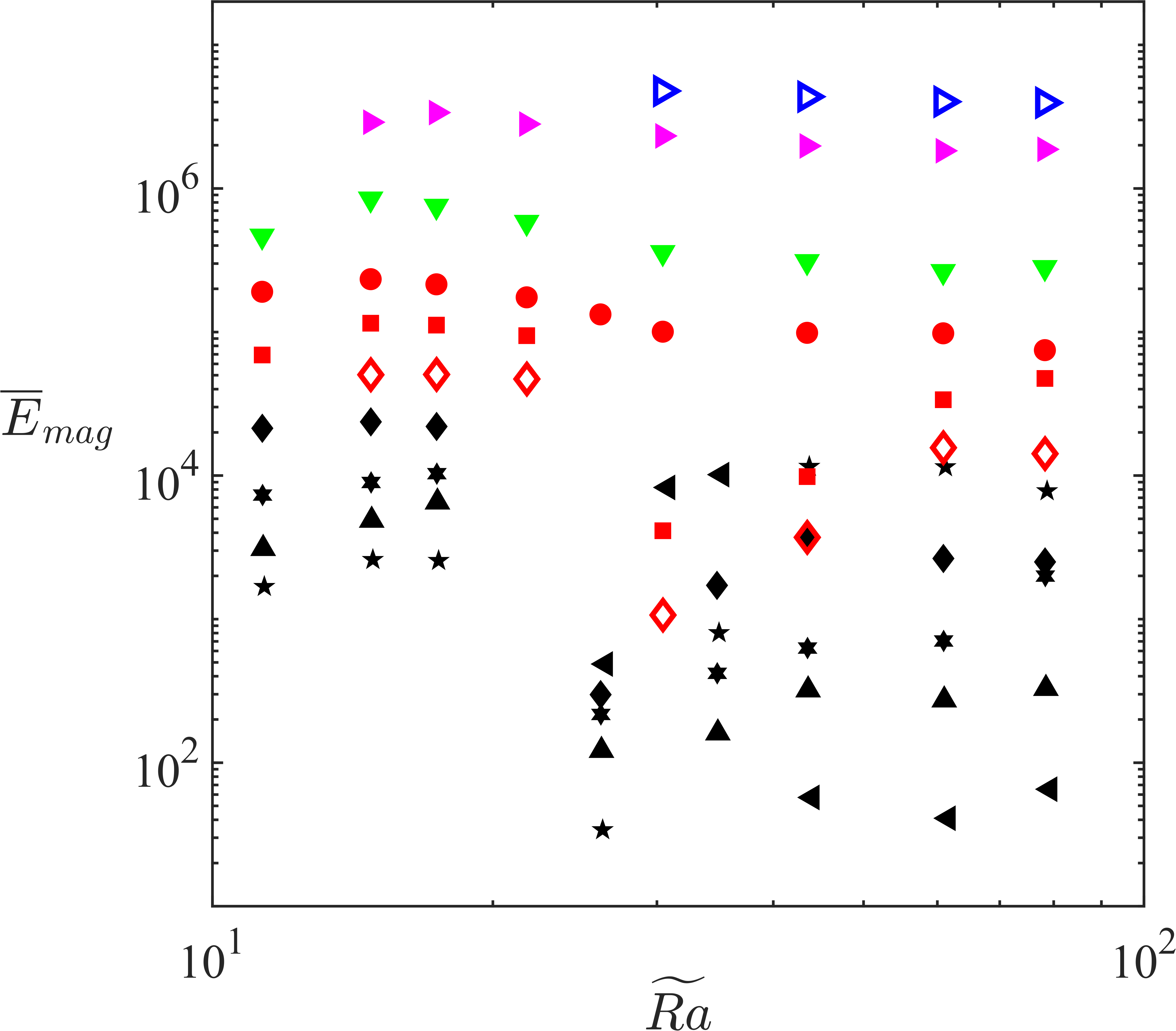} } \quad
                 \subfloat[]{
      \includegraphics[width=0.44\textwidth]{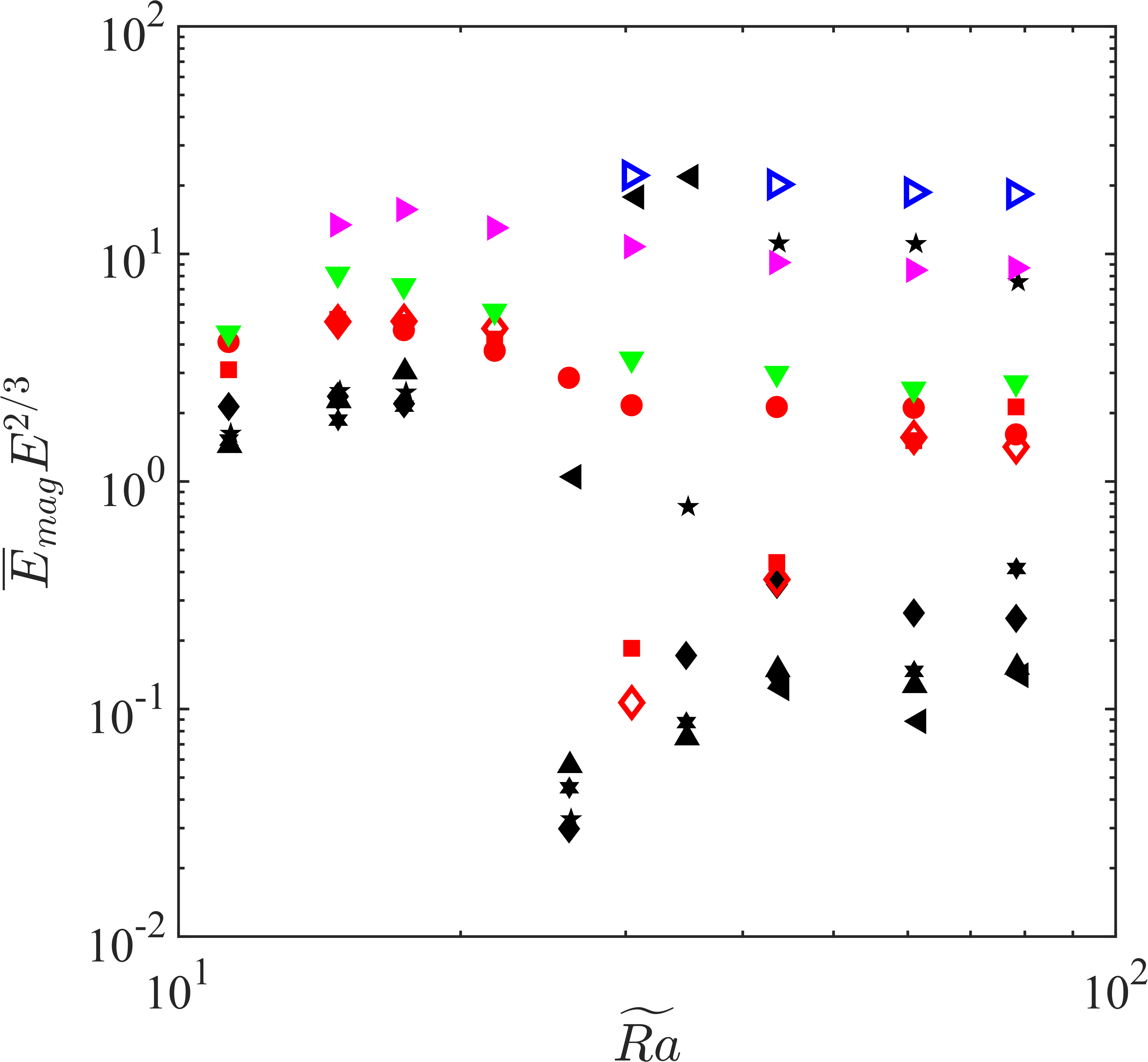} }      
 \end{center}
\caption{Magnetic energy for all simulations: 
 (a) $E_{mag}$ versus $\Rat$; 
 (b) rescaled magnetic energy, $\widetilde{E}_{mag}=E_{mag}E^{2/3}$, versus $\Rat$; 
 (c) mean magnetic energy $\overline{E}_{mag}$ versus $\Rat$; 
  (c) rescaled  mean magnetic energy $\overline{E}_{mag}E^{2/3}$ versus $\Rat$.
Symbol shape represents different values of the Ekman number ($E$) and color represents different values of the magnetic ($Pm$): black indicates $Pm=1$; red indicates $Pm=0.3$; green indicates $Pm=0.2$; magenta indicates $Pm=0.1$; blue indicates $Pm=0.05$.
 \label{F:Em}}
\end{figure*}

The magnetic energy $E_{mag}$ and the rescaled magnetic energy  $\widetilde{E}_{mag} = E_{mag}E^{2/3}$ are shown in Figures \ref{F:Em}(a) and (b), respectively. For the majority of our cases, a smaller value of the Ekman number tends to produce a stronger magnetic field (as measured by the magnetic energy) when $\Rat$ and $Pm$ are fixed. \textcolor{black}{For all cases with $Pm=1$, and cases with $Pm = 0.3$ and $E \ge 3\times 10^{-7}$, there is either a significant drop in magnetic energy or a lack of dynamo action in the approximate range $20 \lesssim \Rat \lesssim 30$. This behavior was also observed in the investigations of \cite{aT12} and \cite{cG17} for $E \ge 10^{-6}$. However, this drop in magnetic energy becomes less significant or is not observed at all for cases with $E \lesssim 1\times 10^{-7}$. Although we have not investigated this effect in detail, it appears that as the Ekman number is reduced, the small scale dynamo has already been fully activated once large scale dynamo action ceases. As suggested in Figure~\ref{F:Em}(b), a factor of $E^{2/3}$ appears to collapse the majority of the data to order unity values, though significant scatter in the data remains due to differences in $Pm$ and the corresponding dynamo behavior.}

The mean magnetic energy is shown in Figure \ref{F:Em}(c); the corresponding asymptotically rescaled data is shown in panel (d). In the large scale dynamo regime, the mean magnetic energy appears to saturate and then decreases slightly as $\Rat$ increases. This behaviour is similar to that observed in spherical dynamo simulations, which show a saturation of the axisymmetric component of the magnetic field \citep{mC21,rO21}. The exact cause of this effect is not currently known, but we speculate that it may be due to the breakdown of the $\alpha^2$-dynamo that occurs at finite $\Rmt$, perhaps related to so-called $\alpha$-quenching \citep[e.g.][]{sV92,fC96}. As shown in previous work \citep[e.g.][]{mS66a,kM19}, dynamos can only be rigorously classified as $\alpha^2$ if the \textcolor{black}{ small scale} magnetic Reynolds number is small. In the rapidly rotating limit in which significant separation exists between the small horizontal convective length scale and the layer depth, the requirement for an $\alpha^2$-dynamo is that $\Rmt \ll 1$ \citep{mC15b}. In contrast, for $\Rmt = O(1)$ there is no rigorous closure relating the large and small scale magnetic field components.

 \begin{figure*}
 \begin{center}
      \includegraphics[width=0.5\textwidth]{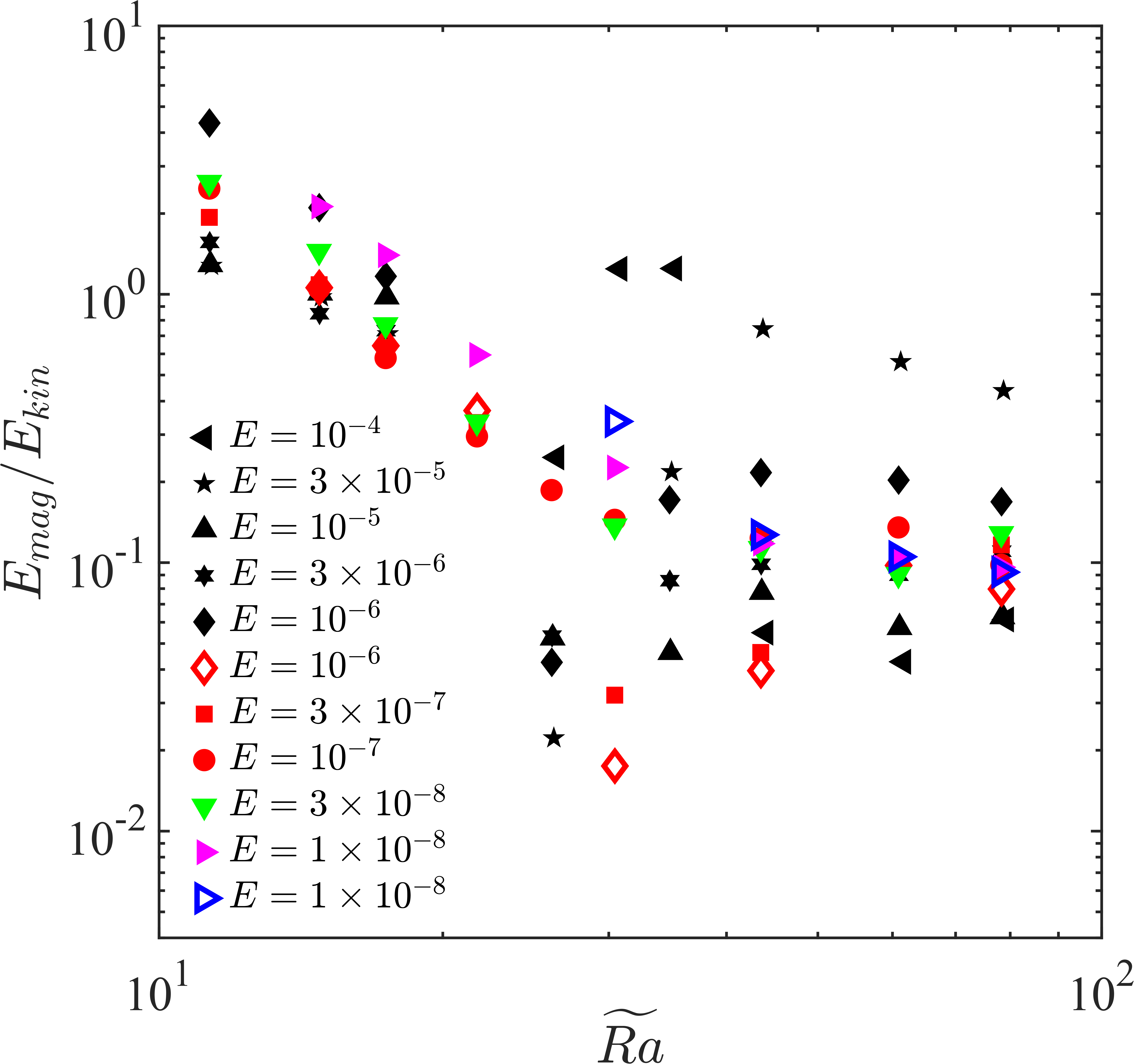} 
         
 \end{center}
\caption{ Ratio of magnetic energy to kinetic energy for all simulations.
Symbol shape represents different values of the Ekman number ($E$) and color represents different values of the magnetic ($Pm$): black indicates $Pm=1$; red indicates $Pm=0.3$; green indicates $Pm=0.2$; magenta indicates $Pm=0.1$; blue indicates $Pm=0.05$.
\label{F:Efraction}}
\end{figure*} 

\textcolor{black}{We can further characterise the dynamos by plotting the ratio of the magnetic energy to the kinetic energy, as shown in Figure \ref{F:Efraction}. We find that the energy ratio is greater than unity only for relatively small Rayleigh numbers, $\Rat < 20$, indicating that kinetic energy dominates in the majority of our simulations. The decrease of this ratio with increasing $\Rat$ suggests that small scale dynamos are more likely to yield smaller magnetic energy relative to kinetic energy. Within this small scale dynamo regime we find many of the simulations have an energy ratio of $O(10^{-1})$. }

\textcolor{black}{The asymptotic theory predicts that the energy ratio becomes large in the limit $\Rmt \rightarrow 0$. While we do find energy ratios that exceed unity, for computational reasons the majority of our simulations are within the regime $\Rmt = O(1)$, so it might be expected that we do not observe large values of the energy ratio. In the Earth's outer core, $\Rmt = O(10^{-2})$, and the energy ratio is thought to be as large as $\approx 10^4$. Spherical dynamo investigations find energy ratios that exceed those observed here, though these values rarely exceed $10$. The reader is referred to \cite{nS17} where a comparison of these values is made for several spherical dynamo investigations. A possible reason for the difference in values of this energy ratio for the two geometries may simply be related to the magnetic field morphology. Whereas spherical dynamos can generate large scale field in both the horizontal and vertical directions, the plane layer is only capable of generating large scale field in the horizontal direction.}

\subsection{Length scales and energy spectra}


In this subsection we quantify the length scales present in both the velocity field and the magnetic field using a combination of energy spectra and Taylor microscales. \textcolor{black}{This procedure is important for understanding how the length scales depend on the non-dimensional parameters.} Both the kinetic and magnetic energy spectra are separated into horizontal and vertical components denoted by superscripts $H$ and $V$, respectively. The spectra are averaged in time and depth.
For example, the (time-averaged) kinetic energy spectra at depth $z$ are defined as
\be
\widehat{E}_{kin}^{H}(k,z)=\frac{1}{2E^2} \sum_{k} (\widehat{u}^*\widehat{u}  +\widehat{v}^*\widehat{v} ),
\ee
and
\be
\widehat{E}_{kin}^{V}(k,z)=\frac{1}{2E^2} \sum_{k} (\widehat{w}^*\widehat{w} ) ,
\ee
where $\widehat{u}$, $\widehat{v}$, and $\widehat{w}$ are the Fourier coefficients of the three velocity field components, and the superscript $(*)$ denotes a complex conjugate. 
\textcolor{black}{The horizontal wavenumber is $\mathbf{k} = (k_x,k_y)$, where the modulus is denoted by $k = \sqrt{k_x^2 + k_y^2}$.}
The corresponding depth-averaged spectra are then computed via
\be
\widehat{E}_{kin}^{H}(k)=\int_{0}^{1} \widehat{E}_{kin}^{H}(k,z) dz,
\label{E:Hkin}
\ee
and 
\be
\widehat{E}_{kin}^{V}(k)=\int_{0}^{1} \widehat{E}_{kin}^{V}(k,z) dz.
\label{E:Vkin}
\ee

\begin{figure*}
 \begin{center}
                  \subfloat[]{
      \includegraphics[width=0.45\textwidth]{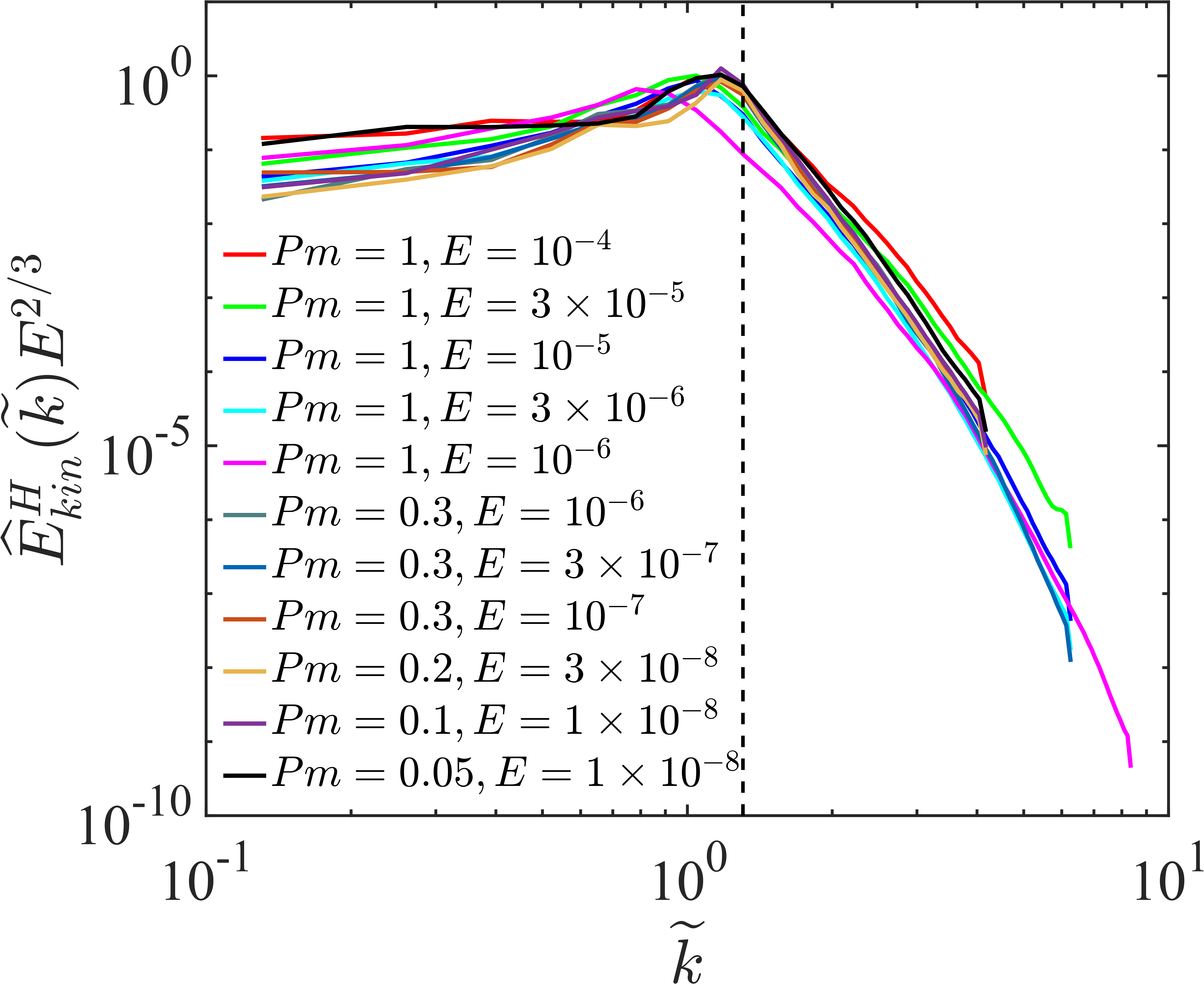} }
      \quad
                    \subfloat[]{
      \includegraphics[width=0.47\textwidth]{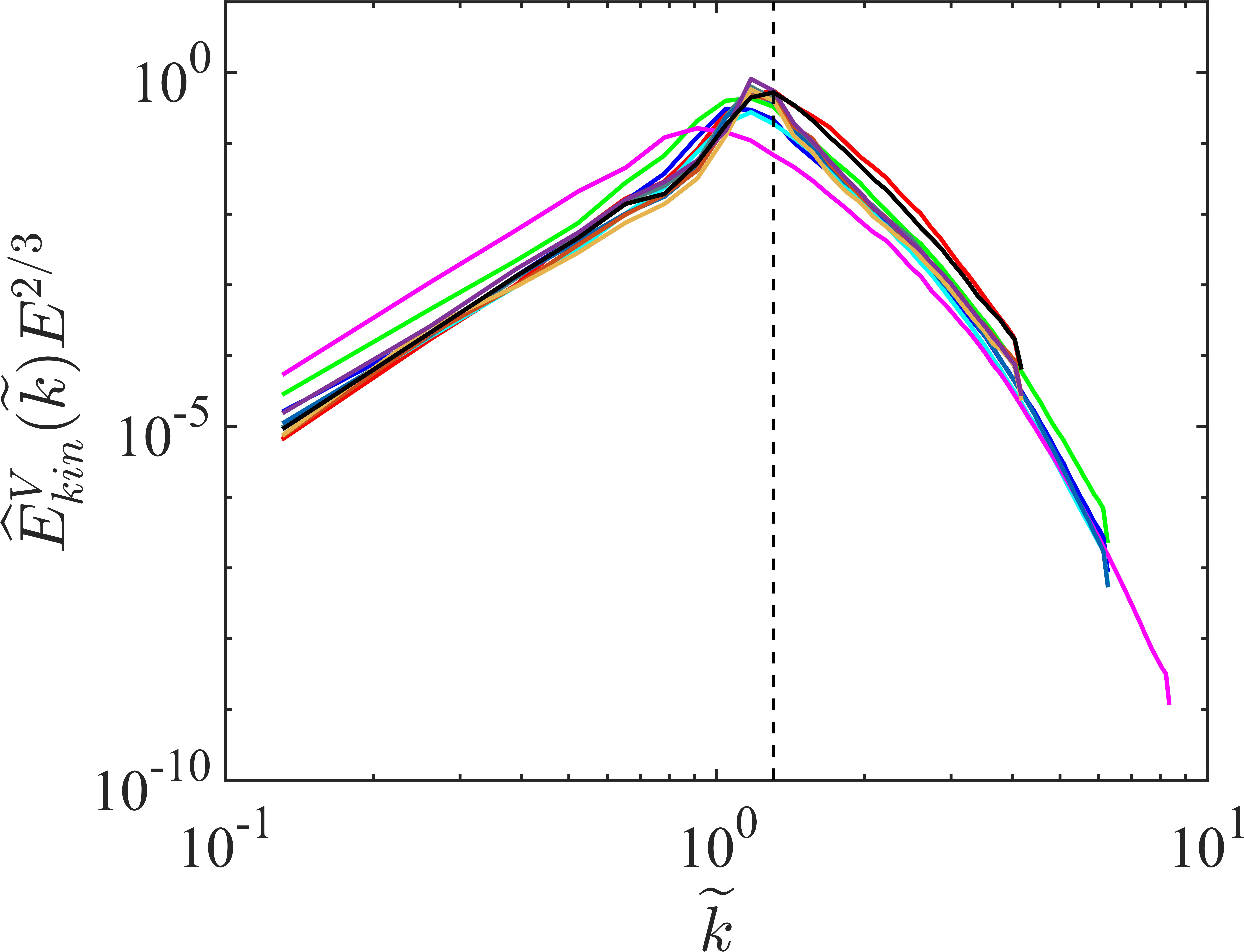} } \\
                  \subfloat[]{
      \includegraphics[width=0.47\textwidth]{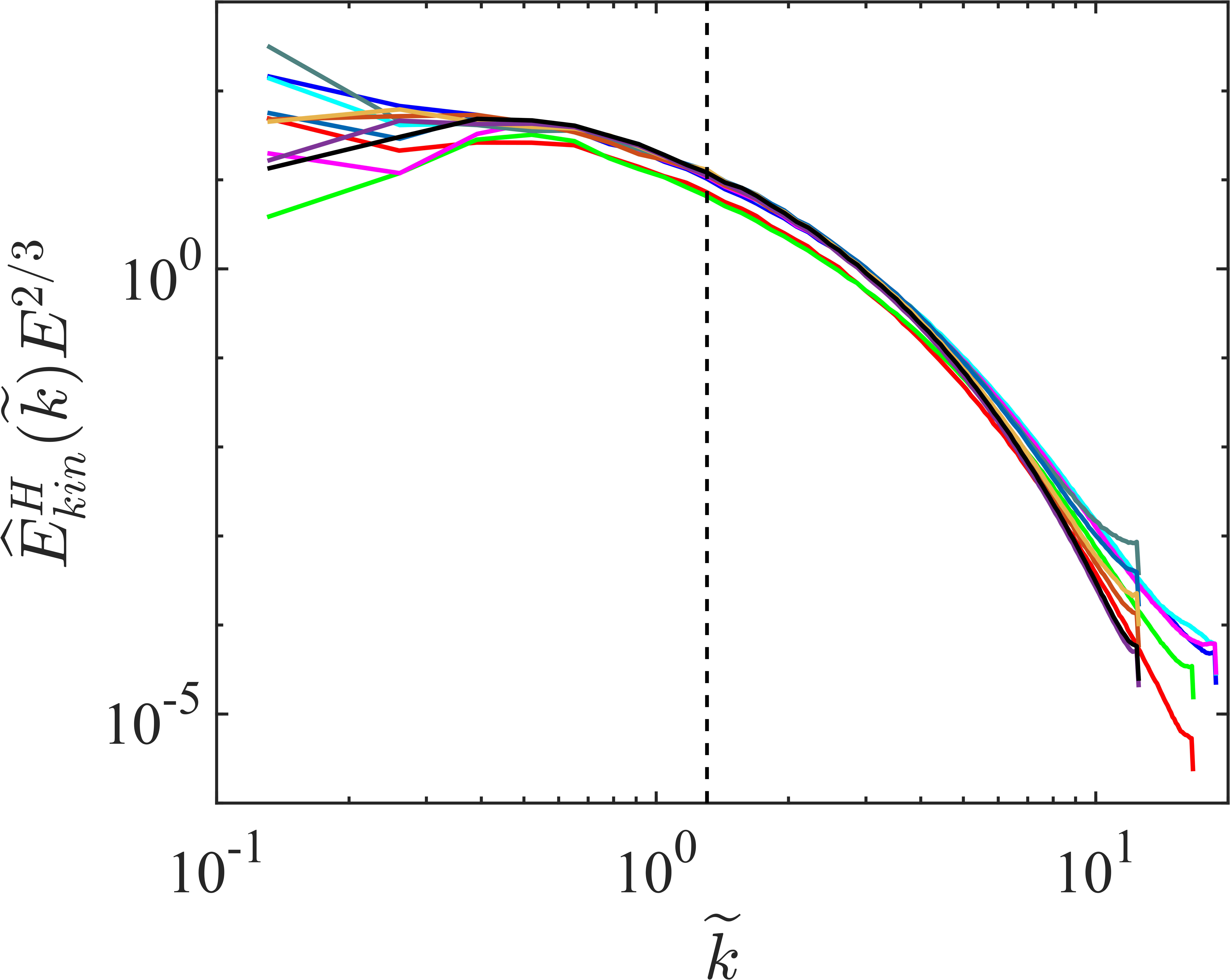} }
      \quad
                    \subfloat[]{
      \includegraphics[width=0.47\textwidth]{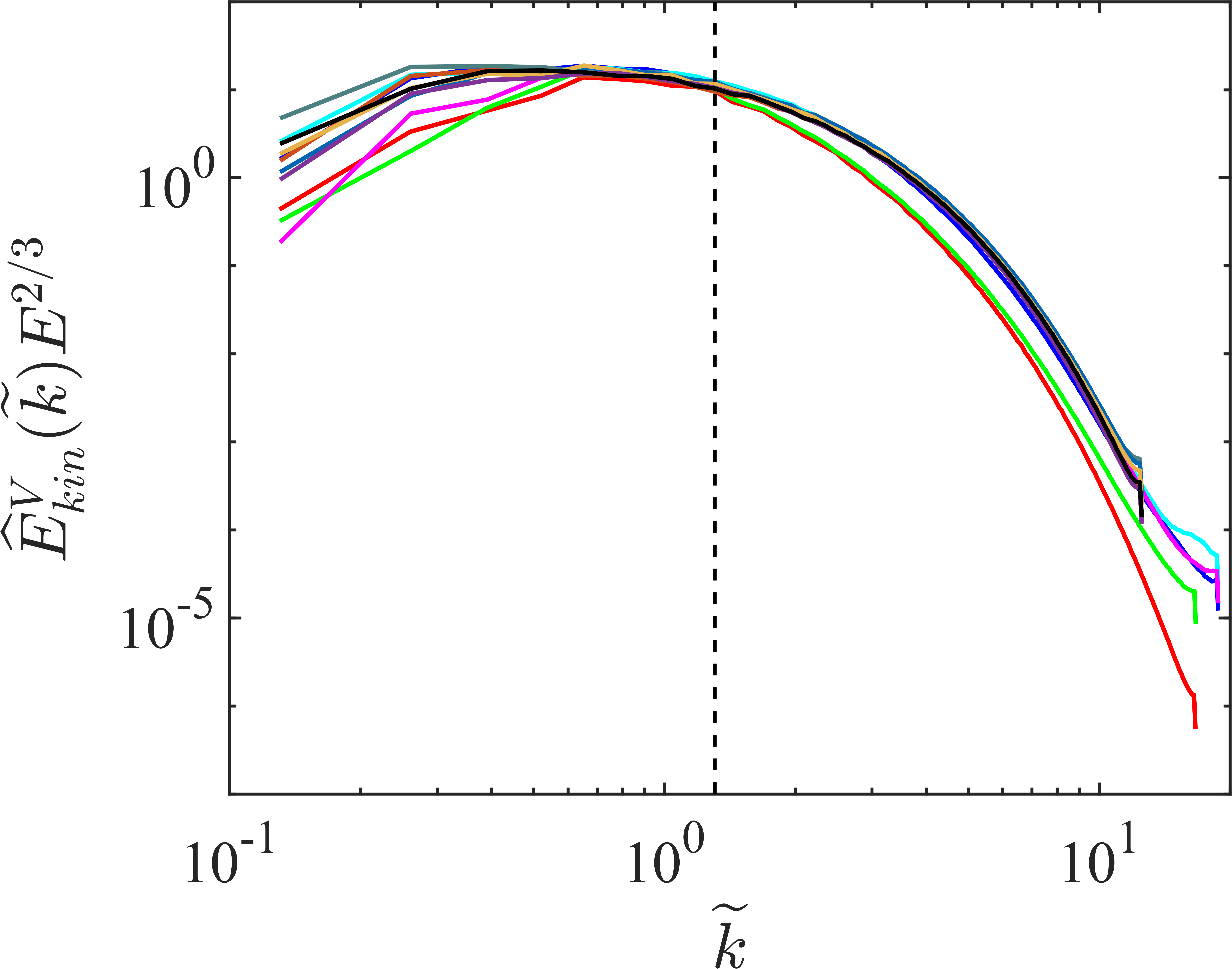} } 
 \end{center}
\caption{ Asymptotically rescaled horizontal (a,c) and vertical (b,d) kinetic energy spectra for (a,b) $Ra=1.7Ra_c$ ($\Rat\approx 15$) and (b,d) $Ra=9Ra_c$ ($\Rat\approx 78$). The asymptotically rescaled horizontal wavenumber is defined by $\kt = k E^{1/3}$.
The vertical dashed line denotes the asymptotically rescaled critical horizontal wavenumber, $\kt_c \approx 1.3048$.
 \label{F:EkSPE}}
\end{figure*}

Asymptotically rescaled kinetic energy spectra are shown in Figure~\ref{F:EkSPE} for two representative Rayleigh numbers: a relatively low value of $Ra=1.7Ra_c$ ($\Rat\approx 15$) is shown in panels (a,b) and the highest Rayleigh number of $Ra=9Ra_c$ ($\Rat\approx 78$) in shown in panels (c,d). As illustrated in Figure~\ref{F:vorticityrendering_E1e8}, cases with $\Rat\approx 15$ can be considered quasi-laminar, whereas cases with $\Rat\approx 78$ are turbulent. The spectra are plotted in terms of the rescaled horizontal wavenumber, $\kt = k E^{1/3}$; the critical value of $\kt_c \approx 1.3048$ is shown by the dashed vertical line. The kinetic energy scaling of $\widehat{E}_{kin}^{H}(k) \sim E^{-2/3}$ (or $\widehat{E}_{kin}^{V}(k) \sim E^{-2/3}$) was given in equation \eqref{E:kin_scale}. We note that all of the cases shown have qualitatively, and to some degree, quantitatively, similar behavior across varying Ekman and magnetic Prandtl numbers. The collapse of the spectra in wavenumber space indicates that all length scales in the flow scale as $E^{1/3}$, even for turbulent flows. These similarities in the spectra are expected given the asymptotic state of the system (i.e.~small Ekman and Rossby numbers). 
As expected, a peak in the spectra is observed near $\kt_c \approx 1.3048$ for  cases near convection onset (i.e.~$\Rat\approx 15$); by comparison the spectra for $\Rat\approx 78$ are flatter in the vicinity of $\kt_c$. Figure~\ref{F:EkSPE}(a) shows that no dominant large scale horizontal motion, as characterised by significant energy in the smallest wavenumbers,  is observed near convection onset. On the other hand, Figure~\ref{F:EkSPE}(b) suggests that strong large scale horizontal flows develop for sufficiently large $\Rat$, as indicated by the peak at the lowest rescaled wavenumber. We do not observe an obvious systematic influence of $Pm$ on the kinetic energy spectra. For instance, whereas some cases with $Pm < 1$ show a tendency for LSV formation with a peak in the horizontal kinetic energy spectra at the smallest rescaled wavenumber, other cases with different Ekman numbers do not exhibit an obvious signature of LSV formation. We note that when $Pm$ and $\Rat$ are fixed, the large scale horizontal motion appears to be suppressed as $E$ decreases. Although we do not investigate this observation in detail, it is likely due the effect of magnetic damping on these flows \citep[e.g.][]{cG17, sM19}.

\begin{figure*}
 \begin{center}
                  \subfloat[]{
      \includegraphics[width=0.43\textwidth]{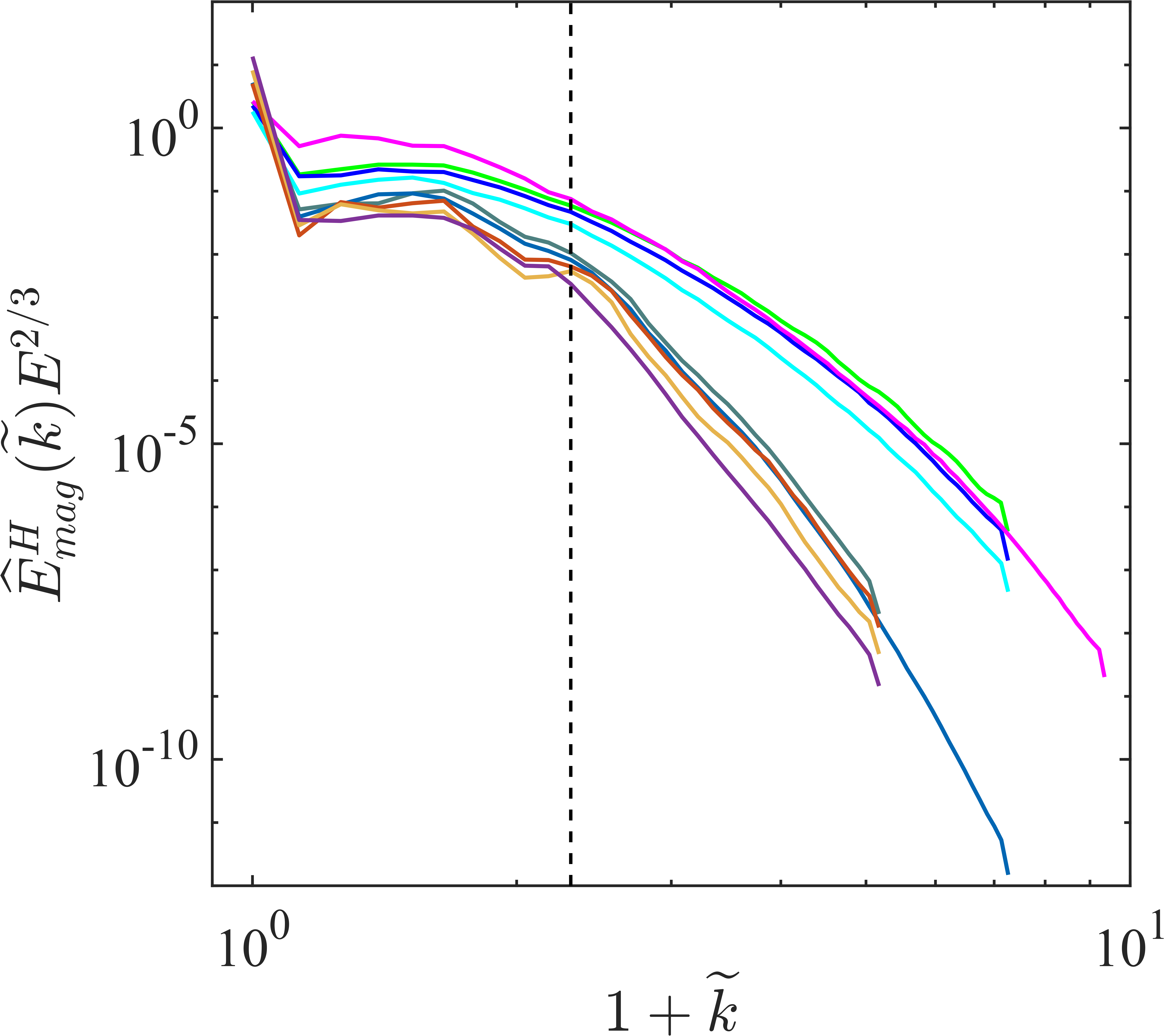} }
      \quad
                    \subfloat[]{
      \includegraphics[width=0.43\textwidth]{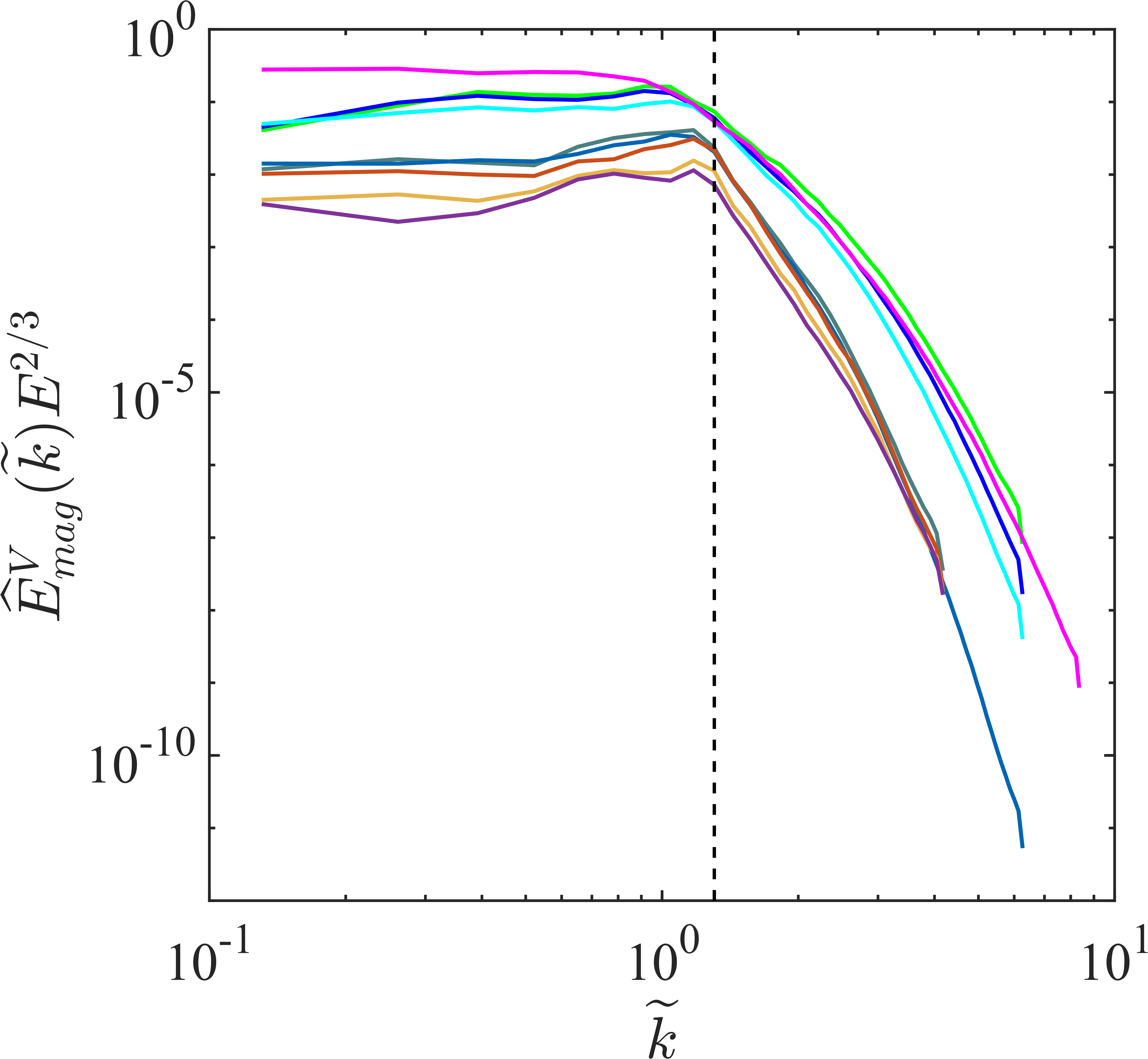} } \\
                  \subfloat[]{
      \includegraphics[width=0.43\textwidth]{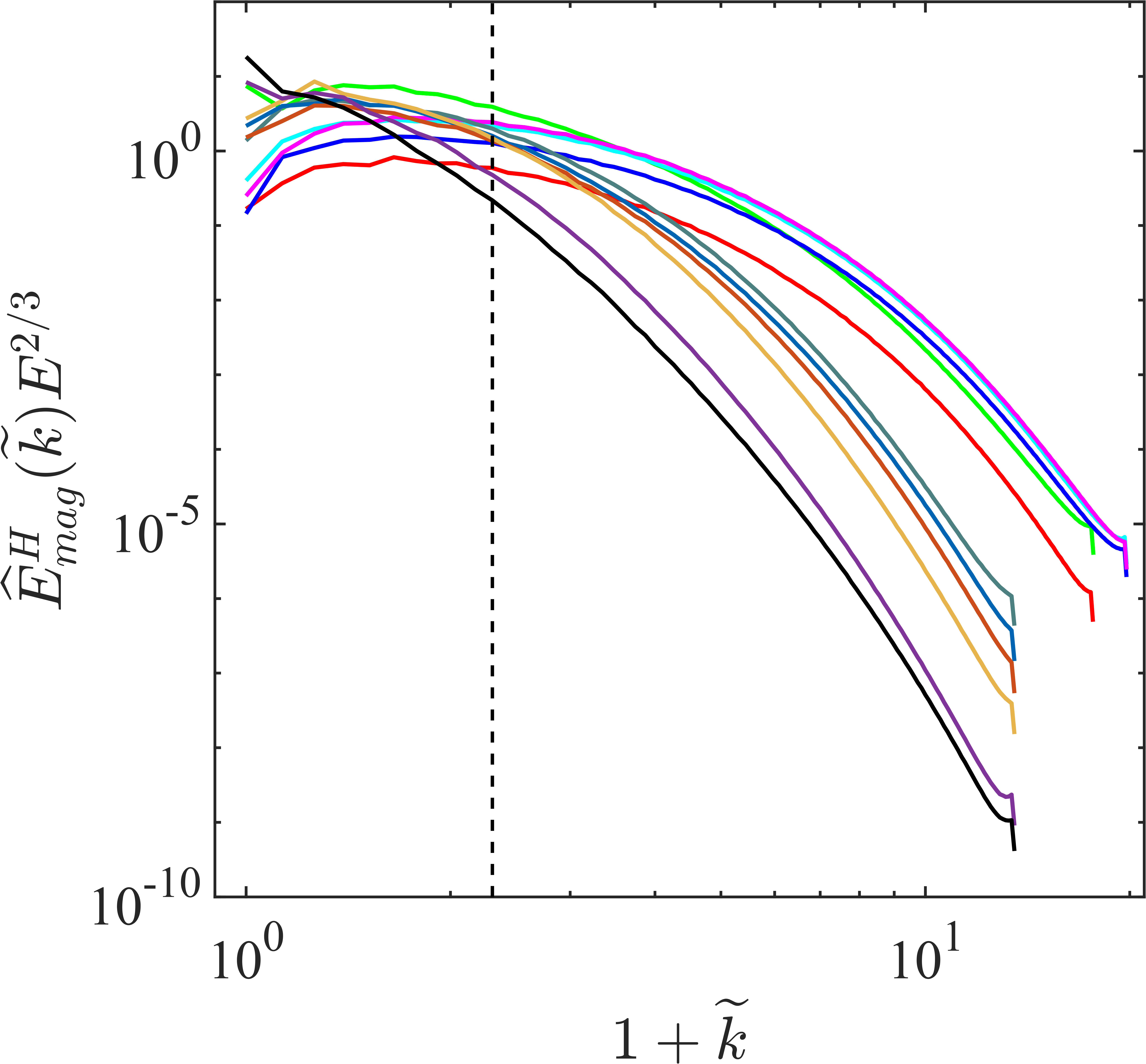} }
      \quad
                    \subfloat[]{
      \includegraphics[width=0.43\textwidth]{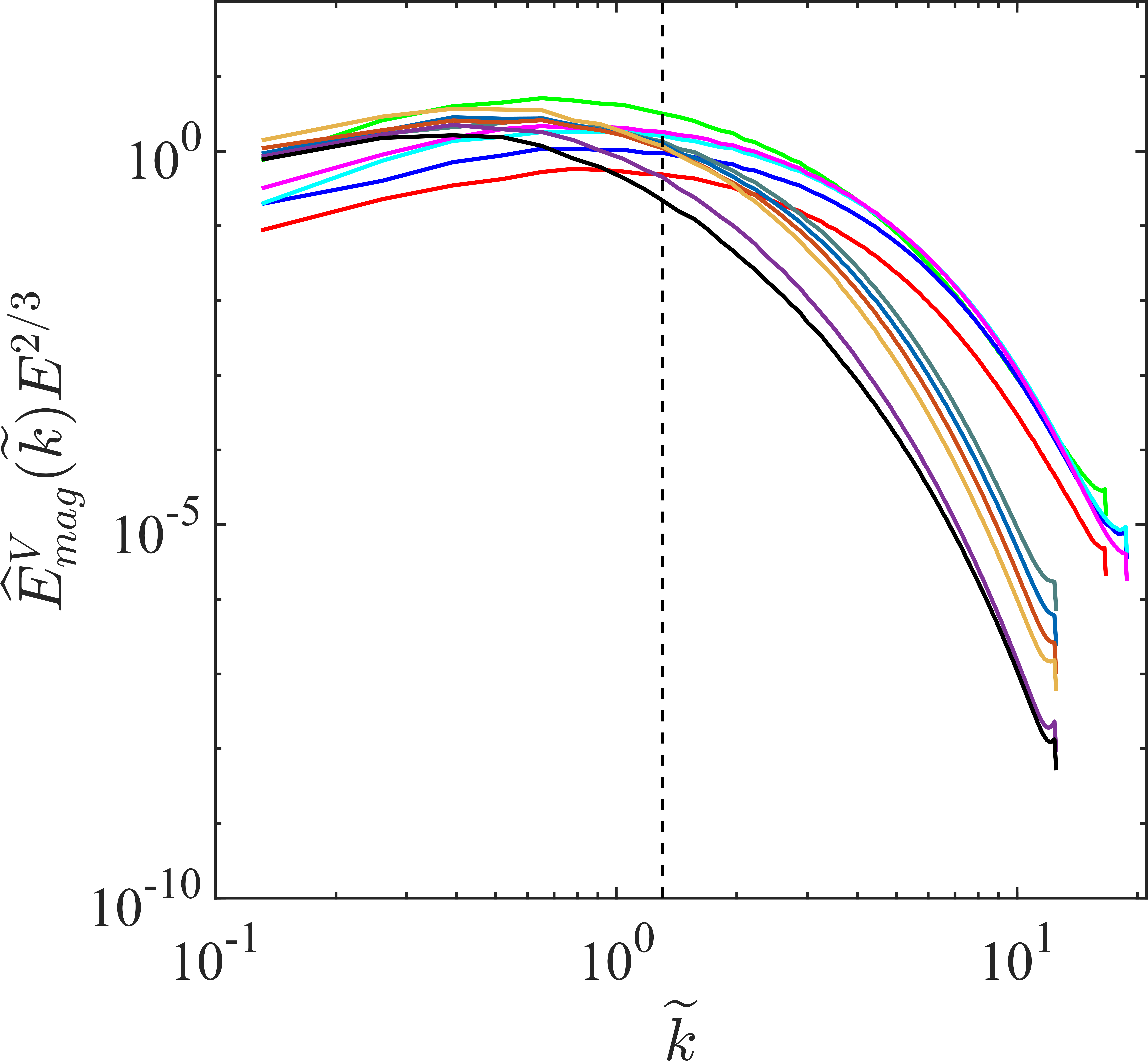} } 
 \end{center}
\caption{ Asymptotically rescaled horizontal (a,c) and vertical (b,d) magnetic energy spectra for (a,b) $Ra=1.7Ra_c$ ($\Rat\approx 15$) and (b,d) $Ra=9Ra_c$ ($\Rat\approx 78$). The asymptotically rescaled horizontal wavenumber is defined by $\kt = k E^{1/3}$.
The vertical dashed line denotes the asymptotically rescaled critical horizontal wavenumber, $\kt_c \approx 1.3048$. The colors have the same meaning as defined in Figure \ref{F:EkSPE}.
 \label{F:EmSPE}}
\end{figure*}

The magnetic energy equivalent of equations \eqref{E:Hkin} and \eqref{E:Vkin} are also computed and shown in Figure~\ref{F:EmSPE}. As for the kinetic energy spectra, we observe a collapse of the spectra with respect to the wavenumber scaling. In contrast to the kinetic energy spectra, there is considerably more spread in the rescaled magnitudes of the spectra, though the most energetic modes rescale approximately to $O(1)$ values. This spread is consistent with data reported in previous subsections. For instance, the relative size of the mean field varies considerably for a given value of $\Rat$ (and $Pm$). We notice that for $\Rat \approx 15$, all of the dynamos show a peak in the horizontal spectra at the smallest value of $\kt$, indicating the presence of a mean magnetic field. We observe two classes of dynamo spectra: cases with $Pm < 1$ show a steep drop in magnitude for $\kt \gtrsim \kt_c$, whereas cases with $Pm = 1$ show a much weaker drop in magnitude. At this same value of $\Rat\approx15$, the corresponding vertical spectra show a nearly flat spectra for $\kt \lesssim \kt_c$, and a rapid drop in magnitude for $\kt \gtrsim \kt_c$. 
This distinction between spectra with $Pm < 1$ and those with $Pm=1$ persists at $\Rat \approx 78$ shown in Figures~\ref{F:EmSPE}(b) and (d), although these higher Rayleigh number spectra are understandably more broadband in structure. The rapid drop in the spectra with increasing $\widetilde{k}$ for $Pm < 1$ is expected when magnetic diffusion and mean field stretching balance in the fluctuating induction equation \citep{gG60,aS07}.

  \begin{figure*}
 \begin{center}
                  \subfloat[]{
      \includegraphics[width=0.43\textwidth]{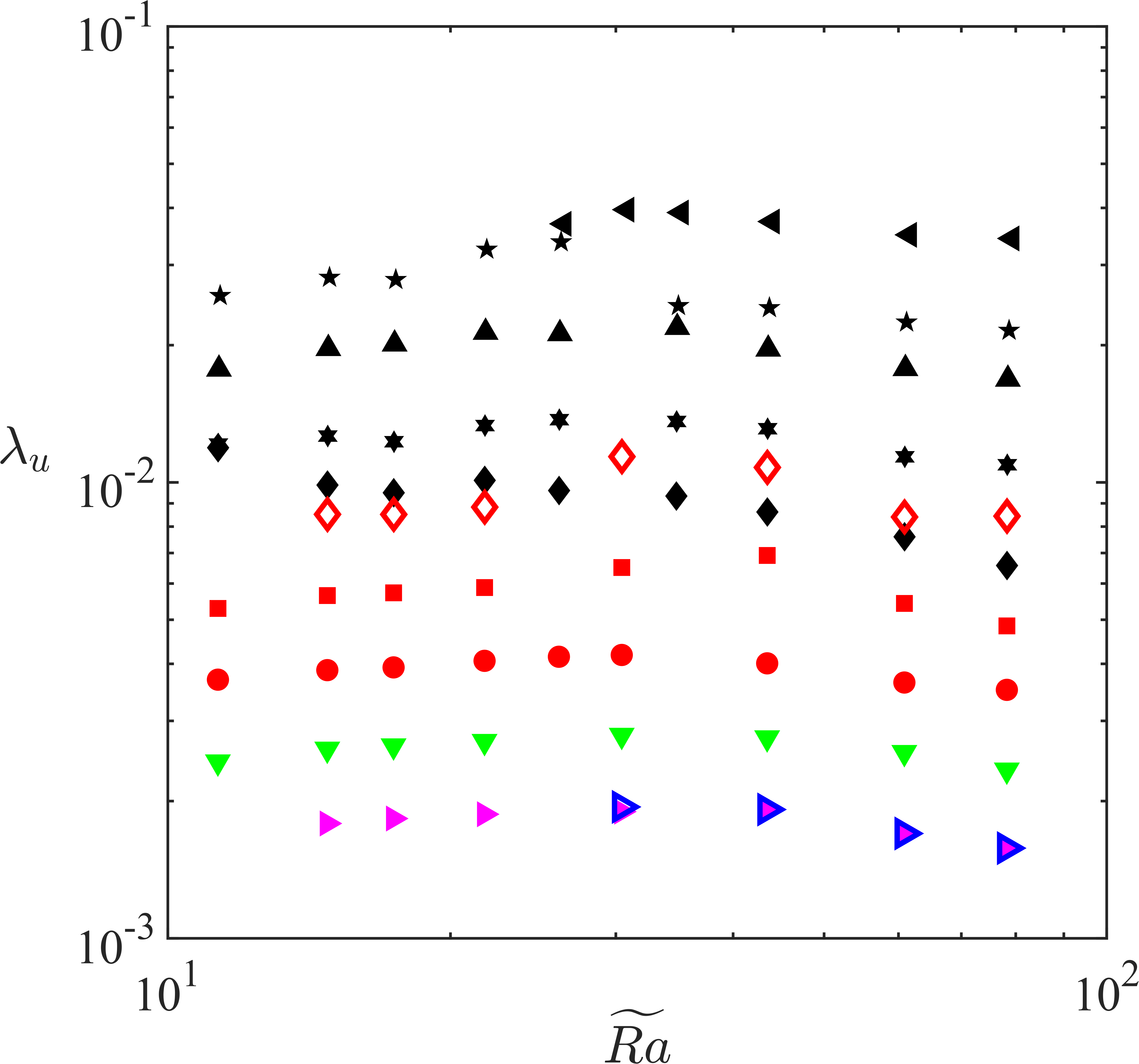} } \quad
                  \subfloat[]{
      \includegraphics[width=0.43\textwidth]{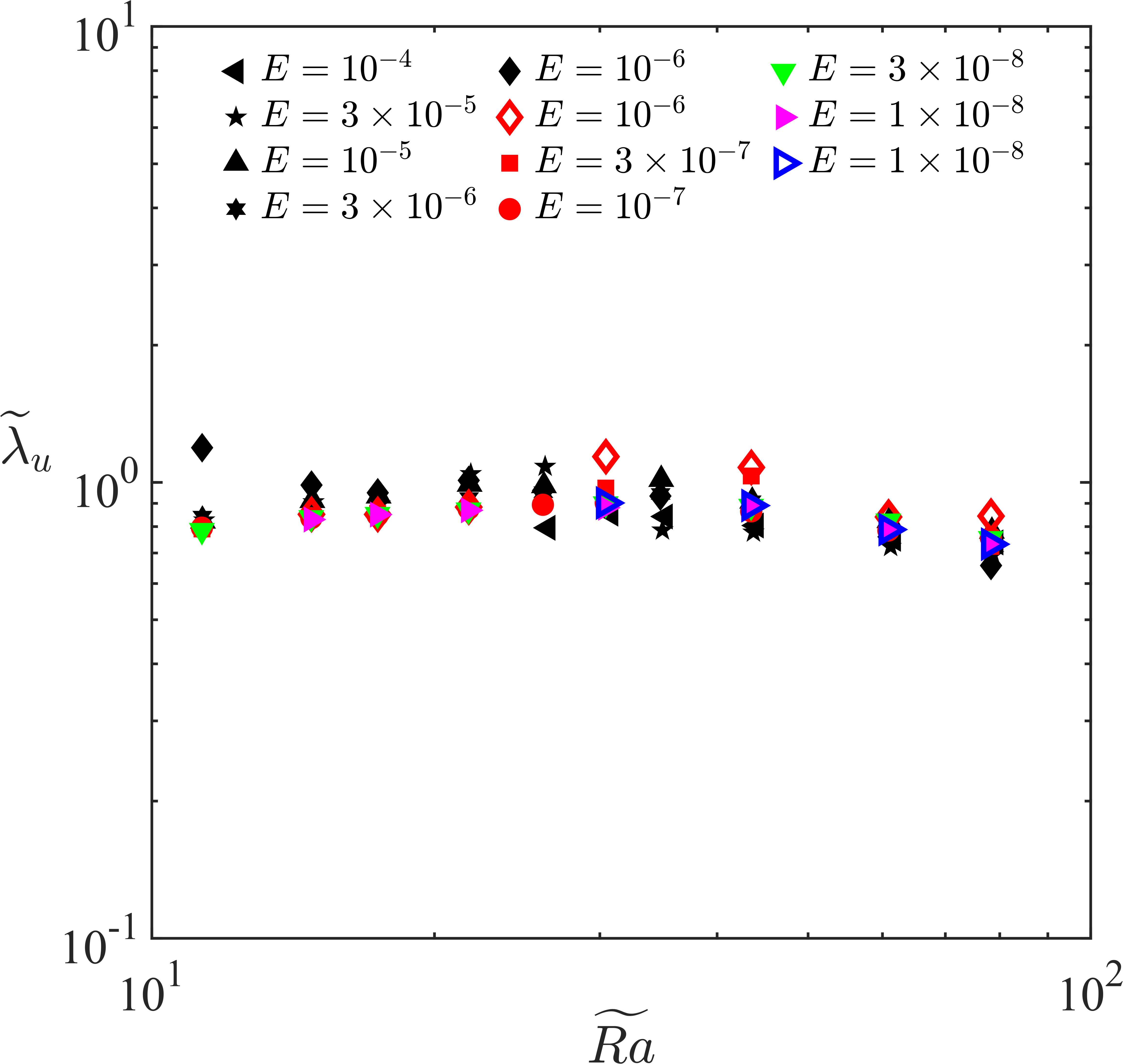} } \\
                  \subfloat[]{
      \includegraphics[width=0.43\textwidth]{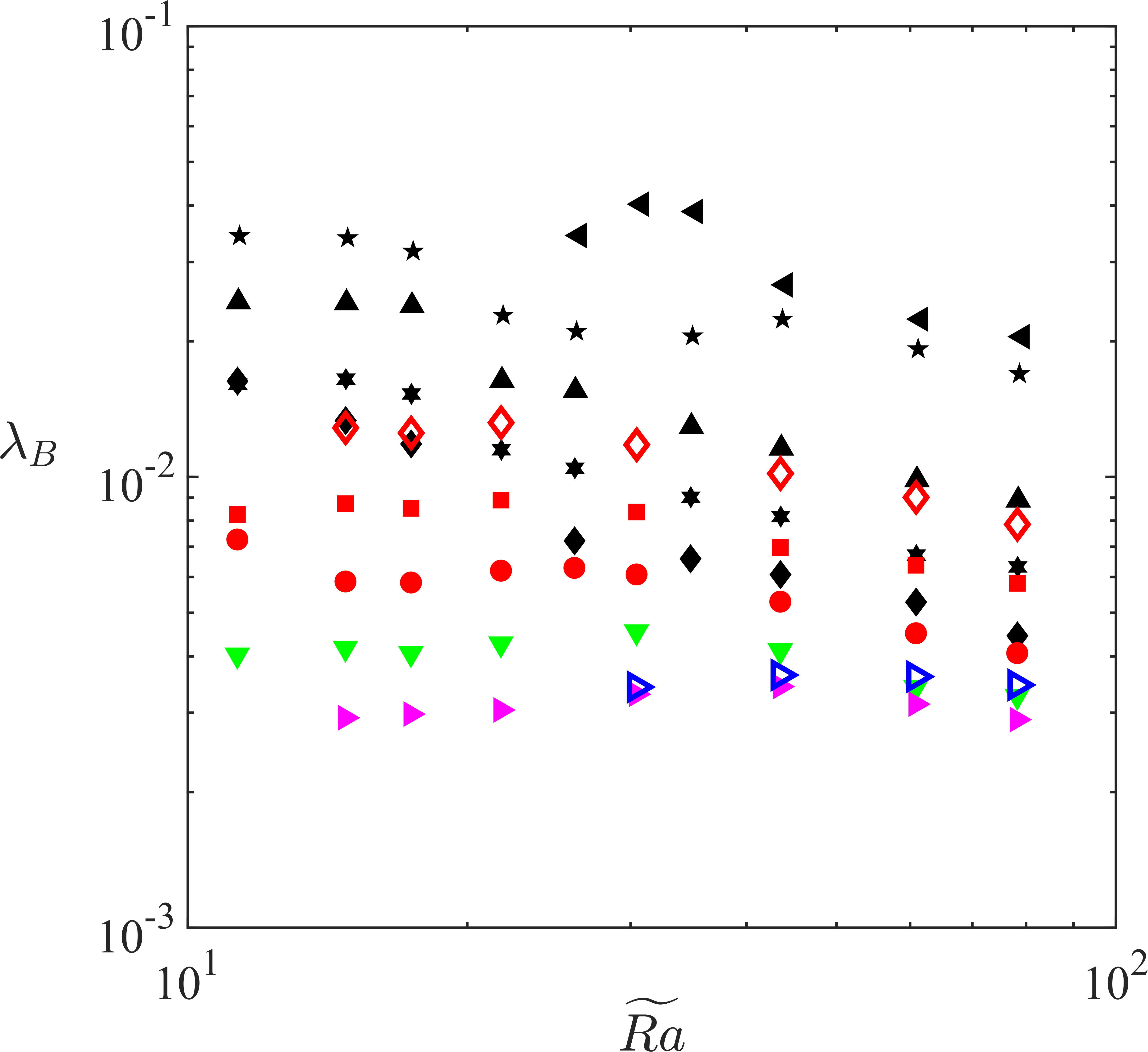} } \quad
                  \subfloat[]{
      \includegraphics[width=0.43\textwidth]{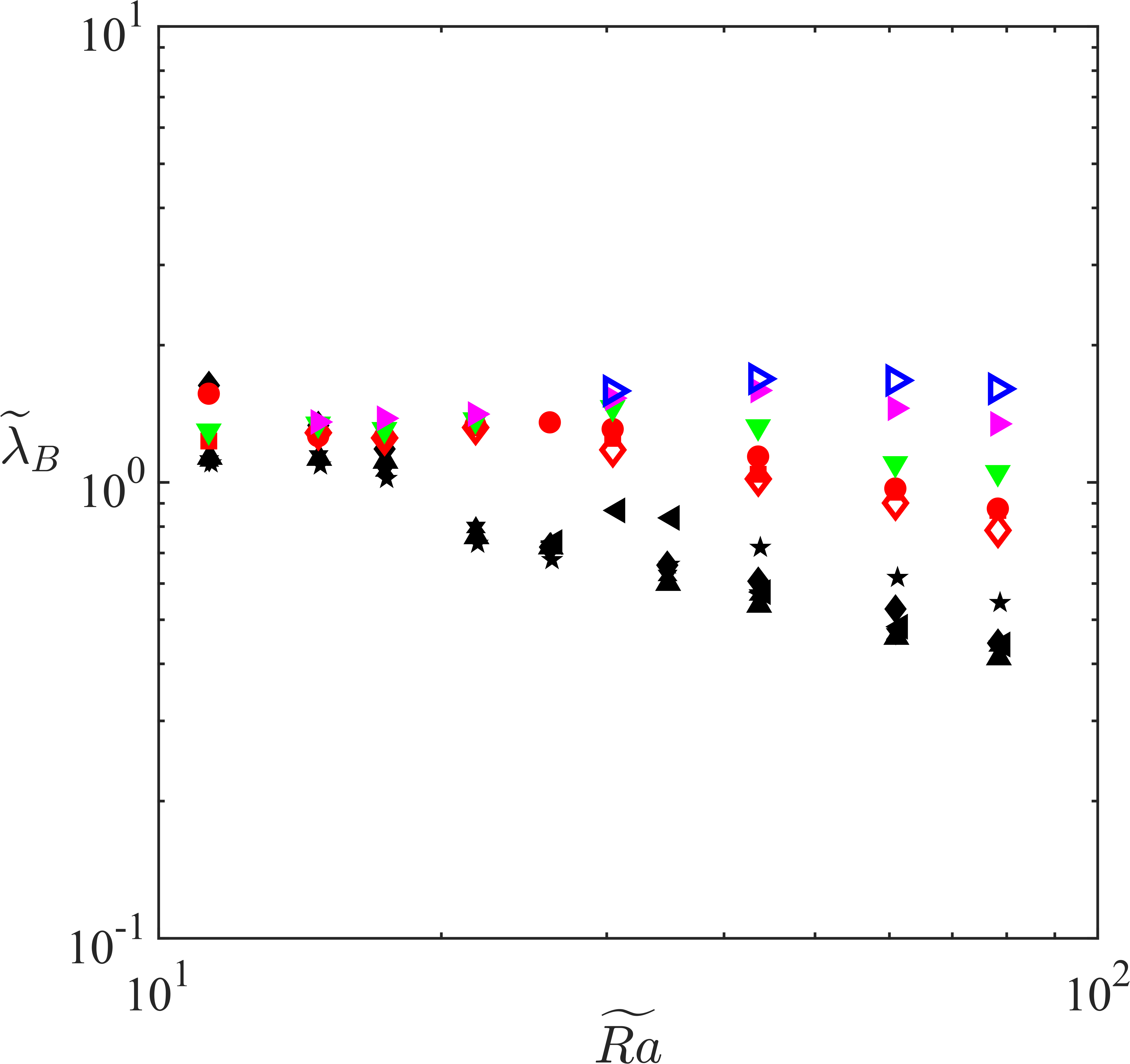} } \\
                  \subfloat[]{
      \includegraphics[width=0.43\textwidth]{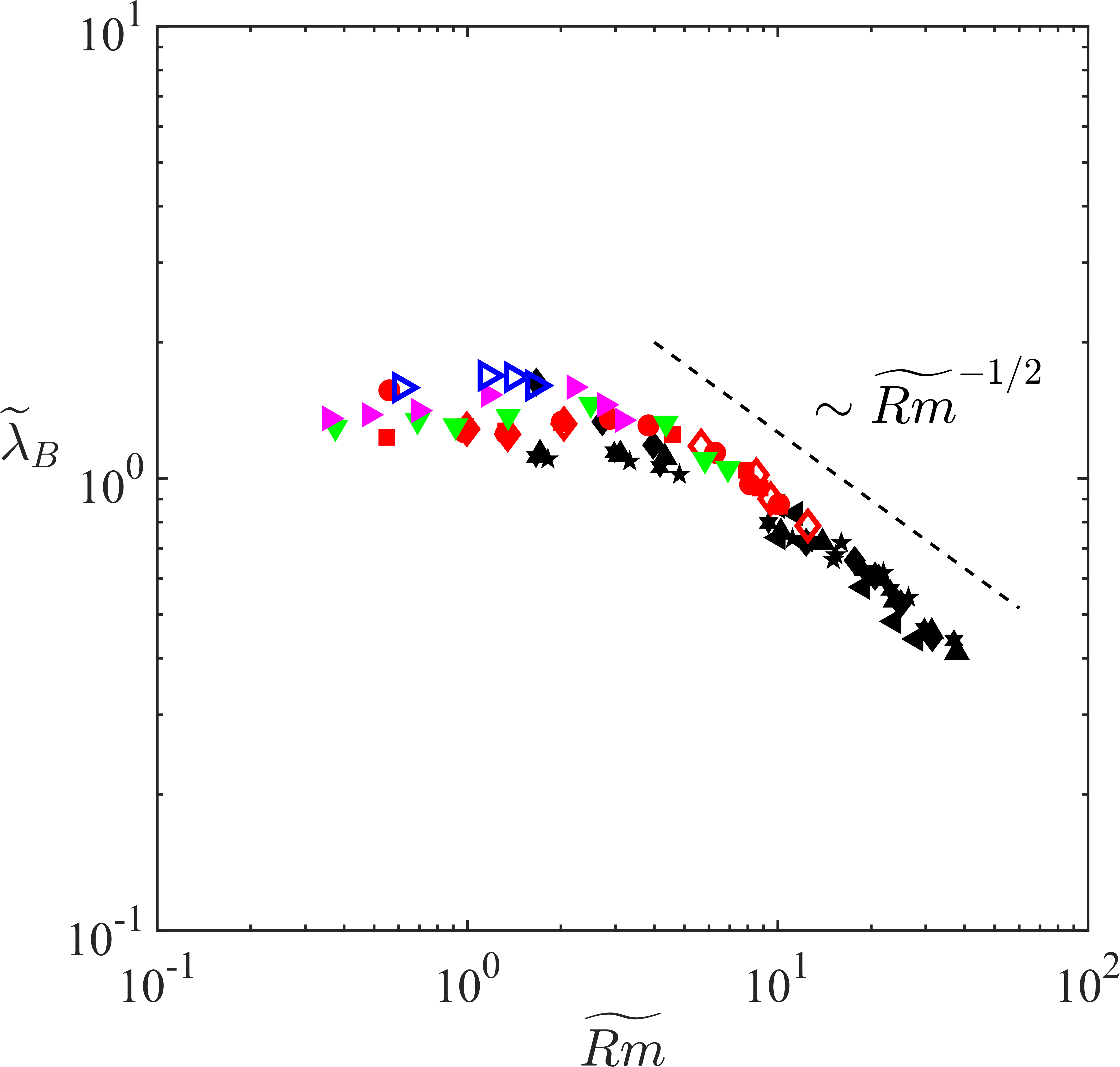} }      
 \end{center}
\caption{ Taylor microscales for the velocity and magnetic fields: (a) velocity microscale versus $\Rat$;
(b) rescaled velocity microscale $(\widetilde{\lambda}_u=\lambda_{u}E^{-1/3})$  versus $\Rat$;
(c) magnetic microscale versus $\Rat$;
(d) rescaled magnetic microscale $(\widetilde{\lambda}_B=\lambda_{B}E^{-1/3})$ versus $\Rat$;
 (e) rescaled magnetic microscale versus $\Rmt$. Symbol shape represents different values of the Ekman number ($E$) and color represents different values of the magnetic ($Pm$): black indicates $Pm=1$; red indicates $Pm=0.3$; green indicates $Pm=0.2$; magenta indicates $Pm=0.1$; blue indicates $Pm=0.05$.
 \label{F:Taylor}}
\end{figure*}

The Taylor microscale for both the velocity field and the magnetic field are defined by, respectively,
\be
\lambda_{u}=\sqrt{	\frac{\langle \mathbf{u}^2\rangle }{	 \langle (\nabla\times \mathbf{u})^2\rangle}} ,
\ee
and
\be
\lambda_{B}=\sqrt{	\frac{\langle\mathbf{B^{\prime}}^2\rangle }{	 \langle (\nabla\times \mathbf{B})^2\rangle}} .
\ee
The corresponding asymptotically rescaled Taylor microscales are defined by
\be
\widetilde{\lambda}_u = \lambda_{u} E^{-1/3}, \quad \widetilde{\lambda}_B = \lambda_{B} E^{-1/3} .
\ee 
The above definitions follow from assuming that the derivatives appearing in the definition of the Taylor microscales are dominated by the horizontal, $E^{-1/3}$ factor.
Note that we use the fluctuating magnetic energy  to  calculate  the magnetic  Taylor microscale because we are interested in the length scale of the fluctuating magnetic field. On the other hand, we find that the dissipation contribution from the mean magnetic field is negligible compared to that from the fluctuating magnetic field. For simplicity we therefore use the total dissipation in  the calculations shown here.

Figures~\ref{F:Taylor}(a) and (c) show the velocity and magnetic Taylor microscales as a function of $\Rat$, respectively; the corresponding rescaled microscales are shown in panels (b) and (d). Good collapse is observed for both microscales when the $E^{-1/3}$ rescaling is applied, although this is to be expected in light of the collapse of the full spectra shown in Figures \ref{F:EkSPE} and \ref{F:EmSPE}. The rescaled velocity microscale is nearly constant over the range of investigated Rayleigh numbers; similar behavior was found for a fixed value of $E$ in a recent experimental study of rotating convection \citep{mM21}. In contrast, we find two different regimes for $\lambda_B$, depending on the particular combination of $Pm$ and $\Rat$, as shown in Figure \ref{F:Taylor}(d). These regimes become more clear when plotted versus the reduced magnetic Reynolds number, $\Rmt$, as shown in Figure \ref{F:Taylor}(e): when $\Rmt \lesssim 3$ the magnetic microscale is nearly constant and we find $\lambda_B \sim \lambda_u$ (or equivalently $
\widetilde{\lambda}_B \sim \widetilde{\lambda}_u$); for $\Rmt \gtrsim 3$ we find $\lambda_B < \lambda_u$. This change in scaling behavior can be attributed to the transition from large scale dynamo action to small scale dynamo action. As previously shown for the mean magnetic energy fraction, and suggested by theory \citep{mC15b}, this transition is well-characterised by the size of $\Rmt$. 

The scaling behavior of $\lambda_B$ can be understood by considering the fluctuating induction equation,
\be
\dst \Bbp + \ub \cdot \nabla \Bbp = \mBb \cdot \nabla \ub + \Bbp \cdot \nabla \ub  + \frac{E}{Pm}\nabla^2 \Bbp .
\label{eq:fluc_induc}
\ee
We assume that in the small scale dynamo regime the large scale magnetic field is asymptotically small relative to the fluctuating magnetic field, $|\mBb| \ll  |\Bbp|$. As the magnetic Reynolds number increases the two terms on the left hand side of equation \eqref{eq:fluc_induc} will tend to dominate, but the stretching and diffusion terms will reach a subdominant balance such that 
\be
\Bbp \cdot \nabla \ub \sim \frac{E}{Pm} \nabla^2 \Bbp \quad \Rightarrow \quad \frac{U}{\lambda_u} \sim \frac{E}{Pm} \frac{1}{\lambda_B^2} .
\ee
Noting that $U$ has units of the large scale Rossby number we can use $U = Re E \sim \Ret E^{2/3}$ such that
\be
\lambda_B \sim E^{1/3} \Rmt^{-1/2},
\ee
which agrees with the observed scaling. These results suggest that the length scale of the magnetic field is controlled by both $E$ and $\Rmt$, depending on the particular regime. In terms of large scale quantities, this suggests that the scaling behaviour of the ohmic dissipation scale in both the large scale dynamo regime and the small scale dynamo regime becomes
\be
\text{Large scale dynamo regime}: \quad \lambda_B \sim E^{1/3}, 
\ee
\be
 \text{Small scale dynamo regime}: \quad \lambda_B \sim Rm^{-1/2} E^{1/6}.
\ee

\subsection{Force balances}


\textcolor{black}{In this subsection we numerically analyze the forces in the simulations. A similar analysis has been completed by \cite{aG21} for the hydrodynamic convection problem, though they did not consider the asymptotic scaling behaviour of the system. In comparison to \cite{aG21}, our parameter space is restricted to the rapidly rotating regime. Here we extend the force balance analysis to the dynamo problem and show that the asymptotic predictions of section \ref{S:theory} are consistent with the numerical data, confirming that dynamos in the plane layer geometry are QG in the limit of rapid rotation.} Each force is denoted according to
\be
\underbrace{\partial_t \bold{u} }_{F_t}= \underbrace{ E\nabla^2 \bold{u} }_{F_v} +\underbrace{(-\bold{u}\cdot\bold{\nabla}\bold{u} )}_{F_a}  + \underbrace{(\bold{B}\cdot\bold{\nabla}\bold{B})}_{F_l} + \underbrace{\frac{RaE^2}{Pr}\theta^{\prime}\bold{\hat{z}}}_{F_b}  +\underbrace{( -\bold{\hat{z}}\times\bold{u})}_{F_c} +\underbrace{( - \nabla_\perp p^{\prime}  -\dsz p^{\prime} \hz\\\\)}_{F_p} .
\label{eq1}
\ee 
The hydrostatic balance in the vertical component of the momentum equation has been removed by defining the fluctuating pressure according to $p^{\prime}(x,y,z,t)  = p(x,y,z,t) -  \overline p(z,t)$. In what follows we report time-averaged global rms values of the above forces.

\begin{figure*}
 \begin{center}
 \subfloat[]{
      \includegraphics[width=0.47\textwidth]{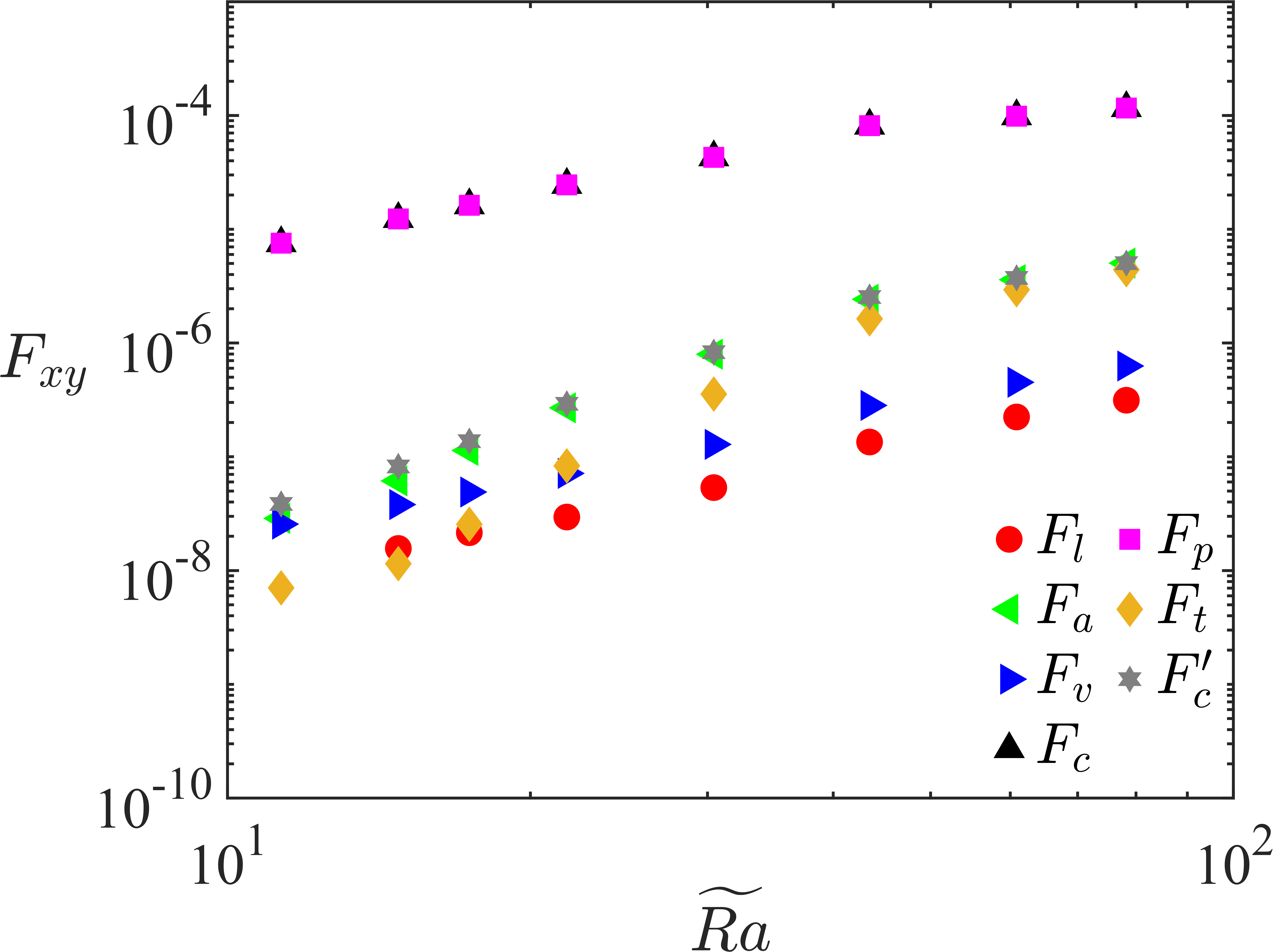} }
      \quad
 \subfloat[]{
      \includegraphics[width=0.45\textwidth]{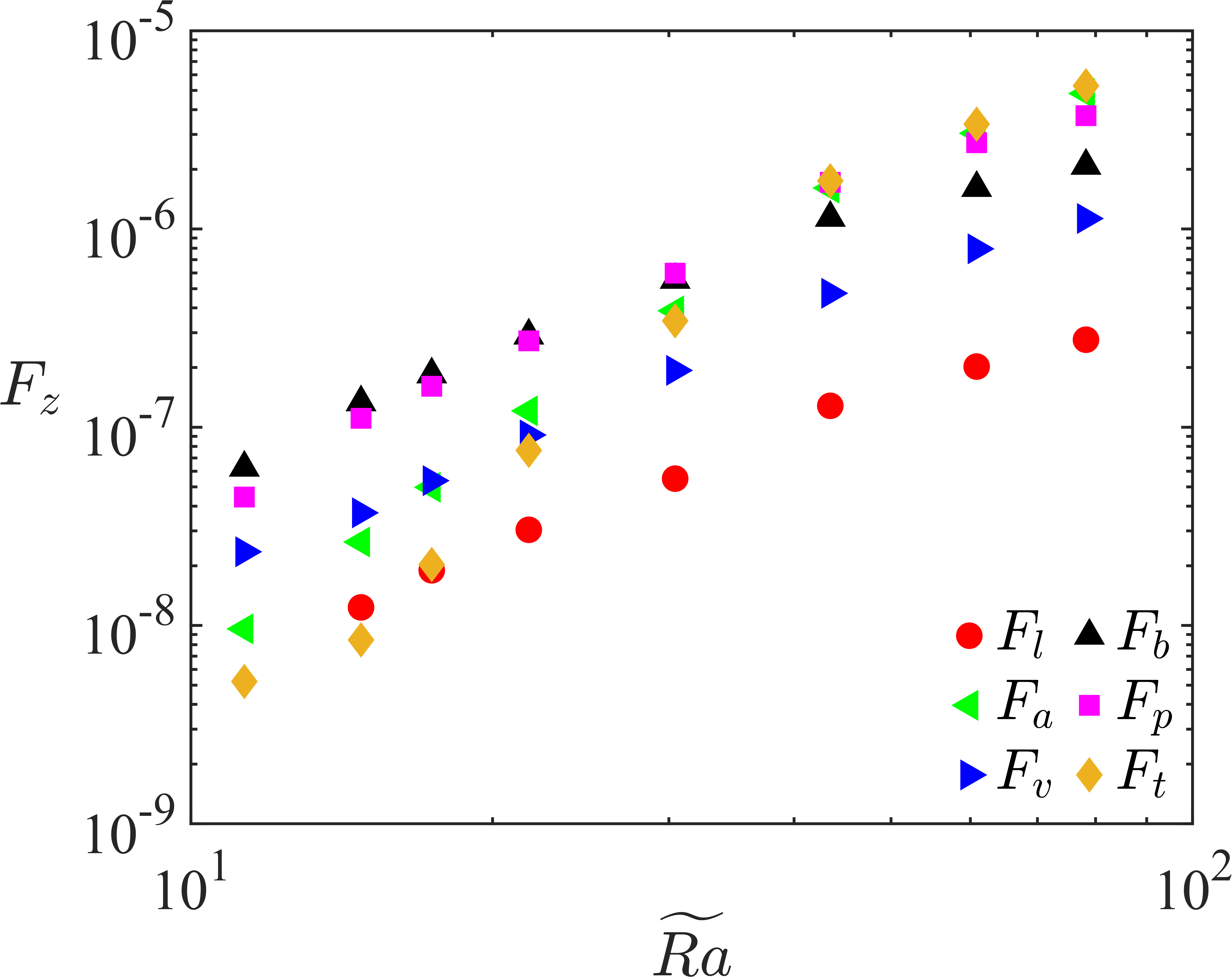}   }
 \end{center}
\caption{Global rms values of all forces for cases with $E=1\times 10^{-8}$ and $Pm=0.1$. 
 (a) rms forces in the horizontal direction versus $\Rat$;  (b) rms forces in the vertical direction versus $\Rat$. Note that the data points for the Coriolis and pressure gradient forces are directly on top of one another in panel (a).
 \label{F:forceRa}}
\end{figure*}

Figure~\ref{F:forceRa} shows rms values of the forces as a function of $\Rat$ for the specific case of $E=1\times 10^{-8}$ and $Pm=0.1$. The horizontal components of the forces are shown in panel (a) and the vertical components are shown in panel (b). Other combinations of non-dimensional parameters were computed and show similar trends with the particular cases shown. For all cases, we find a dominant balance between the Coriolis force and the pressure gradient force in the horizontal dimensions, which suggests that all cases are in the QG dynamo regime. The global rms of the sum of the Coriolis and pressure gradient forces, defined by 
\be
\textcolor{black}{F^{\prime}_c = F_c + F_p,}
\ee 
is also shown and can be considered the ageostrophic component of the Coriolis force. As expected from QG theory this ageostrophic component is comparable in magnitude to all other terms in the (horizontal) momentum equation. Near the onset of convection we find that all of the subdominant terms are of the same order of magnitude; as $\Rat$ is increased we find that advection, inertia, and $F^{\prime}_c$ become larger in magnitude than the viscous force and the Lorentz force. In particular, we find that inertia becomes one of the largest subdominant terms for $\Rat \gtrsim 40$, suggesting that these flows become fully turbulent for Rayleigh numbers larger than this value. \textcolor{black}{These observations suggest that the system enters a state that is well described by the so-called Coriolis-Inertia-Archimedean (CIA) balance \citep[e.g.][]{cJ15}, though further investigation is necessary to confirm if the dominant length scales that arise in the system are consistent with this balance}. The Lorentz force and viscous force remain comparable in magnitude over the investigated range of $\Rat$, though we find the differences are larger in the vertical component of the momentum equation, as shown in panel (b). Non-rotating dynamos also exhibit an approximate balance between the Lorentz and viscous forces \citep{mY21}. The buoyancy force and the vertical pressure gradient force remain the largest terms in the vertical component of the momentum equation for $\Rat \lesssim 40$; for larger $\Rat$ we find that inertia, advection and the pressure gradient force dominate.

\begin{figure*}
 \begin{center}
 \subfloat[]{
      \includegraphics[width=0.47\textwidth]{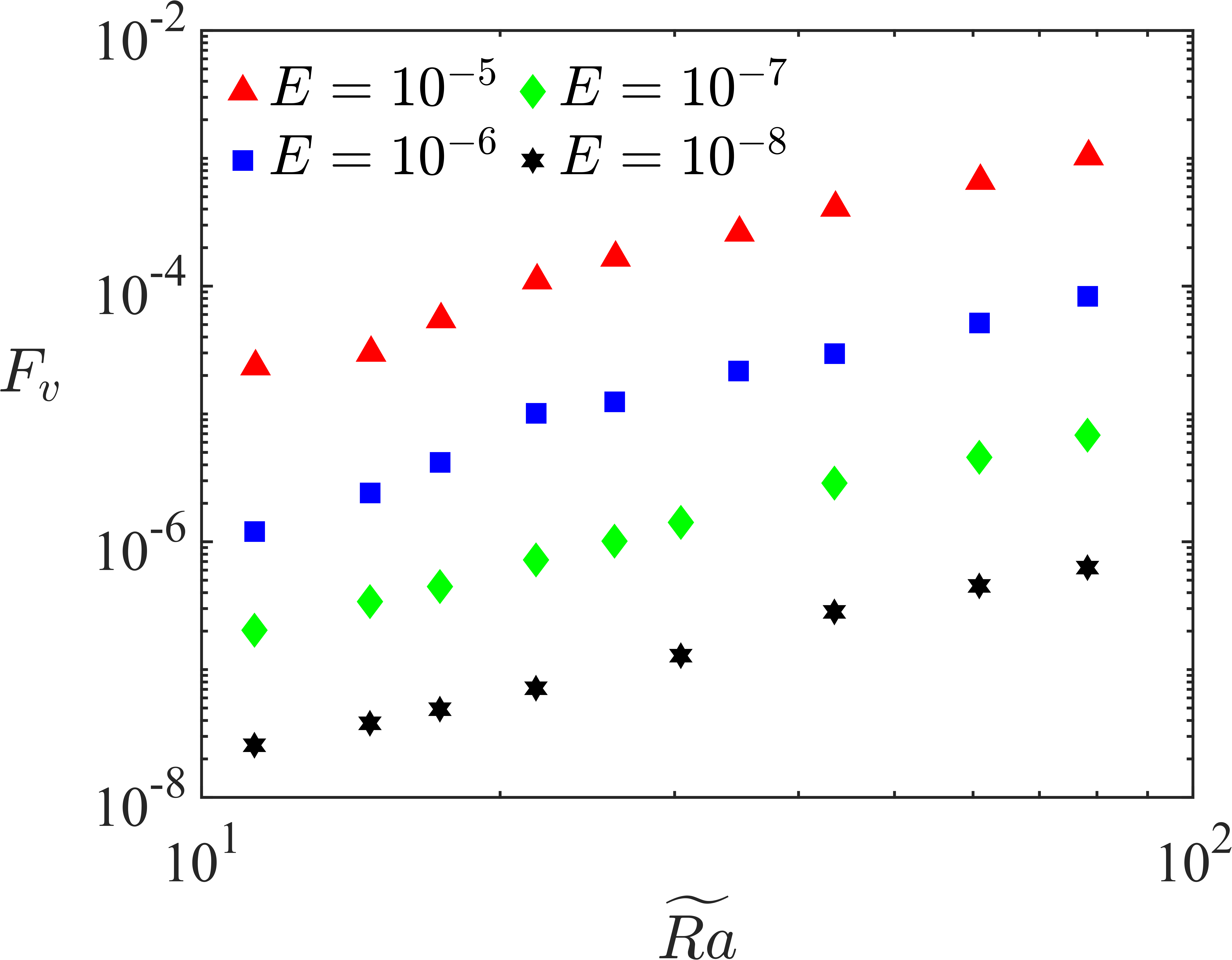} }
      \quad
 \subfloat[]{
      \includegraphics[width=0.47\textwidth]{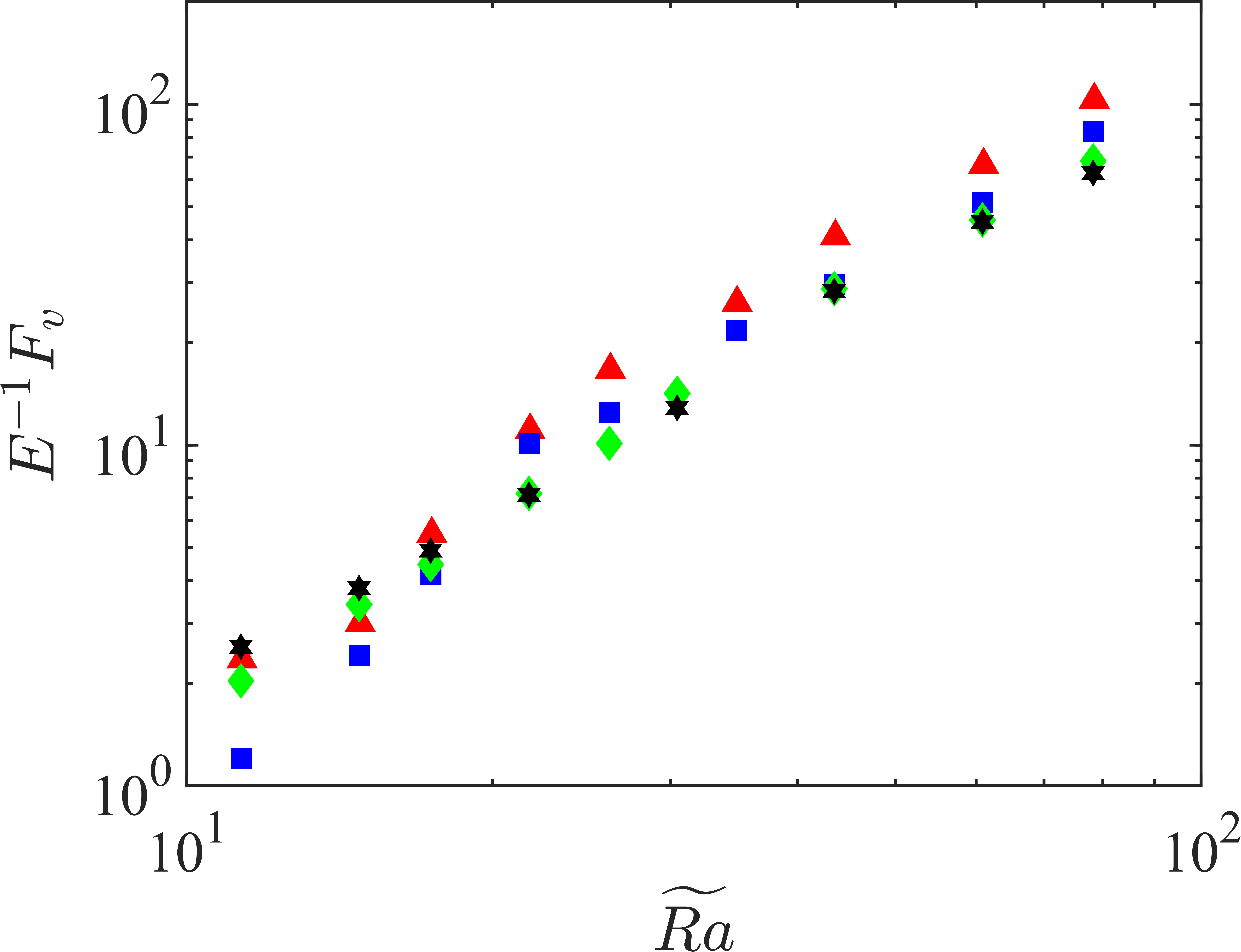} } \\
 \subfloat[]{
      \includegraphics[width=0.47\textwidth]{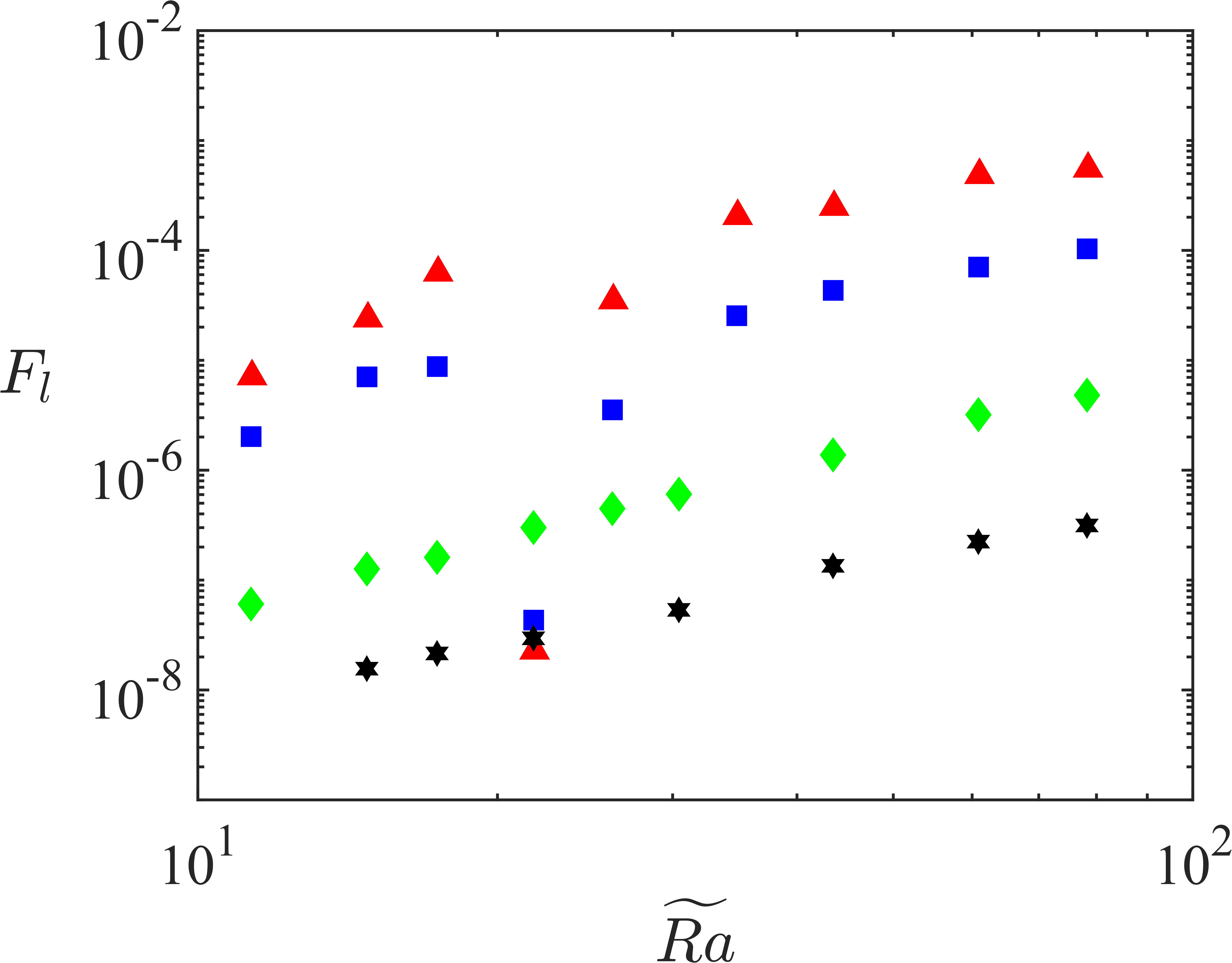} }
      \quad
 \subfloat[]{
      \includegraphics[width=0.47\textwidth]{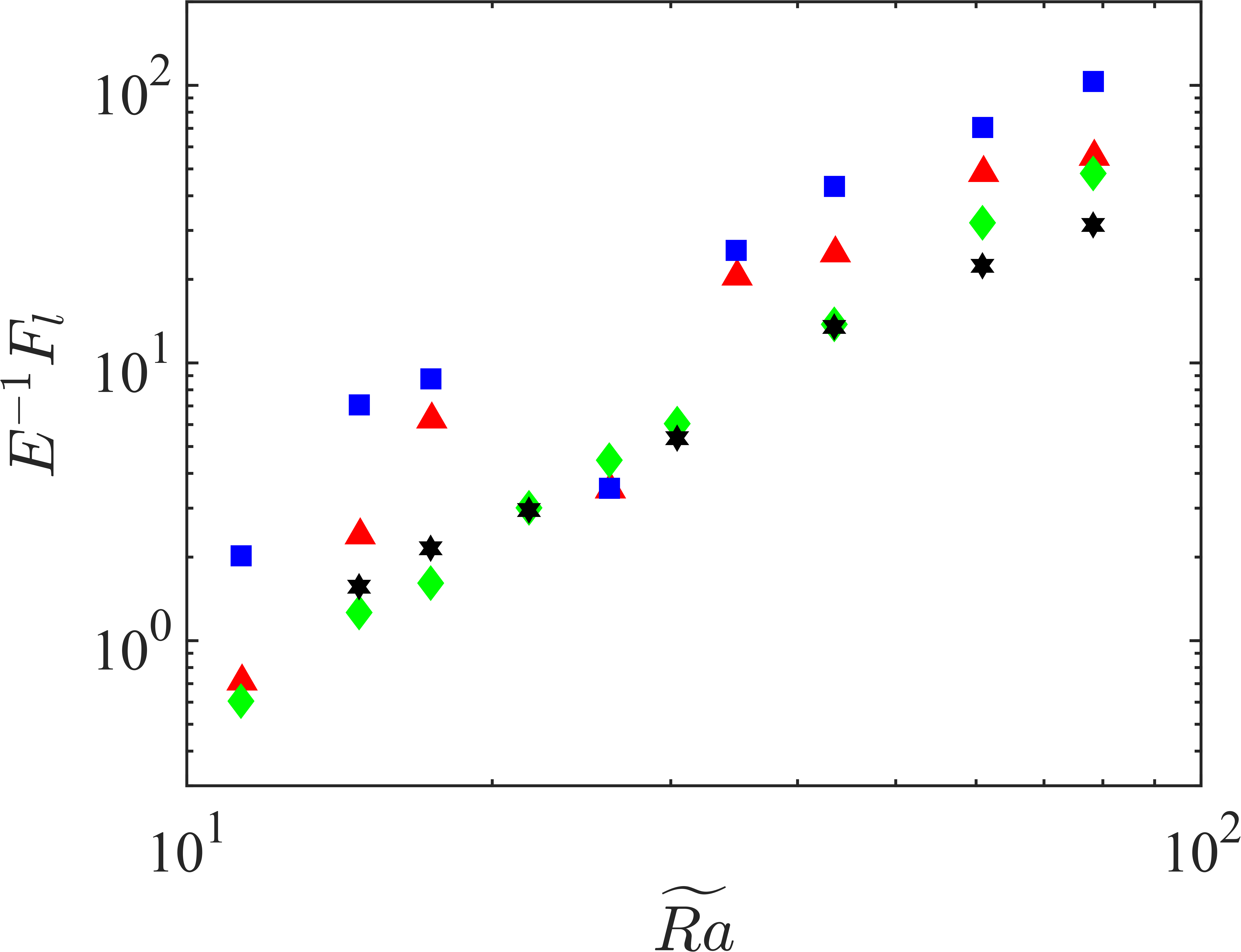} } \\      
 \subfloat[]{
      \includegraphics[width=0.47\textwidth]{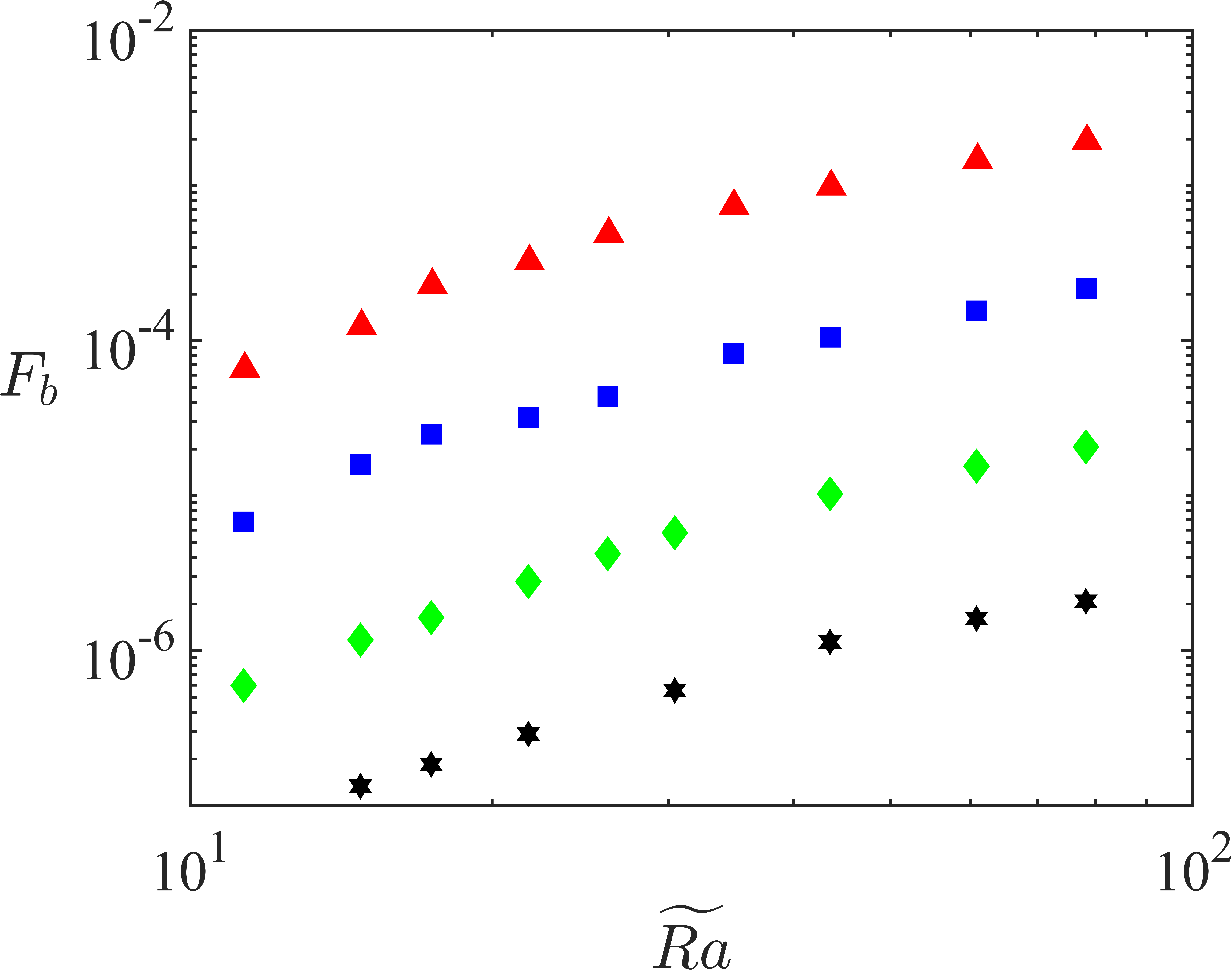} }
      \quad
 \subfloat[]{
      \includegraphics[width=0.47\textwidth]{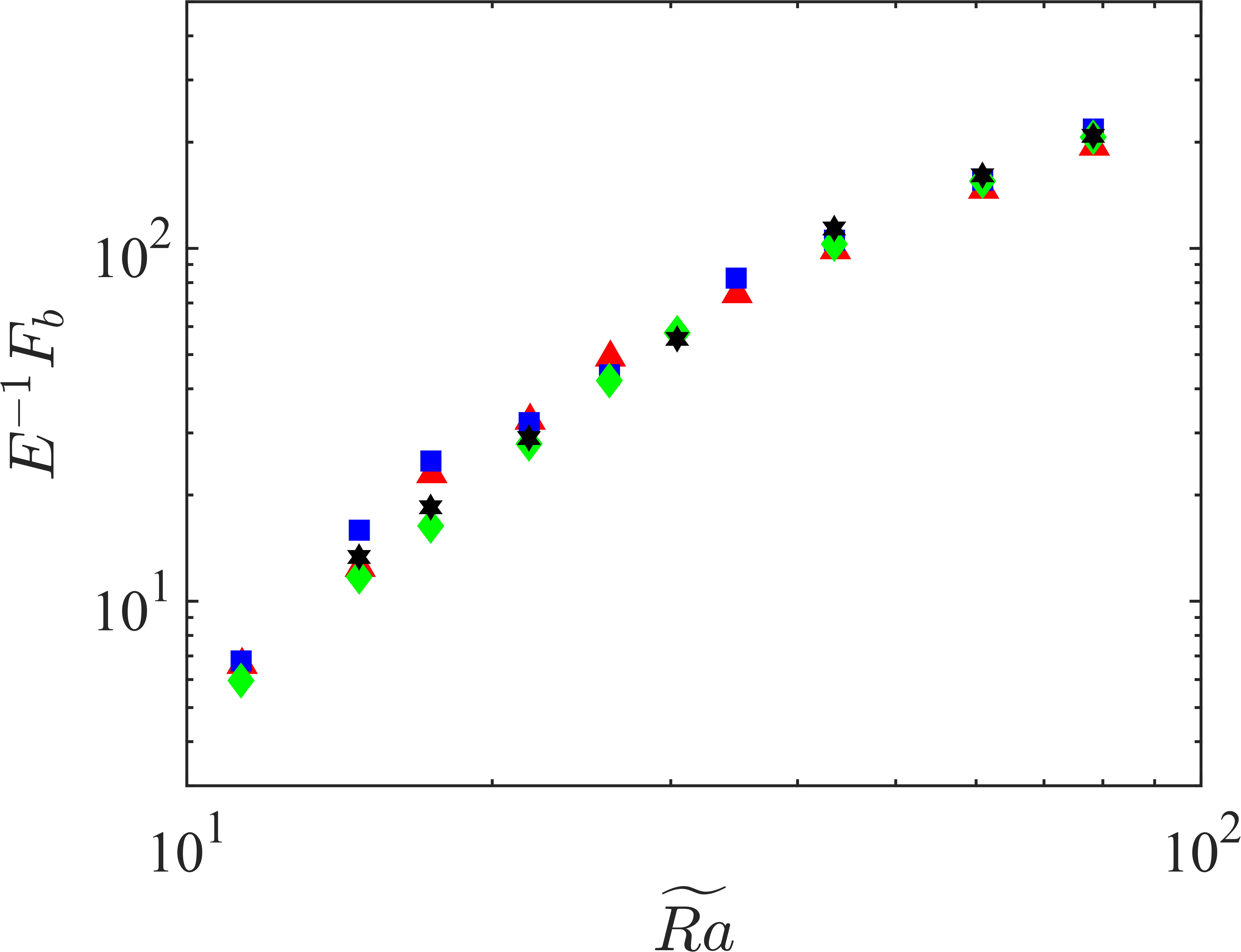} } \\  
\end{center}
\caption{Scaling behaviour of (a,b) the viscous force; (c,d) the Lorentz force; and (e,f) the buoyancy force for different Ekman numbers. The left column (a,c,e) shows the unscaled forces and the right column (b,d,f) shows the asymptotically rescaled forces. The markers have the same meaning in all figures.
 \label{F:force_scaling}}
\end{figure*}

A comparison of forces for a selection of different Ekman numbers (and different $Pm$) is made in Figure \ref{F:force_scaling}. Here we plot (a,b) the viscous force; (c,d) the Lorentz force; and (e,f) the buoyancy force for Ekman numbers $E=10^{-5}$ ($Pm=1$), $10^{-6}$ ($Pm=1$), $10^{-7}$ ($Pm=0.3$), and $10^{-8}$ ($Pm=0.1$). The left column of the figure shows the unscaled data and the right column shows the asymptotically rescaled data. The forces show a good collapse when rescaled via the asymptotic predictions of section 3. Of the three forces shown, the Lorentz force shows the most scatter, which can be attributed to different values of $Pm$ (or, equivalently, $\Rmt$). \textcolor{black}{As shown in panel (f), varying $Pm$ has no significant effect on the scaling behaviour of either the viscous  force or the buoyancy force.}

\section{Discussion} 
\label{S:conclusions}

A systematic numerical investigation of convection-driven dynamos in the rotating plane layer geometry has been carried out for varying Ekman number, Rayleigh number and magnetic Prandtl number. The observed flow regimes ranged from quasi-laminar convection cells to geostrophic turbulence, \textcolor{black}{similar to the regimes found in hydrodynamic rotating convection \citep[e.g.][]{kJ12}. A primary purpose of this investigation was to test the asymptotic theory developed by \cite{mC15b}. In this regard, a variety of physical quantities were computed across the entire range of input parameters to characterise the asymptotic scaling behaviour of the system. In general, excellent agreement between theory and simulation output was found.}



\textcolor{black}{The observed heat transport behavior in the dynamo simulations is broadly similar to that observed in hydrodynamic rotating convection. More specifically, as a function of Rayleigh number, the Nusselt number initially grows quickly, then a more shallow slope is observed once the flow becomes turbulent. When plotted as a function of the reduced Rayleigh number $\Rat = Ra E^{4/3}$, we find that the Nusselt number shows qualitatively, and even quantitatively, similar behavior for all Ekman numbers and magnetic Prandtl numbers. As expected from the energy balance, the viscous and ohmic dissipation also show similar behavior with increasing $\Rat$ and the dissipation data can be collapsed with a scaling of $E^{4/3}$. In the majority of the dynamos simulated, viscous dissipation dominates over ohmic dissipation. In the turbulent regime, the ohmic dissipation fraction is  $f_{ohm} \lesssim 0.5$ for the majority of the cases. This behavior should be contrasted with spherical dynamo studies \citep{uC06}, or magnetoconvection studies \citep{mY19},  where regimes can be found in which ohmic dissipation dominates over viscous dissipation. }

Momentum transport, as characterised by the large scale Reynolds number, scales with the Ekman number according to $E^{-1/3}$, consistent with asymptotic theory \citep{kJ98a,mC15b}. \textcolor{black}{This scaling behaviour is a direct result of the geostrophic balance that occurs on the small, horizontal convective length scale.} Although we do not attempt to provide detailed numerical fits to the data, the Reynolds number increases steeply for small $\Rat$, then transitions to a shallower scaling in the turbulent regime (beyond values of $\Rat \approx 40$), which is roughly consistent with asymptotic studies of non-magnetic rotating convection \citep[e.g.][]{kJ12, mC16b,sM21}.


The magnetic energy of the simulated dynamos shows considerable scatter with the various input parameters, though a scaling of $E^{-2/3}$, as predicted by theory, provides some collapse to the data. \textcolor{black}{We show that a near unity mean magnetic energy fraction, $\overline{E}_{mag}/E_{mag} \approx 1$, is obtainable so long as the Ekman number is small and $Pm < 1$}. It is shown that the asymptotically scaled magnetic Reynolds number, $\Rmt = Rm E^{1/3}$ controls the value $\overline{E}_{mag}/E_{mag}$, and distinguishes dynamos dominated (energetically) by the large scale magnetic field, from those dominated by the small scale magnetic field. \textcolor{black}{It is well known that large scale magnetic fields can be generated when the magnetic Reynolds number is sufficiently small and the flow is coherent, i.e.~laminar \citep[e.g.][]{kM19}. Small values of $\Rmt$ indicate that magnetic diffusion is important on the small convective length scales. Here we find that robust large scale magnetic field can readily be generated even in the presence of incoherent, turbulent flows \citep{mY22}, provided $\Rmt$ remains small. 
In rough agreement with \cite{aT12}, small scale dynamos are consistently achieved when $\Rmt \gtrsim 1$. }


\textcolor{black}{The mean magnetic field shows a saturation with increasing Rayleigh number, despite the fact that the fluctuating magnetic field and associated emf grow with increasing $\Rat$. Similar behavior of the mean magnetic field is observed in spherical dynamos, where the azimuthally averaged component of the magnetic field shows a saturation with Rayleigh number \citep{mC21,rO21}. The exact cause of this saturation is not currently known, but it may be due to a breakdown of the $\alpha^2$-dynamo mechanism that operates only when $\Rmt \lesssim O(1)$ \citep{mC15b}. However, further investigation is necessary to confirm this hypothesis.}


\textcolor{black}{Linear theory predicts that the horizontal length scale at the onset of rotating convection scales with the Ekman number as $E^{1/3}$ in the limit $E\rightarrow 0$ \citep{sC61}. Whether these scales persist in the turbulent regime has remained an open question. Our analysis of both the kinetic and magnetic energy spectra suggests that all length scales in the velocity field and fluctuating magnetic field scale predominantly as $E^{1/3}$, even in the turbulent regime ($\Rat \gtrsim 40$). The viscous dissipation length scale (Taylor microscale) and the ohmic dissipation length scale are computed from these spectra and show good collapse when rescaled with $E^{1/3}$. The viscous dissipation length scale is approximately constant across a broad range of Rayleigh numbers -- this behavior has also been observed in hydrodynamic rotating convection experiments \citep{mM21}. The ohmic dissipation length scale is approximately constant (and of the same order as the viscous dissipation length scale) within the large scale dynamo regime, but transitions to a $\Rmt^{-1/2}$ scaling in the small scale dynamo regime. }

All of the dynamos within our suite of simulations show a leading order geostrophic force balance, with all other forces, including the Lorentz force, acting as perturbations. Thus, the simulated dynamos are within a QG dynamical state. \textcolor{black}{Of course, the relative sizes of the various perturbing forces is fundamental to the resulting QG dynamics.} We find that the Lorentz force tends to be approximately equal in magnitude to the viscous force across the entire range of investigated parameters; this behavior is similar to that observed in non-rotating convection-driven dynamos \citep{mY21}. 

\textcolor{black}{It is often assumed that the amplitude of the magnetic field generated by a rapidly rotating dynamo will saturate once the Lorentz force becomes comparable in magnitude to the Coriolis force \citep[e.g.][]{yF82,pR88a}. For the QG dynamos studied here, such a balance does not occur at leading order, though a higher order balance can take place in which the Lorentz force becomes comparable in magnitude to the ageostrophic component of the Coriolis force \citep[e.g.][]{mC18}. However, since all forces in the momentum equation are of comparable magnitude at this perturbative order of the dynamics, it is less clear whether the comparison of the Lorentz force to only the Coriolis force is meaningful. In particular, for a fixed Ekman number, our simulations show that inertia becomes important in the turbulent regime. Further investigation, perhaps at a fixed small value of the Ekman number, and finer scale variations of the Rayleigh number would help to isolate the effects of these perturbing forces.} 

\textcolor{black}{Several previous studies have examined the relative sizes of the terms in the vorticity equation in both spherical \citep{eD16} and plane layer dynamos \citep{dH16,fC17,dH19}. These studies focused on so-called `strong-field' dynamos in which the curl of the Lorentz force balances the vortex stretching term. In order to reach this balance it is typically necessary to use relatively large values of $Pm$ \citep{eD16} or to neglect inertia in the momentum equation \citep{dH16,fC17,dH19}, since the relative size of inertia will grow with increasing Rayleigh number and eventually dominate the Lorentz force, as observed in the present study. Given the linearity of the momentum equation when inertia is absent, \cite{dH19} decomposed the velocity field into `thermal' and `magnetic' components. Their analysis showed that whereas viscosity was small relative to buoyancy and the Coriolis terms in the thermal component of the vorticity equation, viscosity remained significant in the corresponding magnetic component. This observation is roughly consistent with our own results which show that the viscous force and the Lorentz force are comparable to one another.}




\textcolor{black}{It is well known that in homogenous, isotropic turbulence the largest eddies present in the flow tend to control the viscous dissipation rate such that the Taylor microscale  is a strongly decreasing function of the Reynolds number \citep[e.g.][]{sP00}. Indeed, this behavior is observed in non-rotating Rayleigh-B\'enard convection and the corresponding dynamos \citep[e.g.][]{mY21}. That the dissipation length scales in our simulations, and those in the hydrodynamic study of \cite{mM21}, do not change appreciably with the Rayleigh number indicates that rotationally constrained convective turbulence behaves very differently in comparison to non-rotating convection. This difference may simply be due to the fact that a small, viscous length scale is necessary in order to overcome the constraints imposed by rotation, as is known from linear theory. Provided that the Rossby and Ekman number are both small, the convection remains geostrophically balanced even in the turbulent regime. We stress that the predominance of the $E^{1/3}$ viscous length scale does not require that the viscous force is the largest perturbative force. Indeed, the simulations show that other perturbative forces such as inertia can become larger than the viscous force as the Rayleigh number is increased. However, this perturbation of the geostrophic balance can still occur on the $E^{1/3}$ length scale.
One might argue that the predominance of the $E^{1/3}$ length scale suggests that the Rayleigh numbers (or, equivalently, the Reynolds numbers) accessible in the present simulations are not sufficiently large to access the fully turbulent regime. However, there are two problems with this view. First, to our knowledge, the present set of simulations represent the most extreme DNS of either rotating convection or rotating convection-driven dynamos carried out to date, as based on the combination of Ekman numbers and Reynolds numbers. While the possibility remains that an additional transition in the dynamics might appear at larger Rayleigh numbers (and perhaps smaller Ekman numbers), simulations of the QG model for hydrodynamic rotating convection do not observe such a transition up to $\Rat = 200$ \citep{sM21}.}





\textcolor{black}{The present investigation highlights significant differences between rotating dynamos in the plane geometry and those that occur in spherical geometries. Some of these differences include the observed sequence of force balances and the relative magnitude of the Lorentz force, and the number of asymptotically distinct convective length scales.  It is typical to define the large scale fields in a spherical domain as a zonal average, and small scales as fluctuations away from this average. On the small scales, the leading order force balance is geostrophic in spherical geometries, with the Lorentz force and buoyancy force entering at the next (higher) order in the dynamics, and all other forces are smaller still \citep{rY16}. In contrast, simulations show that the large scale dynamics in spherical dynamos are semi-magnetostrophic with a thermal wind balance in the meridional plane and a Coriolis-Lorentz force balance in the zonal direction \citep{jA05,mC21}. This latter balance allows for saturation of the large scale magnetic field via the Malkus-Proctor mechanism \citep{wM75}. However, no equivalent saturation mechanism for the large scale magnetic field is present in the plane layer geometry. The linear theory of rotating convection in spherical shells shows the presence of two asymptotically distinct convective length scales -- one is the familiar $E^{1/3}$ zonal length and the other is the $E^{2/9}$ radial length scale \citep{eD04}. The influence of different length scales, and the associated anisotropy in the fluctuating velocity field, may play a role in the different sequence of force balances that are observed in the two geometries, though further investigation is necessary to test this hypothesis.}

\section*{Declaration of Interests}
The authors report no conflict of interest.

\section*{Acknowledgements}
This paper benefited greatly from comments made by three anonymous referees.
The authors gratefully acknowledge funding from the National Science Foundation (NSF) through grants EAR-1945270 and SPG-1743852. The computations were carried out on the Summit and Stampede2 supercomputers. Summit is supported by NSF awards ACI-1532235 and ACI-1532236, the University of Colorado Boulder, and Colorado State University. Stampede2 is operated by the Texas Advanced Computing Center (TACC) and was made available through Extreme Science and Engineering Discovery Environment (XSEDE)
allocation PHY180013. Flow visualization was performed with VAPOR \citep{sL19}.

\bibliographystyle{jfm}
\newcommand{\jfm}{J. Fluid Mech.~}\newcommand{\apj}{Astrophys.
  J.~}\newcommand{\jpp}{J. Plasma Phys.~}\newcommand{\mnras}{Mon. Not. Roy.
  Astron. Soc. }\newcommand{\jgr}{J. Geophys. Res.~}\newcommand{\araa}{Annu.
  Rev. Astron. Astrophys.~}\newcommand{\icarus}{Icarus
  }\newcommand{\aap}{Astron. Astrophys.~}\newcommand{\physscr}{Phys. Scripta
  }\newcommand{\ssr}{Space Sci. Rev.~}\newcommand{\pnas}{Proc. Nat. Acad.
  Sci.~}\newcommand{\ncom}{Nat. Comm.~}\newcommand{\njp}{New J.
  Phys.~}\newcommand{\prf}{Phys. Rev. Fluids }\newcommand{\prr}{Phys. Rev. Res.
  }\newcommand{\pre}{Phys. Rev. E }\newcommand{\pepi}{Phys. Earth Planet.
  Int.~}\newcommand{\gji}{Geophys. J. Int.~}

\pagebreak

\section*{Appendix: simulation data}

\newpage

\begin{table}
 \begin{adjustbox}{addcode={\begin{minipage}{\width}}{\caption{%
     Details of the numerical simulations for $Pm = 1$, $E = (1\times 10^{-4}, 3\times 10^{-5})$  cases.  The non-dimensional parameters are the reduced Rayleigh number $\Rat$,   the Nusselt number $Nu$, the Reynolds number $Re$, and the magnetic Reynolds number $Rm$. The spatial resolution is quoted in terms of the de-aliased physical space grid points $N_x \times N_y \times N_z$, where $(N_x,N_y)$ is the horizontal resolution and $N_z$ is the vertical resolution. The numerical timestep size is denoted by $\Delta t$. The magnetic energy is denoted by $E_{mag}$, and the velocity Taylor microscale and magnetic Taylor microscale are denoted by   $\lambda_u$ and $\lambda_B$,  respectively.}\end{minipage}},rotate=90,center}
      \setlength\tabcolsep{6pt}
  \begin{tabular}{lcccccccccc}
  \hline
      $E$   & $\Rat$   & $N_x \times N_y \times N_z$ &$\Delta t$  & $Nu$ & 	$Re$ &               $Rm$ &                $E_{mag}$  &	$\overline{E}_{mag}/E_{mag}$ &	$\lambda_u $ &$\lambda_B$\\
      \hline   
       $1\times 10^{-4}$    &$11.45$    & $96\times 96\times 48$  &    $2$ &	    $ 1.55$&	$ 41.6$ &$ 41.6$ & 	$$ & $$ & $0.0358$& $ $    \\
        $1\times 10^{-4}$    &$14.97$    & $96\times 96\times 48$  &    $0.5$ &	    $ 2.65$&	$ 85.9$ &$  85.9$& 	$		$ & $$ & $0.0373$& $ $    \\
        $1\times 10^{-4}$    &$17.61$    & $96\times 96\times 48$  &    $0.4$ &	    $ 3.87$&	$ 125.8$ &$125.8$& $		$ & $$ & $0.0381$& $ $    \\
       $1\times 10^{-4}$    & $22.01$    & $144\times 144\times 72$  &    $0.2$ &	    $ 6.16$&	$ 205.3$ &$ 205.3$& $		$ & $$ & $0.0416$& $ $    \\
       $1\times 10^{-4}$    & $26.42$    & $144\times 144\times 72$  &    $0.08$ &	    $ 8.61$&	$ 215.9$ &$ 215.9$& $5.8 \times 10^3$ & $0.0846$ & $0.0369$& $0.0343  $    \\
        $1\times 10^{-4}$    &$30.82$    & $192\times 192\times 108$  &    $0.02$ &	    $ 9.25 $&	$ 215.6$ &$  215.6$& $2.9 \times 10^4	$ & $0.2859$ & $0.0397$& $0.0403 $    \\
       $1\times 10^{-4}$    & $35.22$    & $192\times 192\times 108$  &    $0.01$ &	    $ 11.00$&	$ 245.9$ &$ 245.9$& $3.8 \times 10^4$ & $0.2691$ & $0.0391$& $0.0388 $    \\
       $1\times 10^{-4}$    & $44.03$    & $192\times 192\times 144$  &    $0.01$ &	    $ 14.59$&	$ 403.9$ &$ 403.9$& $ 4.5 \times 10^3$ & $0.0128$ & $0.0373$& $ 0.0267 $    \\
       $1\times 10^{-4}$    & $61.64$    & $288\times 288\times 288$  &    $0.005$ &	    $ 18.67$&	$ 509.4$ &$ 509.4$& $5.6 \times 10^3$ & $0.0074$ & $0.0349$& $ 0.0224$    \\
        $1\times 10^{-4}$    &$79.25$    & $384\times 384\times 288$  &    $0.003$ &	    $ 21.99$&	$ 600.5$ &$ 600.5$& $1.1 \times 10^4$ & $0.0059$ & $0.0343$& $ 0.0205$    \\
       $3\times 10^{-5}$    &       $11.37$    & $96\times 96\times 48$  &    $2$ &	    $ 1.49$&	$ 58.4$ &$ 41.6$& 	$ 2.18 \times 10^3		$ & $0.7712$ & $0.0257$& $ 0.0343$    \\
   $3\times 10^{-5}$    & $14.87$    & $144\times 144\times 72$  &    $0.4$ &	    $ 2.24$&	$ 107.2$ &$107.2 $& 	$5.6\times 10^3		$ & $0.4621$ & $0.0282$& $ 0.0339$    \\
   $3\times 10^{-5}$    & $17.49$    & $144\times 144\times 72$  &    $0.2$ &	    $ 3.30$&	$ 155.2$ &$ 155.2$& 	$8.6\times 10^3		$ & $0.2984$ & $0.0278$& $ 0.0317$    \\
   $3\times 10^{-5}$    & $21.86$    & $144\times 144\times 72$  &    $0.15$ &	    $ 6.22$&	$ 358.9$ &$ 358.9$& 	$							$ & $0.0110$ & $0.0325$& $ 	$    \\
   $3\times 10^{-5}$    & $26.23$    & $192\times 192\times 108$  &    $0.08$ &	    $ 9.08$&	$ 495.2$ &$ 495.2$& 	$2.7\times 10^3		$ & $0.0125$ & $0.0337$& $ 0.210$    \\
  $3\times 10^{-5}$    &  $34.98$    & $192\times 192\times 144$  &    $0.015$ &	    $14.65$&	$ 485.8$ &$485.8 $& 	$2.6\times 10^4		$ & $0.0312$ & $0.0244$& $ 0.0205$    \\
  $3\times 10^{-5}$    &  $43.72$    & $216\times 216\times 216$  &    $0.005$ &	    $17.94$&	$ 515.8$ &$ 515.8$& 	$9.9\times 10^4		$ & $0.1170$ & $0.0242$& $ 0.0224$    \\
  $3\times 10^{-5}$    &  $61.21$    & $288\times 288\times 288$  &    $0.002$ &	    $25.27$&	$705.0$  &$705.0 $& 	$1.3\times 10^5		$ & $0.0825$ & $0.0225$& $ 0.0192$    \\
  $3\times 10^{-5}$    &  $78.70$    & $384\times 384\times 324$  &    $0.001$ &	    $31.56$&	$849.1$  &$ 849.1$& 	$1.58\times 10^5		$ &  $0.0496$ & $0.0215$& $ 0.0169$    \\
         \hline
       \end{tabular}
  \end{adjustbox} 
     \label{T:data1}
  \end{table}

  \begin{table}
 \begin{adjustbox}{addcode={\begin{minipage}{\width}}{\caption{%
      Details of the numerical simulations for $Pm = 1$, $E = (1\times 10^{-5},3\times 10^{-6},1\times 10^{-6})$ cases.  }\end{minipage}},rotate=90,center}
      \setlength\tabcolsep{6pt}
  \begin{tabular}{lcccccccccc}
  \hline
   $E$   &    $\Rat$   & $N_x \times N_y \times N_z$ &$\Delta t$  & $Nu$ & 			$Re$ &               $Rm$ &                $E_{mag}$  &	$\overline{E}_{mag}/E_{mag}$ &	$\lambda_u$ &$\lambda_B$\\
      \hline   
  $1\times 10^{-5}$    &  $11.33$    & $96\times 96\times 72$    &    $1$ &	    $ 1.45$&	$ 79.5 $ &$ 79.5 $& 	$4.1\times 10^3$ & $0.7616$ & $0.0177$& $ 0.0245$    \\
  $1\times 10^{-5}$    &  $14.82$    & $144\times 144\times 72$  &    $0.4$ &	    $ 2.07$&	$ 144.3$ &$ 144.3$& 	$1.1\times 10^4		$ & $0.4637$ & $0.0197$& $ 0.0243$    \\
  $1\times 10^{-5}$    &  $17.44$    & $144\times 144\times 72$  &    $0.2$ &	    $ 2.80$&	$ 200.7$ &$ 200.7$& 	$2.0\times 10^4		$ & $0.3320$ & $0.0201$& $ 0.0240$    \\
  $1\times 10^{-5}$    &  $21.80$    & $144\times 144\times 96$  &    $0.2$ &	    $ 5.86$&	$ 474.8$ &$474.8 $& 	$							$ &  $$ & $0.0213$& $ 	$    \\
  $1\times 10^{-5}$    &  $26.16$    & $192\times 192\times 108$  &    $0.1$ &	    $ 9.41$&	$ 646.8$ &$ 646.8$& 	$1.1\times 10^4		$ & $0.0111$ & $0.0212$& $ 0.156$    \\
  $1\times 10^{-5}$    &  $34.88$    & $192\times 192\times 144$  &    $0.015$ &    $15.09$&	$ 982.9$ &$982.9$& 	$2.2\times 10^4		$ & $0.0072$ & $0.0219$& $ 0.0129$    \\
  $1\times 10^{-5}$    &  $43.60$    & $216\times 216\times 216$  &    $0.006$ &    $20.19$&$ 1.11\times 10^3$ &$ 1.11\times 10^3$& 	$4.8\times 10^4		$ & $0.0067$ & $0.0196$& $ 0.0116$    \\
  $1\times 10^{-5}$    &  $61.03$    & $432\times 432\times 384$  &    $0.002$ &    $28.34$&$1.45\times 10^3$ &$1.45\times 10^3$& 	$6.1\times 10^4		$ & $0.0045$ & $0.0178$& $ 0.0099$    \\
  $1\times 10^{-5}$    &  $78.47$    & $432\times 432\times 480$  &    $0.0008$ &   $34.89$&$1.75\times 10^3$ &$ 1.75\times 10^3$& 	$9.7\times 10^4		$ & $0.0034$ & $0.0168$& $ 0.0089$    \\
  $3\times 10^{-6}$    &  $11.32$    & $96\times 96\times 96$  &       $0.5$    &   $ 1.43$&	$ 115.3$ &$ 115.3$& 	$1.0\times 10^4		$ & $0.7055$ & $0.0121$& $ 0.0161$    \\
  $3\times 10^{-6}$    & $14.80$     & $144\times 144\times 96$  &    $0.4$ &	    $ 1.97$&	$ 206.0$ &$ 206.0$& 	$1.8\times 10^4		$ & $0.4986$ & $0.0126$& $ 0.0165$    \\
  $3\times 10^{-6}$    & $17.41$     & $240\times 240\times 96$  &    $0.2$ &	    $ 2.67$&	$ 289.4$ &$ 289.4$& 	$3.1\times 10^4		$ & $0.3363$ & $0.0123$& $ 0.0153$    \\
  $3\times 10^{-6}$    & $21.77$     & $144\times 144\times 108$  &    $0.1$ &	    $ 5.38$&	$ 647.5$ &$647.5 $& 	$							$ & $0.0133$ & $0.0133$& $ 	$    \\
  $3\times 10^{-6}$    & $26.12$     & $192\times 192\times 144$  &    $0.1$ &	    $ 8.71$&	$ 893.5$ & $893.5$& 	$2.1\times 10^4		$ & $0.0102$ & $0.0137$& $ 0.0105$    \\
  $3\times 10^{-6}$    & $34.83$     & $288\times 288\times 288$  &    $0.01$ &	    $15.63$&   $ 1.31\times 10^3$ & $1.31\times 10^3$& 	$7.4\times 10^4		$ & $0.0057$ & $0.0136$& $ 0.0090$    \\
  $3\times 10^{-6}$    & $43.53$     & $288\times 288\times 384$  &    $0.006$ &	    $20.25$& $ 1.60\times 10^3$ &$ 1.60 \times 10^3$& 	$1.3 \times 10^5		$ & $0.0050$ & $0.0131$& $ 0.0082$    \\
  $3\times 10^{-6}$    & $60.94$     & $432\times 432\times 576$  &    $0.002$ &	    $31.19$&  $2.06\times 10^3$ &$2.06\times 10^3$& 	$1.9 \times 10^5		$ & $0.0037$ & $0.0114$& $ 0.0067$    \\
  $3\times 10^{-6}$    & $78.36$     & $432\times 432\times 648$  &    $0.0006$ &	    $41.06$&$2.56\times 10^3$ &$ 1.75\times 10^3$& 	$3.6\times 10^5		$ & $0.0055$ & $0.0109$& $ 0.0063$    \\
   $1\times 10^{-6}$   & $11.31$     & $96\times 96\times 96$  &    $0.5$ &	    $ 1.46$&	$ 166.6$ &$ 166.6$& 	$6.0 \times 10^4		$ & $0.3538$ & $0.0119$& $ 0.0163$    \\
  $1\times 10^{-6}$    &$14.79$      & $192\times 192\times 192$  &    $0.4$ &	    $ 1.93$&	$ 271.9 $ &$ 271.9$& 	$7.8\times 10^4		$ & $0.3047$ & $0.0099$& $ 0.0133$    \\
  $1\times 10^{-6}$    &$17.40$      & $288\times 288\times 192$  &    $0.1$ &	    $ 2.59$&	$ 396.8$ &$ 396.8$& 	$9.2\times 10^4		$ & $0.2393$ & $0.0095$& $ 0.0118$    \\
  $1\times 10^{-6}$    &$21.75$      & $144\times 144\times 192$  &    $0.05$ &	    $ 5.05$&	$ 874.4$ &$874.4$& 	$							$ & $0.0163$ & $0.0101$& $ 	$    \\
  $1\times 10^{-6}$    &$26.10$      & $192\times 192\times 144$  &    $0.1$ &	    $ 7.77$&	$1.23 \times 10^3$ &$   1.23\times 10^3$& 	$3.2 \times 10^4		$ & $0.0092$ & $0.0096$& $ 0.0072$    \\
  $1\times 10^{-6}$    &$34.80$      & $324\times 324\times 384$  &    $0.01$ &	    $14.80$&  $1.77\times 10^3$ &  $1.77\times 10^3$& 	$2.7\times 10^5		$ & $0.0064$ & $0.0093$& $ 0.0066$    \\
  $1\times 10^{-6}$    &$43.50$      & $432\times 432\times 432$  &    $0.006$ &	    $20.58 $& $ 2.06\times 10^3$ &$  2.06\times 10^3$& 	$4.6\times 10^5		$ & $0.0077$ & $0.0086$& $ 0.0061$    \\
  $1\times 10^{-6}$    &$60.90$      & $432\times 432\times 576$  &    $0.0005$ &	    $29.97$&$2.49\times 10^3$ &$2.49\times 10^3$& 	$6.3\times 10^5		$ & $0.0042$ & $0.0076$& $ 0.0053$    \\
  $1\times 10^{-6}$    &$78.31$      & $432\times 432\times 768$  &    $0.0005$ &	    $47.10$&$3.15\times 10^3$ &$ 1.75\times 10^3$& 	$8.3\times 10^5		$ & $0.0030$ & $0.0066$& $ 0.0044$    \\
         \hline
       \end{tabular}
  \end{adjustbox} 
     \label{T:data2}
  \end{table}

    \begin{table}
 \begin{adjustbox}{addcode={\begin{minipage}{\width}}{\caption{%
     Details of the numerical simulations for $Pm = 0.3$, $E = (1\times 10^{-6},3\times 10^{-7},1\times 10^{-7})$ cases.  }\end{minipage}},rotate=90,center}
      \setlength\tabcolsep{6pt}
  \begin{tabular}{lcccccccccc}
  \hline
      $E$   &    $\Rat$   & $N_x \times N_y \times N_z$ &$\Delta t$  & $Nu$ & 			$Re$ &               $Rm$ &                $E_{mag}$  &	$\overline{E}_{mag}/E_{mag}$ &	$\lambda_u$ &$\lambda_B$\\
      \hline   
 $1\times 10^{-6}$    &      $11.31$    & $96\times 96\times 96$  &    $8$ &	    $ 1.57$&	$ 194.7$ &$ 58.4$& 	$						$ & $$ & $0.0077$& $ $    \\
 $1\times 10^{-6}$    & $14.79$    & $96\times 96\times 96$  &    $2$ &	    $2.24$&	$ 331.5$ &$  99.5$& 	$5.8\times 10^4		$ & $0.8667$ & $0.0085$& $ 0.0129$    \\
  $1\times 10^{-6}$    &$17.40$    & $96\times 96\times 96$  &    $1$ &	    $ 2.88$&	$ 449.0$ &$ 134.7$& 	$6.5\times 10^4		$ & $0.7808$ & $0.0085$& $ 0.0125$    \\
  $1\times 10^{-6}$    &$21.75$    & $96\times 96\times 96$  &    $0.5$ &	    $4.41$&	$ 681.6$ &$204.5$& 	$8.5\times 10^4		$ & $0.5493$ & $0.0088$& $ 0.0132	$    \\
  $1\times 10^{-6}$    &$30.45$    & $96\times 96\times 162$  &    $0.1$ &	    $ 10.52$&	$1.89\times 10^3$ &$   566.2$& 	$3.1\times 10^4		$ & $0.0343$ & $0.0114$& $ 0.0118$    \\
 $1\times 10^{-6}$    & $43.50$    & $192\times 192\times 240$  &    $0.01$ &	    $18.62$&$ 2.84\times 10^3$ &$852.9 $& 	$1.6\times 10^5		$ & $0.0232$ & $0.0108$& $ 0.0102$    \\
  $1\times 10^{-6}$    &$60.90$    & $216\times 216\times 432$  &    $0.005$ &	    $30.64$&$3.16\times 10^3$ &$948.9$& 	$4.9\times 10^5		$ & $0.0320$ & $0.0084$& $ 0.0090$    \\
  $1\times 10^{-6}$    &$78.31$    & $288\times 288\times 576$  &    $0.003$ &	    $40.39$&$4.17\times 10^3$ &$ 1.25\times 10^3$& 	$6.9\times 10^5		$ & $0.0205$ & $0.0084$& $ 0.0079$    \\
  $3\times 10^{-7}$    &       $11.31$    & $96\times 96\times 96$  &    $8$ &	    $ 1.53$&	$ 273.5$ &$ 82.0$& 	$	7.2\times 10^4		$ & $0.9563$ & $0.0053$& $ 0.0083$    \\
   $3\times 10^{-7}$    &$14.79$    & $144\times 144\times 144$  &    $2$ &	    $2.20$&	$ 490.0$ &$ 147.0$& 	$1.30\times 10^5		$ & $0.8837$ & $0.0056$& $ 0.0087$    \\
   $3\times 10^{-7}$    &$17.40$    & $144\times 144\times 144$  &    $1$ &	    $ 2.83$&	$ 661.1$ &$ 198.3$& 	$1.39\times 10^5		$ & $0.8037$ & $0.0057$& $ 0.0085$    \\
   $3\times 10^{-7}$    &$21.74$    & $96\times 96\times 192$  &    $0.5$ &	    $4.24$&	$ 1.00\times 10^3$ &$300.5$& 	$1.7\times 10^5		$ & $0.5651$ & $0.0059$& $ 0.0089	$    \\
   $3\times 10^{-7}$    &$30.44$    & $96\times 96\times 216$  &    $0.2$ &	    $ 9.59$&	$  2.28\times 10^3$ &$   683.9$& 	$8.3\times 10^4		$ & $0.0495$ & $0.0065$& $ 0.0084$    \\
   $3\times 10^{-7}$    &$43.49$    & $192\times 192\times 288$  &    $0.06$ &	    $17.67$&$ 3.94\times 10^3$ &$1.18\times 10^3$& 	$3.6\times 10^5		$ & $0.0274$ & $0.0069$& $ 0.0070$    \\
   $3\times 10^{-7}$    &$60.89$    & $216\times 216\times 480$  &    $0.005$ &	    $27.87$&$4.37\times 10^3$ &$1.31\times 10^3$& 	$9.5\times 10^5		$ & $0.0357$ & $0.0054$& $ 0.0064$    \\
   $3\times 10^{-7}$    &$78.28$    & $288\times 288\times 576$  &    $0.003$ &	    $41.72$&$5.05\times 10^3$ &$ 1.51\times 10^3$& 	$1.5\times 10^6		$ & $0.0320$ & $0.0048$& $ 0.0058$    \\
    $1\times 10^{-7}$    &      $11.31$    & $96\times 96\times 240$  &    $4$ &	    $ 1.53$&	$ 402.1$ &$ 120.6$& 	$	2.0\times 10^5		$ & $0.9520$ & $0.0037$& $ 0.0073$    \\
   $1\times 10^{-7}$    &$14.78$    & $96\times 96\times 240$  &    $4$ &	    $2.19$&	$ 703.3$ &$ 211.0$& 	$2.6\times 10^5		$ & $0.9009$ & $0.0039$& $ 0.0059$    \\
   $1\times 10^{-7}$    &$17.39$    & $96\times 96\times 240$  &    $3$ &	    $ 2.83$&	$ 954.9$ &$ 286.5$& 	$2.6\times 10^5		$ & $0.8142$ & $0.0039$& $ 0.0058$    \\
    $1\times 10^{-7}$    & $21.74$    & $96\times 96\times 240$  &    $1$ &   $ 4.21$&	$ 1.45\times 10^3$ &$ 434.3$& 	$3.1\times 10^5		$ & $0.5633$ & $0.0041$& $ 0.0062$    \\
   $1\times 10^{-7}$    &$26.09$    & $96\times 96\times 240$  &    $0.5$ &	    $6.23$&	$  2.07\times 10^3$ &$620.2$& 	$4.0\times 10^5		$ & $0.3330$ & $0.0041$& $ 0.0063	$    \\
   $1\times 10^{-7}$    &$30.44$    & $96\times 96\times 288$  &    $0.2$ &	    $ 8.85$&	$ 2.75 \times 10^3$ &$   825.5$& 	$5.5\times 10^5		$ & $0.1838$ & $0.0042$& $ 0.0061$    \\
   $1\times 10^{-7}$    &$43.48$    & $192\times 192\times 432$  &    $0.05$ &	    $16.08$&$ 4.51\times 10^3$ &$1.35\times 10^3$& 	$1.3 \times 10^6		$ & $0.0786$ & $0.0040$& $ 0.0053$    \\
   $1\times 10^{-7}$    &$60.88$    & $216\times 216\times 540$  &    $0.02$ &	    $26.50$&$5.85\times 10^3$ &$1.75\times 10^3$& 	$2.3\times 10^6		$ & $0.0423$ & $0.0036$& $ 0.0045$    \\
   $1\times 10^{-7}$    &$78.27$    & $288\times 288\times 648$  &    $0.01$ &	    $37.55$&$7.25\times 10^3$ &$ 2.17 \times 10^3$& 	$2.58\times 10^6		$ & $0.0290$ & $0.0035$& $ 0.0041$    \\
         \hline
       \end{tabular}
  \end{adjustbox} 
     \label{T:data3}
  \end{table}

        \begin{table}
 \begin{adjustbox}{addcode={\begin{minipage}{\width}}{\caption{%
      Details of the numerical simulations for $Pm = 0.2$, $E = 3\times 10^{-8}$ cases, $Pm = 0.1$, $E = 1\times 10^{-8}$ cases and $Pm = 0.05$, $E = 1\times 10^{-8}$ cases.  }\end{minipage}},rotate=90,center}
      \setlength\tabcolsep{5pt}
  \begin{tabular}{lccccccccccc}
  \hline
        $Pm$   &  $E$   &  $\Rat$   & $N_x \times N_y \times N_z$ &$\Delta t$  & $Nu$ & 			$Re$ &               $Rm$ &                $E_{mag}$  &	$\overline{E}_{mag}/E_{mag}$ &	$\lambda_u$ &$\lambda_B$\\
      \hline   
    $0.2$    &$3\times 10^{-8}$    &   $11.31$    & $96\times 96\times 288$  &    $20$ &	    $ 1.56$&	$ 603.4$ &$ 120.7$& 	$	4.8 \times 10^5		$ & $0.9795$ & $0.0024$& $ 0.0040$    \\
   $0.2$    &$3\times 10^{-8}$      &$14.78$    & $96\times 96\times 288$  &    $4$ &	    $2.32 $&	$ 1.11 \times 10^3$ &$ 222.0 $& 	$8.9 \times 10^5		$ & $0.9486$ & $0.0026$& $ 0.0042$    \\
   $0.2$    &$3\times 10^{-8}$    &$17.39$    & $96\times 96\times 288$  &    $4$ &	    $ 2.95$&	$ 1.47 \times 10^3$ &$ 294.1 $& 	$8.3 \times 10^5		$ & $0.9029$ & $0.0026$& $ 0.0041$    \\
     $0.2$    &$3\times 10^{-8}$    &$21.74$    & $96\times 96\times 384$  &    $2$ &   $ 4.27 $&	$ 2.18 \times 10^3$ &$ 435.0 $& 	$7.90 \times 10^5		$ & $0.7349$ & $0.0027$& $ 0.0043$    \\
    $0.2$    &$3\times 10^{-8}$    &$30.44$    & $96\times 96\times 384$  &    $0.6$ &	    $ 8.48$&	$  4.02 \times 10^3$ &$   803.3$& 	$1.1\times 10^6		$ & $0.3238$ & $0.0028$& $ 0.0045$    \\
   $0.2$    &$3\times 10^{-8}$    &$43.48$    & $216\times 216\times 480$  &    $0.06$ &	    $15.77 $&$ 7.02 \times 10^3$ &$1.40\times 10^3$& 	$2.8 \times 10^6		$ & $0.1114$ & $0.0028$& $ 0.0041$    \\
   $0.2$    &$3\times 10^{-8}$    &$60.87$    & $192\times 192\times 576$  &    $0.01$ &	    $23.80$&$9.37\times 10^3$ &$1.87\times 10^3$& 	$4.0\times 10^6		$ & $0.0665$ & $0.0026$& $ 0.0034$    \\
   $0.2$    &$3\times 10^{-8}$    &$78.27$    & $288\times 288\times 768$  &    $0.005$ &	    $36.44$&$1.11\times 10^4$ &$ 2.22\times 10^3$& 	$7.9\times 10^6		$ & $0.0355$ & $0.0023$& $ 0.0033$    \\
   $0.1$    &$1\times 10^{-8}$    &      $11.30$    & $96\times 96\times 288$  &    $5$ &	    $ 1.574$&	$ 926.1$ &$ 92.6$& 	$						$ & $$ & $0.0015$& $  $    \\
  $0.1$    &$1\times 10^{-8}$    &$14.78$    & $96\times 96\times 384$  &    $5$ &	    $2.48$&	$ 1.67\times 10^3$ &$ 167.0$& 	$2.9\times 10^6		$ & $0.9781$ & $0.0018$& $ 0.0029$    \\
  $0.1$    &$1\times 10^{-8}$    &$17.39$    & $96\times 96\times 384$  &    $5$ &	    $ 3.16$&	$ 2.24\times 10^3$ &$ 223.8$& 	$3.5\times 10^6		$ & $0.9648$ & $0.0018$& $ 0.0030$    \\
    $0.1$    &$1\times 10^{-8}$    &$21.74$    & $96\times 96\times 384$  &    $2.5$ &   $ 4.47$&	$ 3.22\times 10^3$ &$ 322.0$& 	$3.1\times 10^6		$ & $0.9114$ & $0.0019$& $ 0.0030$    \\
    $0.1$    &$1\times 10^{-8}$    &$30.44$    & $192\times 192\times 540$  &    $0.5$ &	    $ 8.08$&	$  5.43\times 10^3$ &$   543.4$& 	$3.3\times 10^6		$ & $0.6953$ & $0.0019$& $ 0.0033$    \\
 $0.1$    &$1\times 10^{-8}$    & $43.48$    & $192\times 192\times 540$  &    $0.15$ &	    $16.63$&$ 1.02\times 10^4$ &$1.02\times 10^3$& 	$6.2\times 10^6		$ & $0.3206$ & $0.0019$& $ 0.0034$    \\
 $0.1$    &$1\times 10^{-8}$    &$60.87$    & $216\times 216\times 540$  &    $0.04$ &	    $24.34$&$1.28\times 10^4$ &$1.28\times 10^3$& 	$8.7\times 10^6		$ & $0.2111$ & $0.0017$& $ 0.0031$    \\
  $0.1$    &$1\times 10^{-8}$    &$78.26$    & $288\times 288\times 768$  &    $0.04$ &	    $33.73$&$1.46\times 10^4$ &$ 1.46\times 10^3$& 	$1.0\times 10^7		$ & $0.1821$ & $0.0016$& $ 0.0029$    \\
    $0.05$    &$1\times 10^{-8}$    &  $14.78$    & $96\times 96\times 384$  &    $6$ &	    $2.40$&	$1.76\times 10^3$ &$ 87.8\pm 5.4$& 	$  		$ & $$ & $0.0018$& $  $    \\
   $0.05$    &$1\times 10^{-8}$    &$17.39$    & $96\times 96\times 540$  &    $5$ &	    $ 3.19$&	$ 2.52\times 10^3$ &$ 125.7$& 	$ 		$ & $$ & $0.0019$& $  $    \\
    $0.05$    &$1\times 10^{-8}$    & $21.74$    & $96\times 96\times 540$  &    $2$ &   $ 4.73$&	$ 3.96\times 10^3$ &$ 197.9$& 	$ 		$ & $$ & $0.0020$& $  $    \\
    $0.05$    &$1\times 10^{-8}$    & $30.44$    & $96\times 96\times 540$  &    $1$ &	    $ 8.72$&	$  5.69\times 10^3$ &$   284.6$& 	$5.4\times 10^6		$ & $0.8772$ & $0.0019$& $ 0.0034$    \\
 $0.05$    &$1\times 10^{-8}$    &  $43.48$    & $216\times 216\times 576$  &    $0.1$ &	    $16.63$&$ 1.08\times 10^4$ &$ 537.9$& 	$7.4\times 10^6		$ & $0.5926$ & $0.0019$& $ 0.0036$    \\
  $0.05$    &$1\times 10^{-8}$    & $60.87$    & $216\times 216\times 648$  &    $0.015$ &	    $23.38$&$1.31\times 10^4$ &$ 653.4$& 	$9.0\times 10^6		$ & $0.4477$ & $0.0017$& $ 0.0036$    \\
  $0.05$    &$1\times 10^{-8}$    & $78.26$    & $288\times 288\times 864$  &    $0.01$ &	    $32.17$&$1.53\times 10^4$ &$ 765.1$& 	$1.1\times 10^7		$ & $0.3665$ & $0.0016$& $ 0.0035$    \\
         \hline
       \end{tabular}
  \end{adjustbox} 
   \label{T:data4}
  \end{table}

\end{document}